\documentclass[letterpaper]{article} 
\usepackage{aaai2026}  
\usepackage{times}  
\usepackage{helvet}  
\usepackage{courier}  
\usepackage[hyphens]{url}  
\usepackage{graphicx} 
\usepackage{natbib}  
\usepackage{caption} 
\usepackage{algorithm}
\usepackage{algorithmic}

\usepackage{newfloat}
\usepackage{listings}
\DeclareCaptionStyle{ruled}{labelfont=normalfont,labelsep=colon,strut=off} 
\floatstyle{ruled}
\newfloat{listing}{tb}{lst}{}
\floatname{listing}{Listing}
\usepackage{amssymb}
\usepackage{booktabs}
\usepackage{makecell}
\usepackage{multirow}
\usepackage{array}
\usepackage{bbding}
\usepackage{subcaption}
\usepackage{enumitem}
\usepackage{xurl}
\usepackage[most]{tcolorbox}
\usepackage{threeparttable}

\title{DeepTracer: Tracing Stolen Model via Deep Coupled Watermarks}
\author{
    Yunfei Yang\textsuperscript{\rm 1,2,3}, Xiaojun Chen\textsuperscript{\rm 1,2,3}\thanks{Corresponding author. This is the extended version of the paper accepted by AAAI 2026.}, Yuexin Xuan\textsuperscript{\rm 4}, Zhendong Zhao\textsuperscript{\rm 1,2}, Xin Zhao\textsuperscript{\rm 1,2,3}, He Li\textsuperscript{\rm 1,2,3}
}
\affiliations{
    \textsuperscript{\rm 1}Institute of Information Engineering, Chinese Academy of Sciences, Beijing, China\\
    \textsuperscript{\rm 2}State Key Laboratory of Cyberspace Security Defense, Beijing, China\\
    \textsuperscript{\rm 3}School of Cyber Security, University of Chinese Academy of Sciences, Beijing, China\\
    \textsuperscript{\rm 4}PetroChina (Beijing) Digital Intelligent Research Institute Co., Ltd., Beijing, China

    \{yangyunfei, chenxiaojun, zhaozhendong, zhaoxin, lihe2023\}@iie.ac.cn, xuanyuexin@petrochina.com.cn
}

\usepackage{bibentry}

\begin{document}

\maketitle

\begin{abstract}
Model watermarking techniques can embed watermark information into the protected model for ownership declaration by constructing specific input-output pairs. However, existing watermarks are easily removed when facing model stealing attacks, and make it difficult for model owners to effectively verify the copyright of stolen models. In this paper, we analyze the root cause of the failure of current watermarking methods under model stealing scenarios and then explore potential solutions. Specifically, we introduce a robust watermarking framework, DeepTracer, which leverages a novel watermark samples construction method and a same-class coupling loss constraint. DeepTracer can incur a high-coupling model between watermark task and primary task that makes adversaries inevitably learn the hidden watermark task when stealing the primary task functionality. Furthermore, we propose an effective watermark samples filtering mechanism that elaborately select watermark key samples used in model ownership verification to enhance the reliability of watermarks. Extensive experiments across multiple datasets and models demonstrate that our method surpasses existing approaches in defending against various model stealing attacks, as well as watermark attacks, and achieves new state-of-the-art effectiveness and robustness.
\end{abstract}

\begin{links}
    \link{Code}{https://github.com/yangyunfei16/DeepTracer}
\end{links}

\section{Introduction}
Deep learning has been widely adopted to solve real-world problems across various fields. To democratize its use, many companies provide Machine Learning as a Service (MLaaS) by deploying models on the cloud \cite{grigoriadis2023machine}. However, training high-performance models requires extensive data, expert design, and costly computation, making these models valuable intellectual property. In practice, they face two main threats: \textit{external attacks} (e.g., query-based model stealing \cite{papernot2017practical,orekondy2019knockoff,truong2021data,Rosenthal2023DisGUIDEDD}) and \textit{internal attacks} by insiders \cite{he2020towards}, both aiming to replicate the victim model.

These threats have driven research on model copyright protection, where model watermarking \cite{li2021survey} has become a mainstream solution. Watermarks are embedded into a model’s internals or behavior and are verified via \textit{white-box} or \textit{black-box} methods. White-box approaches \cite{uchida2017embedding,chen2019deepmarks,zhao2021structural,xie2021deepmark} rely on internal access, which is often impractical. Black-box watermarking \cite{adi2018turning,jia2021entangled,tan2023deep,lv2024mea}, more practical in cloud scenarios, verifies ownership by querying the suspect model with special watermark samples and observing outputs. Our work follows this black-box paradigm.

While many black-box techniques \cite{zhang2018protecting,adi2018turning,jia2021entangled,kim2023margin,lv2024mea} achieve high watermark success on the original model, they often fail on stolen models. Existing methods fall into \textit{out-of-distribution (OOD)} \cite{zhang2018protecting,adi2018turning,jia2021entangled} and \textit{in-distribution (ID)} \cite{kim2023margin,lv2024mea} watermarking. OOD methods craft watermark samples with artificial patterns or pixel blocks, making them difficult for attackers to replicate during model stealing, but also hard for stolen models to retain. In contrast, ID watermarking uses primary task samples, improving retention in stolen models. Though Margin-based \cite{kim2023margin} and MEA-Defender \cite{lv2024mea} improve ID watermarking, their watermark samples are still weakly coupled with primary task features, and no prior work addresses optimizing this coupling through sample selection and constraint optimization. As a result, watermarks degrade under stronger attacks like hard-label, multi-class, and data-free stealing.

To overcome these limitations, we propose DeepTracer, a robust black-box watermarking framework based on deep coupled watermarking. It strengthens the coupling between primary and watermark tasks through both sample design and loss functions. (1) We construct watermark samples by selecting and combining classes that broadly span the primary feature space, ensuring the watermark task distribution is a subset of the primary task distribution. This forces stolen models to learn the watermark task. (2) We design a same-class coupling loss to align watermark and target class samples in output space without harming accuracy. (3) A two-stage watermark sample filtering process further enhances watermark success. In summary, \textbf{our contributions} are threefold:
\begin{itemize}[itemsep=2pt,topsep=0pt,parsep=0pt]
    \item \textbf{Analysis of the Vulnerability of Watermarking.} We systematically analyze the reasons behind the poor robustness of existing watermarking methods under model stealing attacks and find that the root cause is the independence of the \textit{primary task distribution} and the \textit{watermark task distribution}.
    \item \textbf{Novel Robust Watermarking Framework.} We introduce DeepTracer, a robust watermarking framework comprising four stages: watermark samples construction, coupled watermark embedding, watermark key samples generation and model ownership verification. By carefully designing the watermark samples and embedding loss, we achieve high coupling from the feature to the output space, which enhances the robustness of our watermark. The two-stage watermark sample filtering mechanism further selects the most reliable key samples for subsequent ownership verification.
    \item \textbf{Systematic and Comprehensive Evaluation.} Extensive experiments demonstrate that our method outperforms previous approaches across various tasks in defending against model stealing attacks. Moreover, it also shows superior robustness against popular watermark removal attacks, detection attacks, and adaptive attacks.
\end{itemize}

\section{Related Work}
\subsection{Model Stealing Attacks}
Our work concerns on the scenario where adversary stealing the functionality of victim model. Based on the query data type, existing attacks are classified as: \textit{seed sample-based}, \textit{substitute data-based}, and \textit{data-free}.

\textbf{Seed Sample-Based.} These attacks begin with a small subset of training data (seed set) and expand it via adversarial augmentation. JBDA \cite{papernot2017practical} uses Jacobian-based data augmentation to obtain more queries, enabling effective transferability in stolen models.

\textbf{Substitute Data-Based.} Here, public natural data serve as queries. Knockoff \cite{orekondy2019knockoff} uses reinforcement learning to select a transfer set from a large data pool. ActiveThief \cite{pal2020activethief} employs active learning, and MExMI \cite{xiao2022mexmi} combines model stealing attack with membership inference attack to improve performance.

\textbf{Data-Free.} Without access to real data, adversaries train a generator (e.g., GAN) to synthesize queries, jointly optimizing both the generator and stolen model. DFME \cite{truong2021data} and MAZE \cite{kariyappa2021maze} use gradient estimation to update the generator. DFMS-HL \cite{sanyal2022towards} enhances diversity and performs well under hard-label settings. Recent methods \cite{Rosenthal2023DisGUIDEDD,beetham2022dual} introduce dual-model strategies to reduce query costs.

\subsection{Black-Box Watermarking}
Existing black-box watermarking methods \cite{zhang2018protecting,adi2018turning,jia2021entangled,kim2023margin,lv2024mea} embed watermarks by mixing labeled trigger samples into training data, allowing ownership verification via model outputs on these samples. Given their practicality, black-box approaches have gained popularity. They can be divided into \textit{out-of-distribution} and \textit{in-distribution} watermarking, depending on whether watermark features align with the primary task distribution.

\textbf{Out-Of-Distribution Watermarking.} This watermarking uses features disjoint from the primary task. Abstract \cite{adi2018turning} uses abstract art images mapped to target labels. Zhang et al. \cite{zhang2018protecting} propose Content, Noise, and Unrelated sample constructions. However, these methods primarily address internal threats and are vulnerable to model stealing. EWE \cite{jia2021entangled} improves robustness by entangling watermark and primary task samples using soft nearest neighbor loss but suffers from degraded task performance and suboptimal watermark success rate.

\textbf{In-Distribution Watermarking.} The watermark sample features come from the sample features of primary task. Margin-based watermarking \cite{kim2023margin} randomly relabels original samples and pushes them away from decision boundaries to preserve label prediction, but suffers from slow training and impractical query assumptions. Composite \cite{lin2020composite} blends features from two classes to form trigger samples, and MEA-Defender \cite{lv2024mea} builds on it by designing symbiotic watermarks and aligning output distributions. While effective, MEA-Defender struggles under hard-label settings and suffers from conflicting objectives between watermark and verification losses.

Building upon prior works, we propose an innovative watermarking method that maximizes the coupling between the watermark and primary tasks from feature to output space, which makes it difficult for stolen model to avoid learning watermark task, and achieves superior performance even in hard-label scenarios.

\section{Why Do Stolen Models Forget Watermarks?}
\label{sec: analysis of watermark survival}
Over-parameterization \cite{zou2019improved} describes neural networks with more parameters than training samples, which surprisingly generalize well. Such networks adapt to different tasks by activating distinct neuron regions \cite{jia2021entangled}. Previous watermarking methods introduce external features, making watermark tasks out-of-distribution (OOD) relative to the primary task. Although over-parameterized networks can fit both tasks, the stolen model—trained on adversary queries resembling the original data—tends to forget OOD watermark tasks, causing verification failure.

\begin{figure}[t]
    \centering
    \begin{subfigure}{0.48\textwidth}
        \centering
        \includegraphics[width=1.0\textwidth]{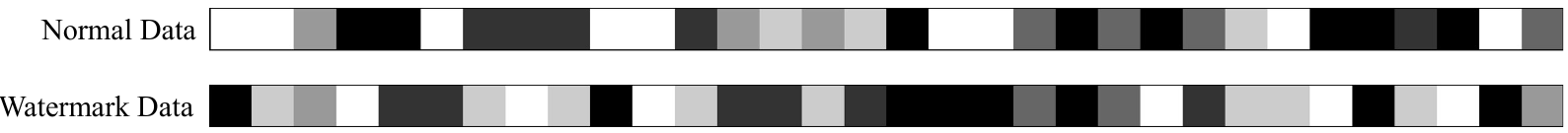}
        \vspace{-0.5cm}
        \caption{Abstract \cite{adi2018turning}}
        \label{fig: activation_Abstract}
    \end{subfigure}
    \vspace{-0.2cm}
    
    \begin{subfigure}{0.48\textwidth}
        \centering
        \includegraphics[width=1.0\textwidth]{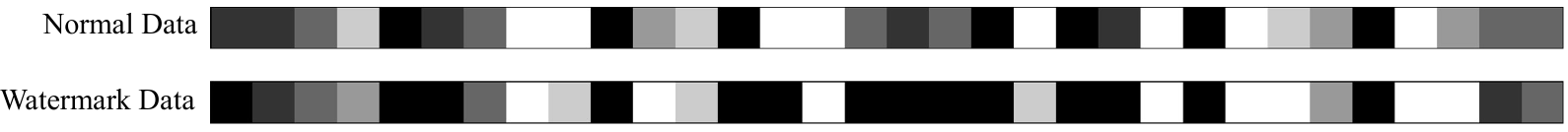}
        \vspace{-0.5cm}
        \caption{MEA-Defender \cite{lv2024mea}}
        \label{fig: activation_MEA-Defender}
    \end{subfigure}
    \vspace{-0.2cm}
    
    \begin{subfigure}{0.48\textwidth}
        \centering
        \includegraphics[width=1.0\textwidth]{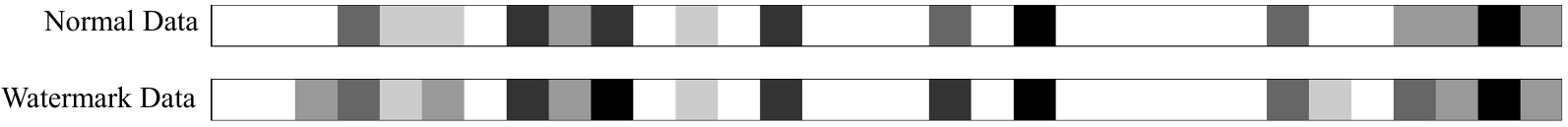}
        \vspace{-0.5cm}
        \caption{DeepTracer (Ours)}
        \label{fig: activation_DeepTracer}
    \end{subfigure}
    \vspace{-0.6cm}
    \caption{Heatmap of activation within the neural network for different watermarking methods. Lighter colors indicate greater activation.}
    \label{fig: heatmap of activation}
    \vspace{-0.3cm}
\end{figure}

To validate that OOD watermarks activate separate neuron regions, we use Abstract \cite{adi2018turning} to embed watermarks into a VGG-like model \cite{lin2020composite} and visualize activations (Figure \ref{fig: activation_Abstract}). Results confirm disjoint activation between clean and watermark samples.

\begin{figure*}[t]
\vspace{-0.0cm}
\centering
\includegraphics[width=0.85\textwidth]{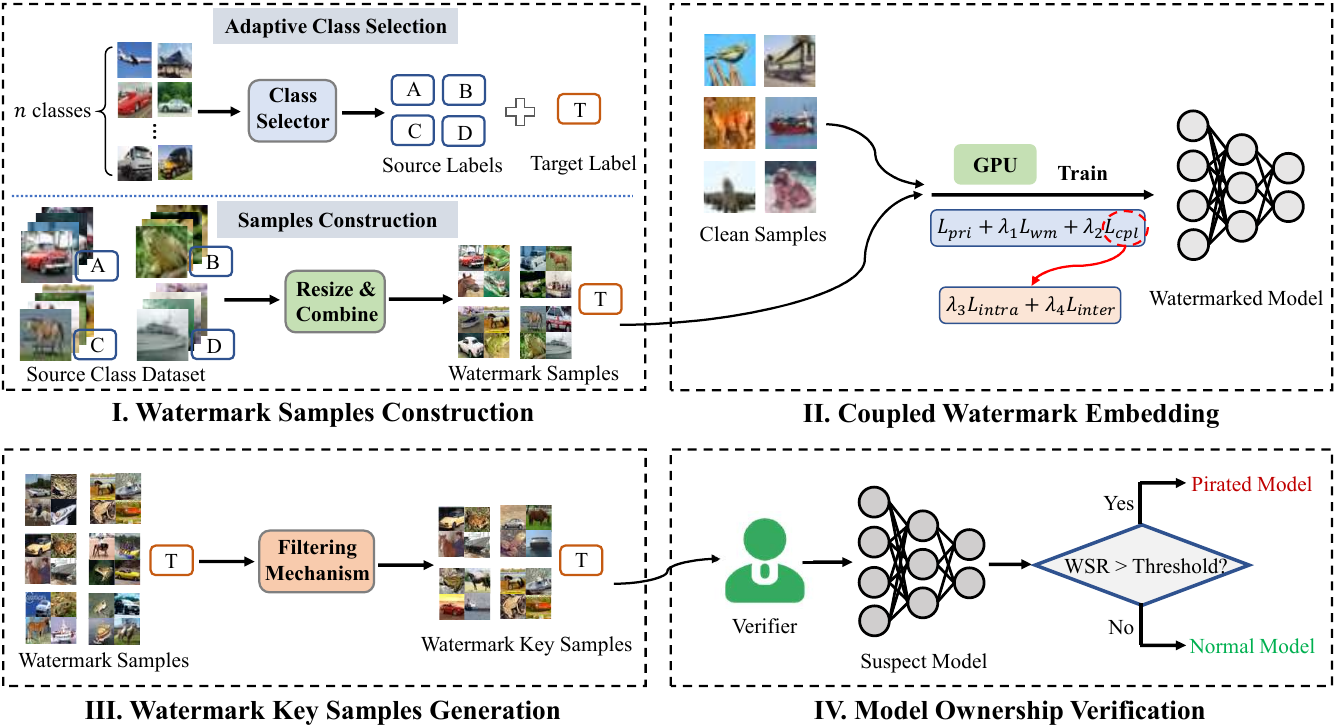}
\vspace{-0.2cm}
\caption{Overview of DeepTracer. The model owner first adaptively selects four source classes and one target label, and then constructs watermark samples and mixes them into the normal dataset for model training. Next, the owner generates a filtered key samples set for the watermarked model and saves it, which is used for future ownership verification of suspect models.}
\label{fig: overview of DeepTracer}
\vspace{-0.3cm}
\end{figure*}

To ensure watermark persistence in stolen models, it is crucial to increase the coupling between the watermark and primary tasks. We propose four ways to improve it: (1) sampling watermark features from the primary distribution, (2) preserving primary features in watermark construction, (3) selecting representative source classes, and (4) enhancing same-label coupling during training. Their detailed analysis is in our Appendix. Our approach, DeepTracer, integrates these strategies. Figure \ref{fig: activation_DeepTracer} shows our watermark samples activate nearly identical neurons as clean samples, outperforming MEA-Defender \cite{lv2024mea} (Figure \ref{fig: activation_MEA-Defender}).

\section{Our Proposed DeepTracer}
\label{sec: proposed method}
\subsection{Threat Model}
\textbf{Adversary.} The adversary is unaware of the victim model's architecture and training data but knowledgeable about its task domain. They can train a stolen model using publicly accessible architectures and data via any known model stealing method, achieving near-original performance within limited queries. These capabilities are consistent with prior work \cite{jia2021entangled,kim2023margin,lv2024mea}.

\textbf{Defender.} The model owner uses proprietary data and advanced training with watermark embedding to create a protected (victim) model. They do not know the attacker’s strategy and must robustly embed watermarks with minimal impact on task performance to ensure successful watermark verification on both the victim and stolen models.

\subsection{Overview}
Figure \ref{fig: overview of DeepTracer} illustrates our framework, consisting of four key stages:

$\bullet$ \textit{\textbf{Watermark Samples Construction.}} We adaptively select four source classes that broadly cover the primary feature space. Samples from these classes are resized and combined, then labeled with a target class (the one with lowest prediction probability by a benign model). This design ensures the watermark distribution is embedded within the primary task distribution.

$\bullet$ \textit{\textbf{Coupled Watermark Embedding.}} While using normal training samples to maintain primary task performance, we design a new watermark embedding loss to strengthen the coupling between watermark and primary task samples. This loss includes a standard watermark classification loss $L_{wm}$ to ensure watermark samples are classified as the target label, and a novel same-class coupling loss $L_{cpl}$, comprising intra-class loss $L_{intra}$ and inter-class loss $L_{inter}$, to enhance label-based coupling.

$\bullet$ \textit{\textbf{Watermark Key Samples Generation.}} A two-tier filtering mechanism is employed: First, we select samples that pass watermark verification on both the victim model and a substitute model (simulated via standard model stealing) but fail on the benign model. Then, we choose the top $M$ samples from this set that maximize the target label classification probability by the substitute model as the final key samples set $S_{K}$ for ownership verification.

$\bullet$ \textit{\textbf{Model Ownership Verification.}} The verifier queries the suspect model with $S_K$ under black-box constraints. If the classification accuracy exceeds a predefined threshold, the model is considered pirated; otherwise, it is deemed normal.

\subsection{Watermark Samples Construction}
\label{sec: watermark samples construction}
To align the \textit{watermark task distribution} with the \textit{primary task distribution}, we select four source classes from the primary task and resize their samples to one-fourth of the original size before combining them, keeping the final watermark sample size unchanged for network input.

Due to the high feature similarity between watermark and primary samples, the model learns both tasks jointly. This coupling ensures that watermark functionality is preserved when the model is stolen. We also find that strategically selecting source classes and target labels outperforms random choices.

\subsubsection{Source Classes Selection}
To ensure broad coverage of primary task distribution, we propose an adaptive source class selection strategy. Specifically, we cluster training sample features and select the class nearest to each cluster center as a representative. The detailed procedure is:

\textbf{(1) Feature Extraction and Class Centroid Calculation.}
First, for each sample $x_{i}$ (with label $y_{i}$), we extract its feature vector using a pre-trained benign model. Then, we calculate the feature centroid $c_{j}$ for each class $j$. Assuming class $j$ has $N_{j}$ samples, the feature centroid is calculated as follows:
\begin{equation}
c_{j}=\frac{1}{N_{j}}\sum\limits_{i=1}^{N_{j}}f_{i}^{j},
\end{equation}
where $f_{i}^{j}$ is the feature vector of sample $x_{i}$ in class $j$.

\textbf{(2) K-Means Clustering.}
Next, we apply K-Means algorithm to cluster all class centroids $c_{j}$ into $K$ clusters (in our method, $K\!=\!4$), each with a centroid $m_{k}$ (where $k$ is cluster index). The objective of clustering is to minimize the following objective function:
\begin{equation}
\min\limits_{M}\sum\limits_{k=1}^{K}\sum\limits_{c_{j} \in C_{k}}\left \| c_{j}-m_{k} \right \|_{2}^{2},
\end{equation}
where $M$ is the set of all cluster centroids, and $C_{k}$ is the set of class centroids in cluster $k$.

\textbf{(3) Selecting Classes Closest to Cluster Centers.}
For each cluster center $m_{k}$ generated by K-Means method, we first calculate the Euclidean distance between all class centroids $c_{j}$ and their cluster centroids, and then select the class with the smallest distance as the representative class of the corresponding cluster. Specifically, the formula is $j^{*} = \mathop{\arg\min}\limits_{j}\left \| c_{j}-m_{k} \right \|_{2}$, where $j^{*}$ is the label of the class closest to cluster center $m_{k}$.

\subsubsection{Target Label Selection}
To avoid false positives, such as watermark detection in independently trained models using datasets with the same distribution as the primary task of the victim model, we set the watermark target label $y^{w}$ as the least likely class for the watermark samples set $X^{w}=\{x_{1},x_{2},...,x_{n}\}$ when classified by the benign model $F_{B}$. The detailed process is as follows:

First, for each watermark sample $x_{i}$, we compute its prediction probability $p_{i}$ using the benign model $F_{B}$:
\begin{equation}
p_{i} = softmax(F_{B}(x_{i})),
\end{equation}
where $softmax(\cdot)$ is the softmax activation function that normalizes the logits to a probability vector summing to one.

Next, we compute the average probability $P^{j}$ for each class $j$ across all $n$ watermark samples:
\begin{equation}
P^{j} = \frac{1}{n}\sum\limits_{i=1}^{n}p_{i}^{j}, \forall j \in \{1,2,...,C\}.
\end{equation}

Finally, we identify the class with the lowest average probability as the watermark target label, i.e., $y^{w} = \mathop{\arg\min}\limits_{j}P^{j}$.

With the source classes and target label as well as combination method established, we can construct sufficient watermark samples for watermark embedding and verification.

\subsection{Coupled Watermark Embedding}
To strengthen output-space coupling, we introduce the same-class coupling loss $L_{cpl}$, which promotes intra-class compactness and inter-class separation using class centroids for efficient and stable optimization. This ensures strong alignment between watermark and target outputs without harming task accuracy.

Specifically, $L_{intra}$ minimizes distances to the corresponding class centroid, while $L_{inter}$ maximizes distances to other class centroids. Mathematically, they are
\begin{equation}
\begin{aligned}
L_{intra} =& \frac{1}{N}\sum\limits_{i=1}^{N} \left \| f_{i}-c_{y_{i}} \right \|_{2}^{2}, \\
L_{inter}\!=\!\frac{1}{N}\!\sum\limits_{i=1}^{N}\sum\limits_{j=1,j\neq y_{i}}^{C}&\!\max(0, margin\!-\!\left \| f_{i}\!-\!c_{j} \right \|_{2})^{2},
\end{aligned}
\end{equation}
where $N$ and $C$ are the number of samples and the number of classes, respectively. $f_{i}$ is output feature vector of sample $x_{i}$ at the last layer of model, $c_{y_{i}}$ is centroid of the class $y_{i}$ corresponding to sample $x_{i}$, and $c_{j}$ is centroid of class $j$. $margin$ is a threshold that penalizes samples within the margin of their class centroid to ensure they move away.

Consequently, the overall training loss for our model is:
\begin{equation}
\label{equ: total loss}
\begin{aligned}
L = L_{pri} + \lambda_{1}L_{wm} + \lambda_{2}L_{cpl},\\
L_{cpl} = \lambda_{3}L_{intra} + \lambda_{4}L_{inter},
\end{aligned}
\end{equation}
where $L_{pri}$ is the primary task loss, and $L_{wm}$ is the watermark classification loss. $\lambda_{1}$, $\lambda_{2}$, $\lambda_{3}$ and $\lambda_{4}$ are coefficients.

\subsection{Watermark Key Samples Generation}
\label{sec: Watermark Key Samples Generation}
While the construction method in previous section enables generating ample watermark samples, not all are equally effective. Ideal watermark key samples should yield high success rates on victim and stolen models, while remaining undetectable on benign models. To this end, we propose a two-stage filtering mechanism.

\textbf{Stage 1}: Inspired by \cite{tan2023deep}, we start with an initial watermark sample set $S_0$ and filter it to $S_1$ based on three criteria: (a) passes verification on the victim model, (b) passes on a surrogate model (trained via simulated model stealing using victim predictions), and (c) fails on a benign model (trained without watermarks). Formally:
\begin{equation}
\begin{aligned}
S_{1}=\{(x^{w},y^{w}) \mid (&x^{w},y^{w}) \in S_{0}, F_{V}(x^{w};\theta_{V})=y^{w},\\
F_{S}(x^{w}&;\theta_{S})=y^{w}, F_{B}(x^{w};\theta_{B}) \neq y^{w}\},
\end{aligned}
\end{equation}
where $F_{V}$, $F_{S}$ and $F_{B}$ are victim, surrogate, and benign models parameterized by $\theta_{V}$, $\theta_{S}$ and $\theta_{B}$, respectively.

\textbf{Stage 2}: To further boost success on real-world stolen models, we select the top $M$ samples from $S_1$ most confidently predicted as the target label $y^w$ by the surrogate model $F_S$, forming the final key set $S_K$:
\begin{equation}
S_{K}=TopK(S_{1},M),
\end{equation}
where $TopK(\cdot,\cdot)$ is a function that can select samples that meet the aforementioned requirements. Finally, $S_{K}$ is saved for all subsequent model copyright verification.

\subsection{Model Ownership Verification}
Using $S_K$, the verifier can conduct black-box copyright checks with hard-label queries to a suspect model $M_{suspect}$. If the top-1 accuracy exceeds a predefined threshold, the model is deemed stolen; otherwise, it is not.

\section{Experiments}
\label{sec: Experiments}

\begin{table}[t]
    \centering
    \footnotesize
    \resizebox{0.48\textwidth}{!}{
    \begin{tabular}{llcccc}
        \toprule
        \multirow{2}{*}{\textbf{\scriptsize Dataset}} & \multirow{2}{*}{\textbf{\scriptsize Method}} & \multicolumn{2}{c}{\textbf{\scriptsize Benign Model}} & \multicolumn{2}{c}{\textbf{\scriptsize Watermarked Model}}\\
        \cmidrule(lr){3-4} \cmidrule(lr){5-6}
        & & \scriptsize Acc & \scriptsize WSR & \scriptsize Acc ($\Delta$Acc) & \scriptsize WSR \\
        \midrule
        \multirow{10}{*}{\scriptsize FMNIST} & \scriptsize BadNets & 91.69 & 10.47 & 91.13 (-0.56) & 99.99 \\
        & \scriptsize Composite & 91.53 & 0.55 & 89.78(-1.75) & 95.38 \\
        & \scriptsize Abstract & 91.59 & 12.62 & 91.32(-0.27) & \textbf{100.00} \\
        & \scriptsize Content & 91.51 & 7.66 & 91.41(-0.10) & 99.82 \\
        & \scriptsize Noise & 91.53 & 8.89 & 91.28(-0.25) & 99.53 \\
        & \scriptsize Unrelated	& 91.60 & 1.10 & 91.28(-0.32) & \textbf{100.00} \\
        & \scriptsize EWE & 91.50 & 0.07 & 86.33(-5.17) & \textbf{100.00} \\
        & \scriptsize Margin-based & 89.68 & 4.00 & 92.34(\textbf{+2.66}) & \textbf{100.00} \\
        & \scriptsize MEA-Defender & 91.71 & 0.65 & 88.86(-2.85) & 95.01 \\
        & \textbf{\scriptsize DeepTracer} & 91.55 & \textbf{0.00} & 91.49(-0.06) & \textbf{100.00} \\
        \midrule
        \multirow{10}{*}{\scriptsize CIFAR10} & \scriptsize BadNets & 85.16 & 9.40 & 85.32 (+0.16) & \textbf{100.00} \\
        & \scriptsize Composite & 84.33 & 2.36 & 83.97(-0.36) & 86.82 \\
        & \scriptsize Abstract & 85.10 & 9.66 & 84.66(-0.44) & \textbf{100.00} \\
        & \scriptsize Content & 85.05 & 6.39 & 85.69(+0.64) & 99.22 \\
        & \scriptsize Noise	& 85.16 & 9.48 & 85.38(+0.22) & 99.97 \\
        & \scriptsize Unrelated	& 85.26 & 18.56 & 84.48(-0.78) & \textbf{100.00} \\
        & \scriptsize EWE & 85.12 & 0.91 & 80.98(-4.14) & 19.44 \\
        & \scriptsize Margin-based & 82.34 & 10.24 & 85.99(\textbf{+3.65}) & \textbf{100.00} \\
        & \scriptsize MEA-Defender & 84.26 & 2.01 & 83.44(-0.82) & 91.82 \\
        & \textbf{\scriptsize DeepTracer} & 85.31 & \textbf{0.00} & 85.59(+0.28) & \textbf{100.00} \\
        \midrule
        \multirow{10}{*}{\scriptsize CIFAR100} & \scriptsize BadNets & 51.66 & 1.30 & 48.39(-3.27) & \textbf{100.00} \\
        & \scriptsize Composite & 52.08 & 1.63 & 46.89(-5.19) & 96.87 \\
        & \scriptsize Abstract & 51.51 & \textbf{0.00} & 48.29(-3.22) & \textbf{100.00} \\
        & \scriptsize Content & 51.64 & 1.16 & 49.06(-2.58) & 98.84 \\
        & \scriptsize Noise	& 51.92 & 1.10 & 47.63(-4.29) & 96.21 \\
        & \scriptsize Unrelated & 51.67 & 1.08 & 48.00(-3.67) & \textbf{100.00} \\
        & \scriptsize EWE & 51.67 & \textbf{0.00} & 59.09(\textbf{+7.42}) & 64.51 \\
        & \scriptsize Margin-based & 51.59 & \textbf{0.00} & 41.20(-10.39) & \textbf{100.00} \\
        & \scriptsize MEA-Defender & 51.66 & 1.75 & 47.68(-3.98) & 98.80 \\
        & \textbf{\scriptsize DeepTracer} & 51.67 & \textbf{0.00} & 50.72(-0.95) & \textbf{100.00} \\
        \bottomrule
    \end{tabular}
    }
    \vspace{-0.2cm}
    \caption{Comparison of harmlessness and effectiveness (\%) with other watermarking methods.}
    \label{tab: Comparison of harmlessness and effectiveness}
    \vspace{-0.3cm}
\end{table}

\begin{table*}[t]
    \centering
    \footnotesize
    \resizebox{\textwidth}{!}{
    \begin{tabular}{llcccccccccccc}
        \toprule
        \multirow{3}{*}{\textbf{Dataset}} & \multirow{3}{*}{\textbf{Method}} & \multicolumn{4}{c}{\textbf{JBDA}} & \multicolumn{4}{c}{\textbf{Knockoff}} & \multicolumn{4}{c}{\textbf{DFME}} \\
        \cmidrule(lr){3-6} \cmidrule(lr){7-10} \cmidrule(lr){11-14}
        & & \multicolumn{2}{c}{Soft Label} & \multicolumn{2}{c}{Hard Label} & \multicolumn{2}{c}{Soft Label} & \multicolumn{2}{c}{Hard Label} & \multicolumn{2}{c}{Soft Label} & \multicolumn{2}{c}{Hard Label} \\
        \cmidrule(lr){3-4} \cmidrule(lr){5-6} \cmidrule(lr){7-8} \cmidrule(lr){9-10} \cmidrule(lr){11-12} \cmidrule(lr){13-14}
        & & Acc & WSR & Acc & WSR & Acc & WSR & Acc & WSR & Acc & WSR & Acc & WSR \\
        \midrule
        \multirow{10}{*}{FMNIST} & BadNets	& 86.56 & 25.61 & 81.48 & 13.51 & 83.05 & 38.91 & 57.22 & 11.60 & 68.35 & 98.66 & 57.48 & 85.18 \\
        & Composite	& 85.59 & 27.97 & 81.26 & 3.38 & 48.65 & 96.14 & 42.23 & 80.61 & 34.99 & 0.00 & 46.42 & 0.00 \\
        & Abstract & 84.41 & 19.04 & 84.27 & 18.30 & 72.07 & 39.98 & 55.05 & 18.26 & 57.60 & 17.16 & 65.10 & 16.38 \\
        & Content & 87.49 & 11.51 & 82.63 & 12.24 & 83.21 & 18.05 & 54.16 & 5.58 & 68.13 & 9.39 & 69.92 & 5.85 \\
        & Noise	& 84.86 & 12.51 & 81.72 & 13.13 & 68.85 & 1.79 & 38.41 & 1.91 & 18.31 & 88.21 & 25.25 & 49.53 \\
        & Unrelated	& 86.63 & \textbf{96.17} & 83.75 & \textbf{92.66} & 80.55 & 26.54 & 48.73 & 5.15 & 58.36 & 94.31 & 61.43 & 96.03 \\
        & EWE & 80.75 & 63.74 & 83.58 & 56.96 & 49.08 & 0.00 & 38.62 & 0.00 & 29.36 & 0.00 & 54.46 & 0.00 \\
        & Margin-based & 86.05 & 49.92 & 80.73 & 41.12 & 79.63 & 10.40 & 43.92 & 12.96 & 63.20 & 5.12 & 66.00 & 5.28 \\
        & MEA-Defender & 84.66 & 46.17 & 82.42 & 8.61 & 64.49 & 91.22 & 59.86 & 26.35 & 44.00 & 0.00 & 46.92 & 0.00 \\
        & \textbf{DeepTracer} & 85.02 & 91.65 & 80.69 & 86.90 & 87.16 & \textbf{100.00} & 64.57 & \textbf{100.00} & 70.03 & \textbf{100.00} & 68.21 & \textbf{100.00} \\
        \midrule
        \multirow{10}{*}{CIFAR10} & BadNets & 68.01 & 10.36 & 52.88 & 9.25 & 80.81 & 9.06 & 74.76 & 9.00 & 59.57 & 31.83 & 46.63 & 54.84 \\
        & Composite & 69.87 & 30.64 & 53.11 & 9.15 & 78.28 & 54.69 & 71.37 & 35.74 & 18.57 & 33.46 & 21.76 & 10.06 \\
        & Abstract & 62.80 & 10.26 & 56.04 & 9.46 & 79.67 & 43.78 & 74.26 & 25.18 & 24.83 & 18.64 & 18.68 & 12.90 \\
        & Content & 65.24 & 14.90 & 55.99 & 6.69 & 81.01 & 8.00 & 73.75 & 8.15 & 20.82 & 33.07 & 40.00 & 9.62 \\
        & Noise	& 65.57 & 11.28 & 55.05 & 5.95 & 80.90 & 12.78 & 73.72 & 10.63 & 23.81 & 29.99 & 16.48 & 10.30 \\
        & Unrelated	& 66.87 & 31.80 & 57.01 & 19.93 & 79.10 & 56.89 & 72.75 & 47.17 & 17.12 & 53.16 & 23.03 & 48.69 \\
        & EWE & 53.97 & 79.75 & 51.72 & 71.18 & 66.45 & 13.02 & 58.69 & 4.67 & 17.75 & 0.98 & 18.70 & 34.72 \\
        & Margin-based & 63.08 & 6.40 & 51.91 & 1.28 & 80.54 & 21.76 & 74.04 & 18.08 & 22.15 & 15.36 & 23.15 & 37.28 \\
        & MEA-Defender & 64.15 & 80.12 & 53.52 & 54.50 & 69.50 & 92.15 & 67.38 & 36.17 & 23.86 & 38.29 & 22.69 & 17.02 \\
        & \textbf{DeepTracer} & 66.32 & \textbf{82.05} & 57.98 & \textbf{77.10} & 75.87 & \textbf{98.75} & 67.01 & \textbf{74.70} & 25.85 & \textbf{100.00} & 16.69 & \textbf{97.15} \\
        \midrule
        \multirow{10}{*}{CIFAR100} & BadNets & 28.36 & 0.38 & 13.07 & 0.33 & 39.15 & 1.43 & 31.03 & 1.13 & 5.00 & 7.71 & 6.82 & 21.66 \\
        & Composite & 28.43 & 0.00 & 13.78 & 0.00 & 33.08 & 0.07 & 33.60 & 0.00 & 2.26 & 0.00 & 3.05 & 0.00 \\
        & Abstract & 29.02 & 0.00 & 14.72 & 0.88 & 38.26 & 1.52 & 33.25 & 1.08 & 3.15 & 2.40 & 3.27 & 2.98 \\
        & Content & 27.72 & 2.26 & 13.94 & 0.47 & 38.80 & 2.35 & 32.64 & 1.94 & 8.47 & 3.80 & 4.55 & 6.96 \\
        & Noise	& 28.61 & 10.03 & 13.19 & 0.89 & 37.94 & 4.89 & 30.80 & 2.45 & 4.76 & 0.31 & 3.14 & 0.00 \\
        & Unrelated & 27.97 & 1.00 & 14.33 & 0.00 & 37.64 & 4.47 & 32.26 & 6.48 & 5.52 & 0.40 & 3.94 & 0.00 \\
        & EWE & 19.34 & 0.00 & 14.91 & 0.00 & 24.64 & 27.60 & 31.58 & 1.04 & 2.64 & 0.26 & 2.28 & 0.00 \\
        & Margin-based & 22.76 & 8.96 & 9.97 & 0.00 & 16.14 & 5.12 & 18.72 & 1.28 & 5.65 & 3.84 & 3.41 & 3.84 \\
        & MEA-Defender & 16.70 & 0.00 & 13.43 & 0.00 & 21.61 & 0.00 & 25.29 & 0.00 & 2.58 & 0.00 & 3.09 & 0.00 \\
        & \textbf{DeepTracer} & 25.71 & \textbf{37.25} & 14.07 & \textbf{31.50} & 31.44 & \textbf{100.00} & 34.18 & \textbf{100.00} & 3.53 & \textbf{34.55} & 5.18 & \textbf{75.60} \\
        \bottomrule
    \end{tabular}
    }
    \caption{Comparison of robustness (\%) against soft and hard label stealing attacks from JBDA, Knockoff, and DFME.}
    \label{tab: Comparison of robustness against three stealing attacks}
\end{table*}

\subsection{Experimental Setup}
\textbf{Datasets.} We evaluate DeepTracer on four popular datasets: Fashion MNIST \cite{xiao2017fashion}, CIFAR10 \cite{krizhevsky2009learning}, CIFAR100 \cite{krizhevsky2009learning}, and ImageNet \cite{deng2009imagenet}. 

\noindent\textbf{Model Architectures.} For the victim models: (1) On Fashion MNIST, we use a simple CNN named NaiveNet, which consists of two convolutional layers and two fully connected layers. (2) On CIFAR10, we use a VGG-like model with four convolutional layers and three fully connected layers. (3) On CIFAR100, we employ ResNet18 \cite{he2016deep}. (4) On ImageNet, we use three different networks: ResNet50 \cite{he2016deep}, DenseNet161 \cite{huang2017densely}, and EfficientNetB2 \cite{tan2019efficientnet}.

The benign (clean) models have the same architecture as the victim models across all datasets. For the surrogate model used in the two-stage filtering: (1) On ImageNet, we use ResNet18 \cite{he2016deep}. (2) On other datasets, we use AlexNet \cite{krizhevsky2012imagenet}. Note that in all experiments, the training set for surrogate model is TinyImageNet \cite{le2015tiny}.

\noindent\textbf{Compared Watermarking Methods.} We compare our method extensively with existing state-of-the-art watermarking methods in terms of robustness against model stealing attacks. These methods include Abstract \cite{adi2018turning}, Content \cite{zhang2018protecting}, Noise \cite{zhang2018protecting}, Unrelated \cite{zhang2018protecting}, EWE \cite{jia2021entangled}, Margin-based \cite{kim2023margin} and MEA-Defender \cite{lv2024mea}. In addition, we also transform two backdoor attack methods BadNets \cite{gu2017badnets} and Composite \cite{lin2020composite} to watermarking methods. 

\noindent\textbf{Evaluation Metrics.} (1) Accuracy (Acc): the proportion of clean test data correctly classified to their ground-truth labels. Higher Acc of victim model indicates lesser impact of our watermark on original functionality. (2) Watermark Success Rate (WSR): the proportion of watermark data classified to target label. Higher WSR on victim and stolen models indicates better effectiveness and robustness of our watermark. Lower WSR on benign models indicates a lower likelihood of false positives on non-watermarked models.

\noindent\textbf{Implementation Details.} We train all victim models in two 100-epoch phases using the Adam optimizer (initial learning rate 0.001, halved every 10 epochs). Watermark samples account for 1\% and 10\% of training data in the first and second phases, respectively, with the second phase fine-tuning from the first. Loss weights are $\lambda_{1}=1.0$, $\lambda_{2}=1.0$, $\lambda_{3}=0.01$, $\lambda_{4}=3.0$, and we set $M=2000$ in the second stage of filtering. Settings for prior watermarking and attack methods are tuned per their original papers. We evaluate numerous watermarked and clean models to set a robust ownership verification threshold, concluding that a 20\% watermark success rate effectively separates the two, ensuring our method’s reliability.

\subsection{Harmlessness and Effectiveness}
\label{sec: Harmlessness and Effectiveness Evaluation}
Table \ref{tab: Comparison of harmlessness and effectiveness} shows that our DeepTracer achieves 0\% WSR on benign models across all datasets, indicating no false positives and outperforming all baselines. This is due to our target label selection and watermark sample generation, which tailor effective keys that maximize WSR on victim/stolen models while minimizing it on benign ones. In contrast, prior methods suffer from false positives due to lack of such design.

Our method also preserves model utility, with only minor accuracy changes: +0.28\% on CIFAR10, -0.06\% on Fashion MNIST, and -0.95\% on CIFAR100. Prior works show more degradation. This is because (1) our watermark samples are derived from task data, avoiding external noise, and (2) the same-class coupling loss improves feature separability.

In terms of effectiveness on victim model, DeepTracer achieves 100\% watermark success on all datasets, outperforming previous methods that only reach 90\% $\sim$ 100\% on some datasets.

\begin{figure*}[t]
    \centering
    \vspace{-0.1cm}
    \begin{subfigure}{0.24\textwidth}
        \centering
        \includegraphics[width=1.0\textwidth]{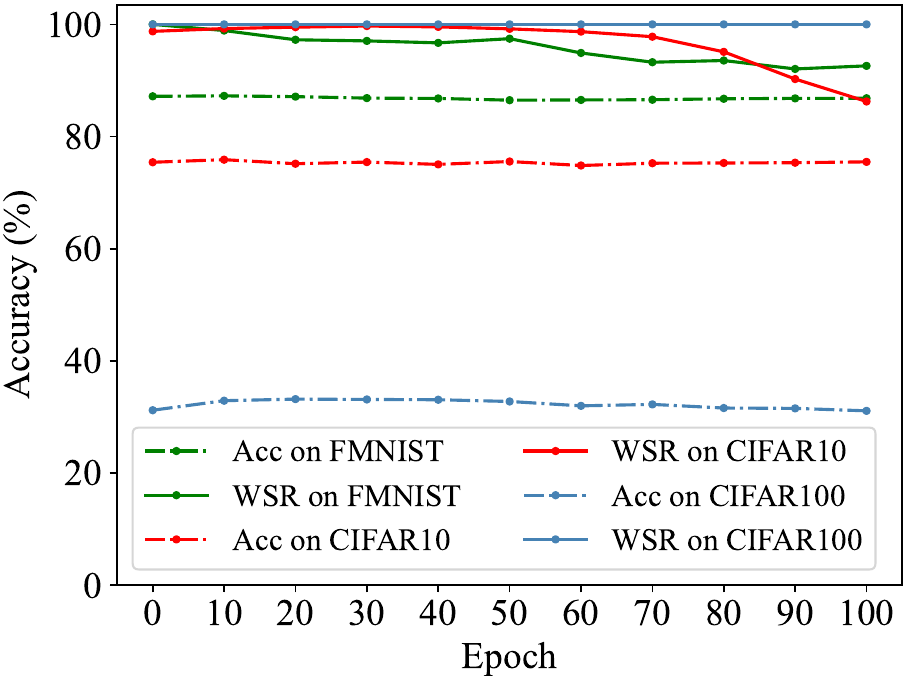}
        \caption{FTLL on Stolen Model}
        \label{fig: FTLL on Stolen Model}
    \end{subfigure}
    \centering
    \begin{subfigure}{0.24\textwidth}
        \centering
        \includegraphics[width=1.0\textwidth]{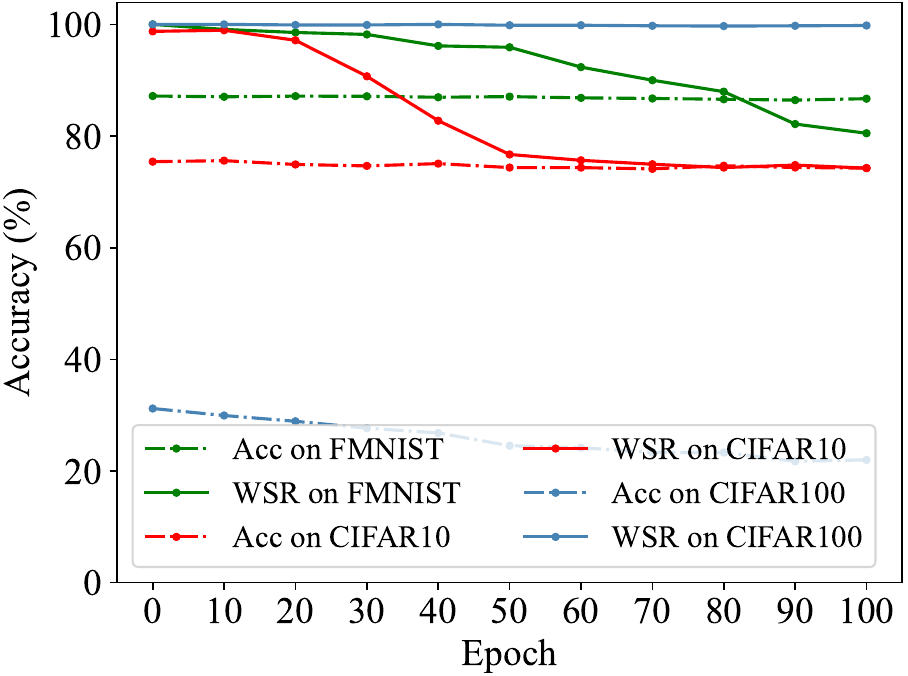}
        \caption{FTAL on Stolen Model}
        \label{fig: FTAL on Stolen Model}
    \end{subfigure}
    \centering
    \begin{subfigure}{0.24\textwidth}
        \centering
        \includegraphics[width=1.0\textwidth]{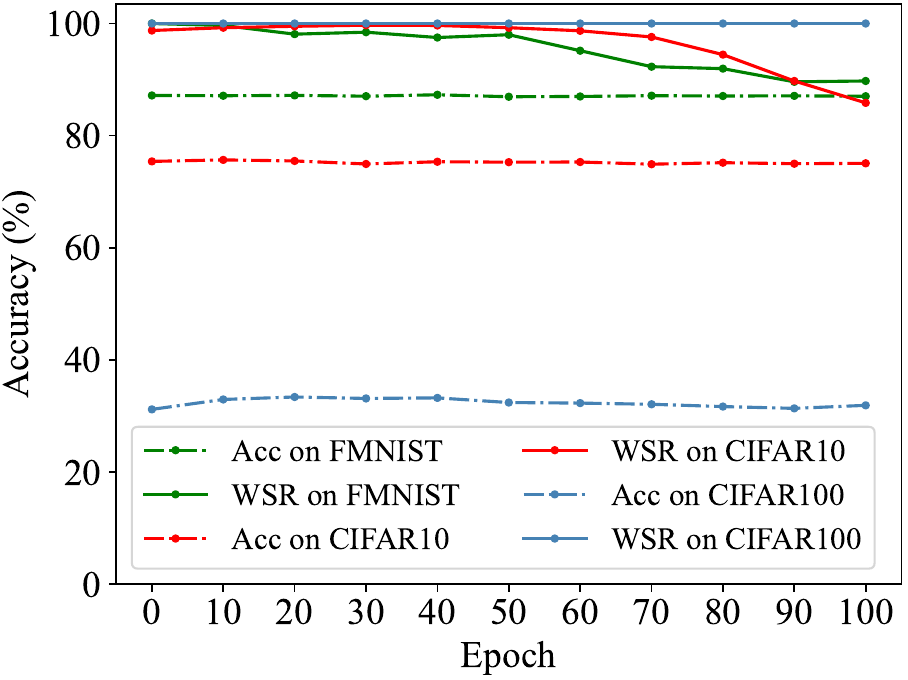}
        \caption{RTLL on Stolen Model}
        \label{fig: RTLL on Stolen Model}
    \end{subfigure}
    \centering
    \begin{subfigure}{0.24\textwidth}
        \centering
        \includegraphics[width=1.0\textwidth]{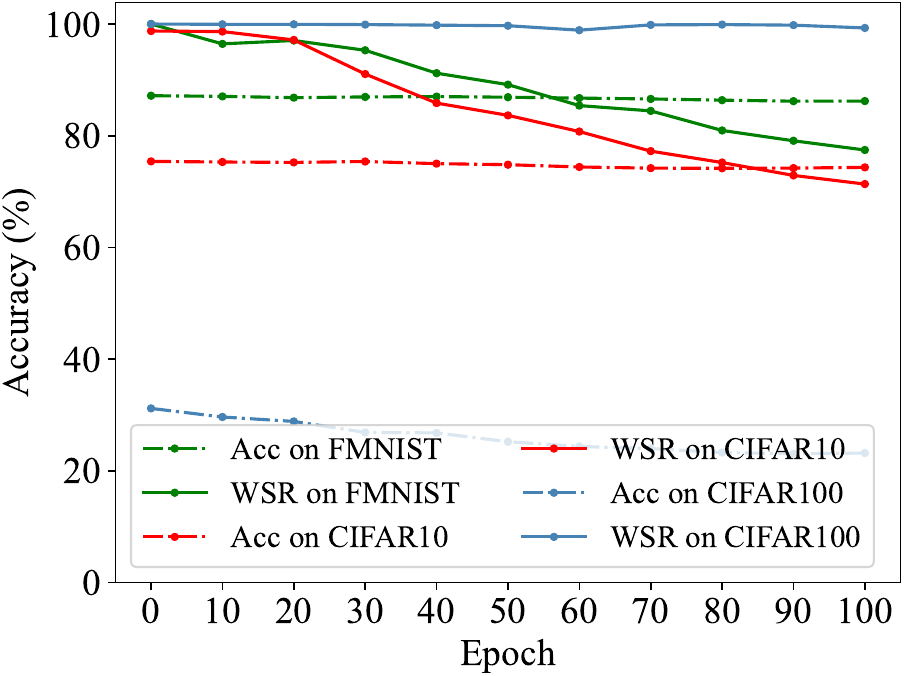}
        \caption{RTAL on Stolen Model}
        \label{fig: RTAL on Stolen Model}
    \end{subfigure}
    \vspace{-0.2cm}
    \caption{Robustness against four fine-tuning attacks on stolen model.}
    \label{fig: fine-tuning attacks on stolen model}
    \vspace{-0.3cm}
\end{figure*}

\begin{figure}[t]
\centering
\includegraphics[width=0.25\textwidth]{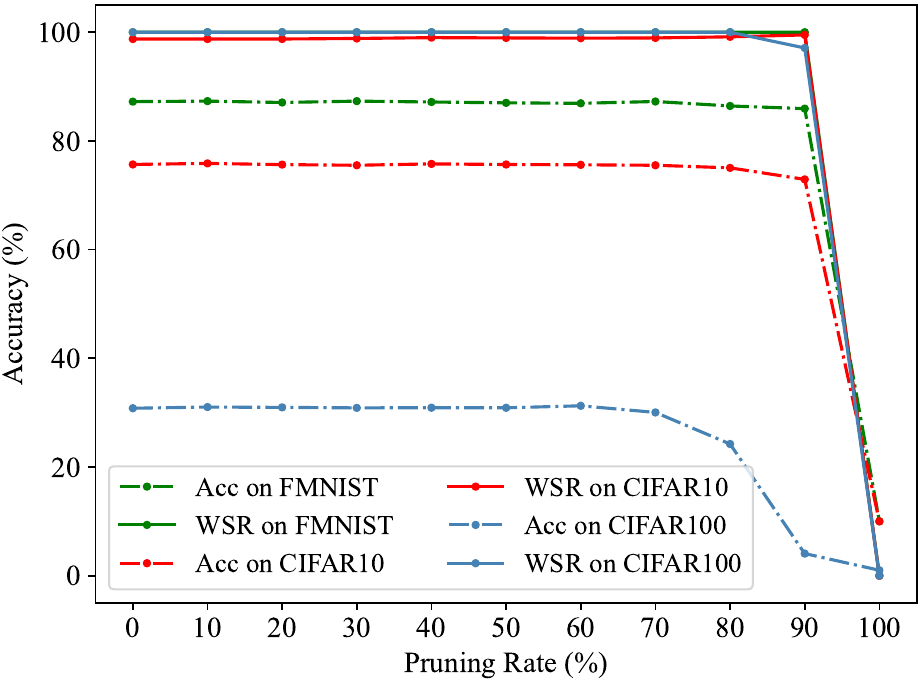}
\vspace{-0.2cm}
\caption{Robustness against pruning attack on stolen model.}
\label{fig: pruning attack on stolen model}
\vspace{-0.1cm}
\end{figure}

\begin{table}[t]
    \centering
    \footnotesize
    \begin{tabular}{
        m{0.7cm}<{\centering}
        m{0.7cm}<{\centering}
        m{0.9cm}<{\centering}
        m{0.7cm}<{\centering}
        m{0.9cm}<{\centering}
        m{0.7cm}<{\centering}
        m{0.9cm}<{\centering}}
        \toprule
        \multirow{2}{*}{\textbf{\scriptsize Bit Size}} & \multicolumn{2}{c}{\textbf{\scriptsize FMNIST}} & \multicolumn{2}{c}{\textbf{\scriptsize CIFAR10}} & \multicolumn{2}{c}{\textbf{\scriptsize CIFAR100}}\\
        \cmidrule(lr){2-3} \cmidrule(lr){4-5} \cmidrule(lr){6-7}
        & \scriptsize Acc & \scriptsize WSR & \scriptsize Acc & \scriptsize WSR & \scriptsize Acc & \scriptsize WSR \\
        \midrule
        16 & 87.23 & 100.00 & 75.67 & 98.75 & 30.82 & 100.00 \\
        8 & 87.39 & 100.00 & 75.80 & 98.60 & 31.11 & 100.00 \\
        6 & 86.59 & 100.00 & 75.25 & 98.70 & 30.83 & 100.00 \\
        4 & 53.78 & 98.70 & 70.35 & 99.40 & 24.15 & 100.00 \\
        3 & 10.00 & 0.00 & 11.99 & 37.55 & 1.44 & 29.20 \\
        2 & 10.00 & 0.00 & 10.00 & 0.00 & 1.00 & 0.00 \\
        1 & 10.00 & 0.00 & 10.00 & 0.00 & 1.00 & 0.00 \\
        \bottomrule
    \end{tabular}
    \vspace{-0.2cm}
    \caption{Quantization attack on stolen model.}
    \label{tab: quantization attack on stolen model}
    \vspace{-0.4cm}
\end{table}

\subsection{Robustness against Model Stealing Attacks}
\label{sec: Robustness against Model Stealing Attacks}
We extensively evaluate our method against a range of model stealing attacks, including JBDA \cite{papernot2017practical}, Knockoff \cite{orekondy2019knockoff}, DFME \cite{truong2021data}, Hard Label, Cross-Dataset, Cross-Architecture, Cross-Dataset\&Cross-Architecture, Distillation-Based, and Double-Stage Stealing. The details of these attacks and the experimental results under the last five attacks can be found in the Appendix.

Here, we evaluate the robustness against JBDA, Knockoff, and DFME under both soft and hard label scenarios on Fashion MNIST, CIFAR10, and CIFAR100. As shown in Table \ref{tab: Comparison of robustness against three stealing attacks}, our method achieves the best performance in nearly all cases, except JBDA on Fashion MNIST. It consistently reaches 100\% or near-100\% WSR on Knockoff and DFME, even on complex tasks like CIFAR100. Notably, in hard-label attacks, our WSR remains above the 20\% threshold, confirming ownership. This highlights the resilience of our deep coupled watermark, while prior methods often fail under strong or complex attacks.


\subsection{Robustness against Removal Attacks}
\label{sec: Robustness against Watermark Removal Attacks}
In addition to model stealing attacks, various popular watermark removal techniques exist (e.g., Fine-Tuning \cite{adi2018turning}, Pruning \cite{uchida2017embedding}, Quantization \cite{lukas2022sok}, Transfer Learning \cite{lukas2022sok}), and the Appendix provides their detailed introduction. Figure \ref{fig: fine-tuning attacks on stolen model}, Figure \ref{fig: pruning attack on stolen model}, and Table \ref{tab: quantization attack on stolen model} respectively show the robustness evaluation for fine-tuning, pruning, and quantization on stolen model, which represents scenarios where the adversary executes a watermark removal attack after obtaining the stolen model. It can be observed that when the accuracy of the model decreases within an acceptable range and the model still has usability, our watermarking method has satisfactory robustness against these removal attacks, and it has sufficient confidence to verify the ownership of the model.

\subsection{Ablation Studies}

\begin{figure}[t]
    \centering
    \begin{subfigure}{0.23\textwidth}
        \centering
        \includegraphics[width=1.0\textwidth]{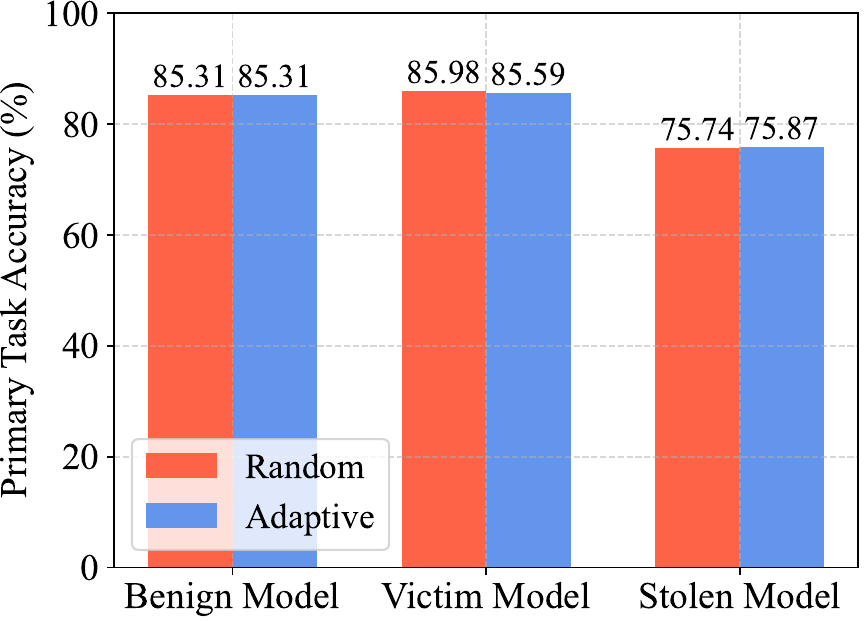}
    \end{subfigure}
    \centering
    \begin{subfigure}{0.23\textwidth}
        \centering
        \includegraphics[width=1.0\textwidth]{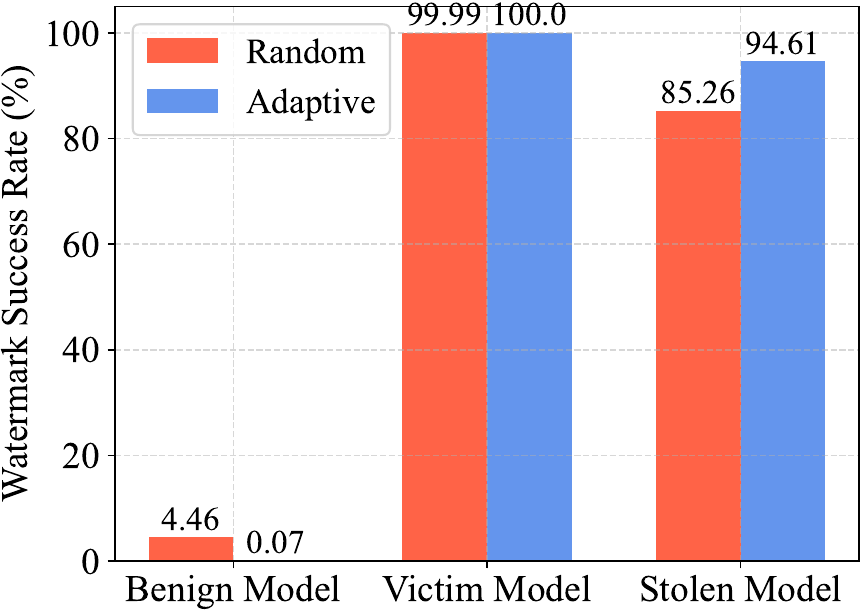}
    \end{subfigure}
    \vspace{-0.2cm}
    \caption{Effect of class selection strategy.}
    \label{fig: ablation on class selection strategy}
    \vspace{-0.2cm}
\end{figure}

\textbf{Watermark Source Class and Target Label Selection Strategy.}
Figure \ref{fig: ablation on class selection strategy} shows our adaptive class selection strategy outperforms random selection by reducing WSR on benign models by 4.39\% and increasing it on stolen models by 9.35\%.

\begin{table}[t]
    \centering
    \footnotesize
    \vspace{-0.0cm}
    \begin{tabular}{
        m{3.2cm}
        m{0.7cm}<{\centering}
        m{0.9cm}<{\centering}
        m{0.7cm}<{\centering}
        m{0.7cm}<{\centering}}
        \toprule
        \multirow{2}{*}{\textbf{\scriptsize Training Loss}} & \multicolumn{2}{c}{\textbf{\scriptsize Victim Model}} & \multicolumn{2}{c}{\textbf{\scriptsize Stolen Model}}\\
        \cmidrule(lr){2-3} \cmidrule(lr){4-5}
        & \scriptsize Acc & \scriptsize WSR & \scriptsize Acc & \scriptsize WSR \\
        \midrule
        $L_{wm}$ only & 84.94 & 99.99 & 75.07 & 82.30 \\
        $L_{intra}$ only & 84.36 & 56.86 & 75.83 & 1.71 \\
        $L_{inter}$ only & 82.22 & 0.06 & 76.47 & 0.10 \\
        $L_{wm}$ \& $L_{intra}$	& 85.41 & 99.97 & 75.78 & 89.62 \\
        $L_{wm}$ \& $L_{inter}$ & 84.94 & 99.99 & 75.07 & 84.35 \\
        $L_{intra}$ \& $L_{inter}$ & 84.16 & 64.38 & 75.88 & 5.58 \\
        $L_{wm}$ \& $L_{intra}$ \& $L_{inter}$ & 85.59 & 100.00 & 75.87 & 94.61 \\
        \bottomrule
    \end{tabular}
    \vspace{-0.2cm}
    \caption{Performance under different loss components.}
    \label{tab: different loss components}
    \vspace{-0.1cm}
\end{table}

\textbf{Training Loss Design.}
Table \ref{tab: different loss components} reveals several insights: (a) $L_{wm}$ is essential—its removal drastically lowers WSR on both victim and stolen models. (b) Adding intra- or inter-class loss boosts WSR on stolen models by 7.32\% and 2.05\%, respectively; both combined yield a 12.31\% gain. (c) Intra-class loss alone yields 56.86\% WSR on victim models, but just 1.71\% on stolen ones without $L_{wm}$.

\begin{figure}[t]
\centering
\includegraphics[width=0.26\textwidth]{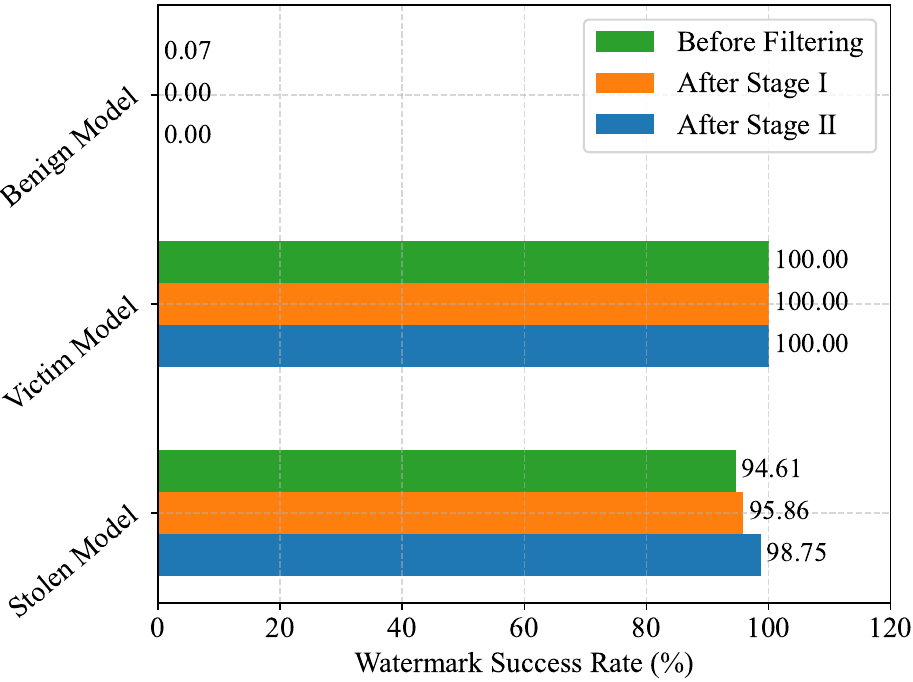}
\vspace{-0.2cm}
\caption{Effectiveness of two-stage filtering mechanism.}
\label{fig: two-stage filtering mechanism}
\vspace{-0.3cm}
\end{figure}

\textbf{Watermark Sample Filtering Mechanism.}
As shown in Figure \ref{fig: two-stage filtering mechanism}, stage one reduces benign WSR from 0.07\% to 0\% and raises stolen WSR from 94.61\% to 95.86\%; stage two further improves it to 98.75\%. This two-stage filter effectively selects optimal watermark samples for final ownership verification.

\section{Conclusion}
In this paper, we first systematically analyze the reason why existing watermarks are susceptible to removal by model stealing attacks and propose potential solutions. We then introduce DeepTracer, a deep coupled watermarking scheme that tightly integrates the watermark and primary tasks through a tailored sample construction method and novel embedding loss. Experiments across multiple benchmarks show that DeepTracer enables reliable ownership verification under model stealing and remains robust against various watermark attacks. We hope this work promotes further research in safeguarding AI model intellectual property and fostering a more secure ecosystem.

\clearpage

\section{Acknowledgments}
We thank all the anonymous reviewers for their constructive feedback. This research is supported by Beijing Municipal Science \& Technology Commission: New Generation of Information and Communication Technology Innovation - Research and Demonstration Application of Key Technologies for Privacy Protection of Massive Data for Large Model Training and Application (Z231100005923047).

\bibliography{aaai2026}

\clearpage

\appendix

\section{\huge Appendix}

\begin{figure*}[t]
\vspace{-0.1cm}
\centering
\includegraphics[width=0.85\textwidth]{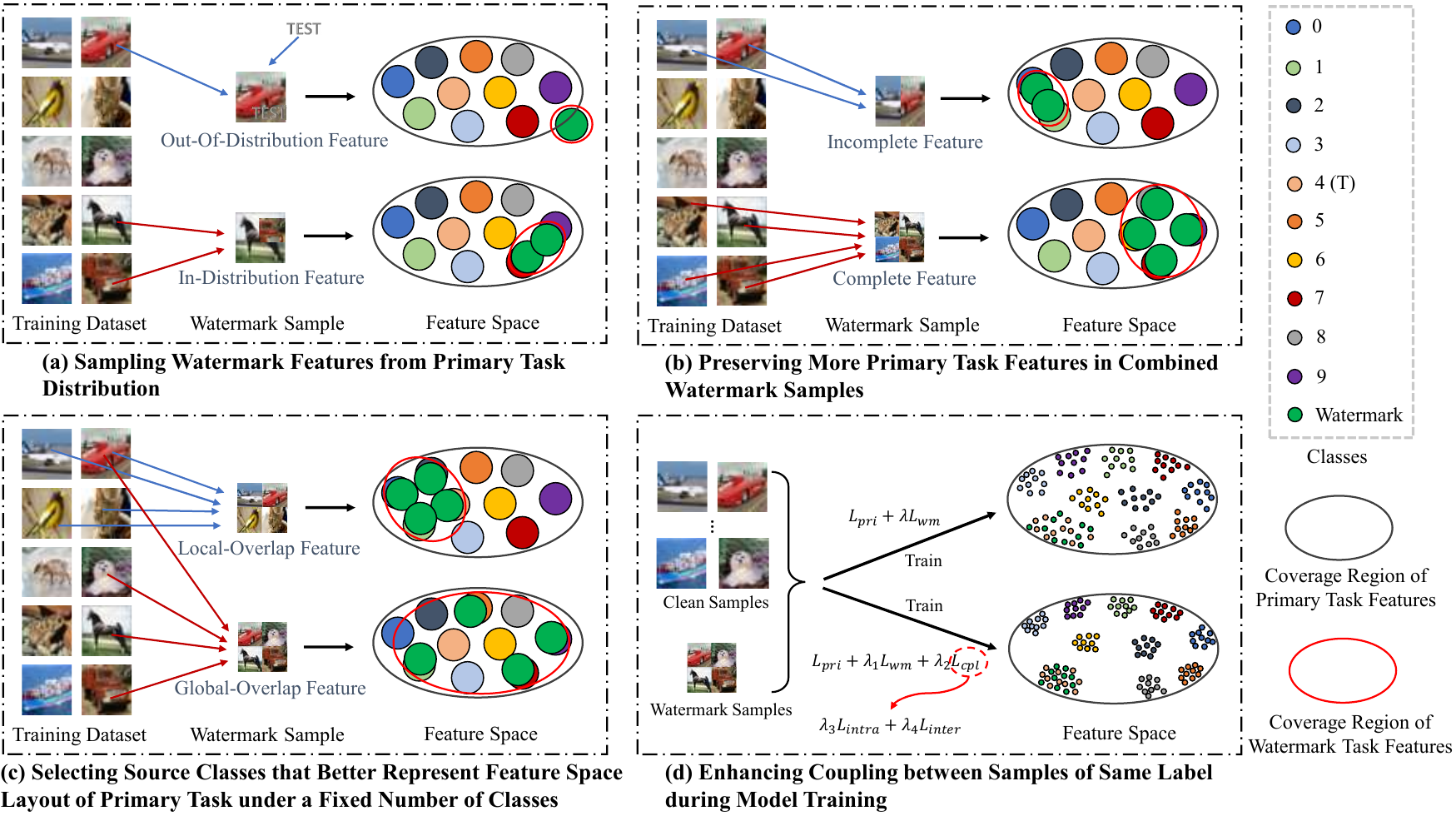}
\vspace{-0.2cm}
\caption{Some feasible solutions to increase the coupling degree between watermark task and primary task. The black ellipse and the red ellipse represent the feature coverage areas of primary task samples and watermark samples, respectively, and the larger overlap area of the two, the higher coupling between watermark distribution and primary task distribution.}
\label{fig: increase coupling degree}
\vspace{-0.3cm}
\end{figure*}

\vspace{0.5cm}
\section{Increasing the Coupling Degree between Watermark Task and Primary Task}
\label{sec: Increasing the Coupling Degree}

$\bullet$ \textit{\textbf{Sampling Watermark Features from Primary Task Distribution}}: The straightforward approach is to sample the watermark features from primary task distribution, like in \cite{lin2020composite} and \cite{lv2024mea}. As shown in Figure \ref{fig: increase coupling degree}a, compared to methods that collect watermark features from an external distribution (e.g., using the content “TEST” as watermark), composing a watermark sample from two training data classes of primary task ensures that it contains the features of the primary task, creating a significant overlap in the feature space. Figure \ref{fig: activation_MEA-Defender} illustrates that when in-distribution watermark samples are used, as in the method proposed by \cite{lv2024mea}, the consistency between the neural activation regions of watermark and primary task samples improves significantly compared to OOD watermarking.

$\bullet$ \textit{\textbf{Preserving More Primary Task Features in Combined Watermark Samples}}: Although \cite{lin2020composite} and \cite{lv2024mea} sample watermark features from primary task distribution, they only select two source classes, and their methods of combining watermark samples (e.g., image cropping and pasting, and stripe area combination) lead to incomplete sample features. These limit the feature overlap between watermark and primary task. In contrast, as illustrated in the lower part of Figure \ref{fig: increase coupling degree}b, we use samples from four source classes and resize them first before combining, which preserves more complete primary task features. It is worth noting that a larger number of source classes does not necessarily result in better performance. This is because an excessive number of images can significantly reduce image resolution and cause the loss of features. As seen from the results in Appendix \ref{sec: The Effect of Different Numbers of Source Classes}, four source classes are optimal.

$\bullet$ \textit{\textbf{Selecting Source Classes that Better Represent Feature Space Layout of Primary Task under a Fixed Number of Classes}}: Even when sampling from four source classes to construct watermark samples, different source classes will produce different results. As depicted in Figure \ref{fig: increase coupling degree}c, random selection of source classes may result in clustering of selected classes into local region of the feature distribution of primary task. More optimally, we select the four source classes that better represent the entire feature space layout of primary task with higher feature coverage than random selection, which can make watermark task more coupled with primary task.

$\bullet$ \textit{\textbf{Enhancing Coupling between Samples of Same Label during Model Training}}: To tightly couple watermark samples with the target class samples in the output feature space, we design a loss that optimize the model to bring same-label samples closer and push different-label samples further apart. This firmly binds the watermark pattern to the target label (T) and makes it harder to be removed, as presented in Figure \ref{fig: increase coupling degree}d.

\section{Details of Experimental Setup}
\label{sec: Details of Experimental Setup}
\textbf{(1) Datasets}

$\bullet$ Fashion MNIST: This dataset consists of 60,000 training images and 10,000 test images, each of which is a 28$\times$28 grayscale image representing fashion items across 10 classes.

$\bullet$ CIFAR10: Comprising 50,000 training images and 10,000 test images. This dataset includes 32$\times$32 color images of animals and vehicles across 10 classes.

$\bullet$ CIFAR100: Similar to CIFAR10, but with 100 classes, each containing 500 training images and 100 test images. The images are also 32$\times$32 color images.

$\bullet$ ImageNet: A challenging real-world dataset with 1,000 classes, approximately 1.3 million training images, and 50,000 test images (we use the original validation set as the test set). Each image is resized to 224$\times$224 color images for our experiments.

\noindent\textbf{(2) Compared Watermarking Methods}

$\bullet$ BadNets \cite{gu2017badnets}: Modifies the backdoor attack method BadNets to a watermarking method, which involves attaching a small pixel block to the bottom-right corner of the training sample as a watermark.

$\bullet$ Composite \cite{lin2020composite}: Modifies the backdoor attack method Composite Backdoor to a watermarking method, which selects two classes of training data and combines their features to create a watermark.

$\bullet$ Abstract \cite{adi2018turning}: Uses abstract art images with different target labels as watermarks.

$\bullet$ Content \cite{zhang2018protecting}: Embeds specific text content, such as "TEST" into some training images as watermarks.

$\bullet$ Noise \cite{zhang2018protecting}: Adds specific Gaussian noise pattern to a portion of training images as watermarks.

$\bullet$ Unrelated \cite{zhang2018protecting}: Uses unrelated OOD samples as watermark samples, but with a same target label.

$\bullet$ EWE \cite{jia2021entangled}: Uses a soft nearest neighbor loss to entangle the representations of watermark samples and training samples for defending against model stealing.

$\bullet$ Margin-based \cite{kim2023margin}: Selects a few samples from the original training set and assigns them random labels as watermark samples, then maximizes their distance to the decision boundary using projected gradient ascent, which makes the watermark harder to be removed.

$\bullet$ MEA-Defender \cite{lv2024mea}: Similar to Composite \cite{lin2020composite} but introduces a combination loss to constrain output distribution of watermark samples to be similar to that of two source classes.

\noindent\textbf{(3) Model Stealing Attack Methods}

$\bullet$ JBDA (Jacobian-based Dataset Augmentation) \cite{papernot2017practical}: Assumes the adversary has access to a small subset of the original training set as seed samples, then generates additional query samples using Jacobian-based data augmentation.

$\bullet$ Knockoff \cite{orekondy2019knockoff}: Utilizes reinforcement learning to efficiently select samples from a large data pool for querying.

$\bullet$ DFME (Data-Free Model Extraction) \cite{truong2021data}: A classic data-free model stealing method where query samples are synthesized by a generator.

$\bullet$ Hard Label Stealing: Assumes adversary can only obtain the top-1 label from victim model's output. We extend JBDA, Knockoff, and DFME to this hard label stealing scenario.

$\bullet$ Cross-Dataset Stealing: The adversary uses a dataset different from primary task distribution to query victim model.

$\bullet$ Cross-Architecture Stealing: The adversary employs a neural network architecture different from that of the victim model for the stolen model.

$\bullet$ Cross-Dataset and Cross-Architecture Stealing: Both the query dataset and the stolen model architecture differ from those used by the victim.

$\bullet$ Distillation-Based Stealing: Assumes the adversary has access to the full original training set but without labels, and it uses the soft labels predicted by the victim model for training, with the KL divergence as the loss function.

$\bullet$ Double-Stage Stealing: Involves performing a second stealing attack on the already extracted model to obtain a new stolen model.

\noindent\textbf{(4) Details of Hyperparameter Selection}

$\bullet$ Considerations for Loss Function Hyperparameters: The hyperparameter $\lambda_{1}$ controls the relative weight of watermark task. A larger value of $\lambda_{1}$ can degrade the performance of primary task, while a smaller value makes watermark task more difficult to learn. To balance the two tasks, we set $\lambda_{1}=1.0$, ensuring both tasks are optimized with equal weight, consistent with prior watermarking methods. The hyperparameter $\lambda_{2}$ serves as a scaling factor for $\lambda_{3}$ and $\lambda_{4}$. For simplicity, we set $\lambda_{2}=1.0$, leaving $\lambda_{3}$ and $\lambda_{4}$ to be determined. Since $\lambda_{3}$ and $\lambda_{4}$ are two components of same-class coupling loss we propose, we evaluate a wide range of $(\lambda_{3}, \lambda_{4})$ combinations while fixing $\lambda_{1}$ and $\lambda_{2}$. We find that the performance is optimal when $(\lambda_{3}, \lambda_{4})=(0.01, 3.0)$.

$\bullet$ Considerations for the Number of Samples $M$ Retained by Watermark Samples Filtering Mechanism: Larger values of $M$ risk retaining unreliable verification samples, which can undermine the filtering mechanism’s effectiveness. On the other hand, smaller values of $M$ lead to a smaller final verification set, potentially causing instability and reducing the reliability of the watermark validation results. Based on these considerations, we empirically set $M=2000$, striking a balance that ensures the watermark success rate is precise to two decimal places.

\section{More Evaluations of Robustness}

\subsection{Robustness against More Model Stealing Attacks}
\label{sec: Robustness against More Model Stealing Attacks}

\begin{table}[t]
    \centering
    \footnotesize
    \begin{tabular}{
        m{0.45cm}<{\centering}
        m{0.8cm}<{\centering}
        m{0.9cm}<{\centering}
        m{1.0cm}<{\centering}
        m{0.8cm}<{\centering}
        m{0.4cm}<{\centering}
        m{1.2cm}<{\centering}}
        \toprule
        \multirow{2}{*}{} & \multirow{2}{*}{\textbf{\scriptsize Victim}} & \multicolumn{5}{c}{\textbf{\scriptsize Stolen Model}} \\
        \cmidrule(lr){3-7}
        & \textbf{\scriptsize Model} & \scriptsize CIFAR10 & \scriptsize CIFAR100 & \scriptsize FMNIST & \scriptsize NICO & \scriptsize TinyImageNet \\
        \midrule
        Acc & 85.71 & 83.49 & 80.09 & 31.88 & 75.83 & 81.37 \\
        WSR & 100.00 & 83.45 & 99.65 & 51.35 & 99.75 & 99.85 \\
        \bottomrule
    \end{tabular}
    \vspace{-0.2cm}
    \caption{Robustness (\%) against cross-dataset stealing attacks.}
    \label{tab: cross-dataset model stealing attacks}
    \vspace{-0.0cm}
\end{table}

\begin{table}[t]
    \centering
    \footnotesize
    \begin{tabular}{
        m{0.4cm}<{\centering}
        m{0.8cm}<{\centering}
        m{0.98cm}<{\centering}
        m{0.75cm}<{\centering}
        m{0.7cm}<{\centering}
        m{0.8cm}<{\centering}
        m{1.1cm}<{\centering}}
        \toprule
        \multirow{2}{*}{} & \multirow{2}{*}{\textbf{\scriptsize Victim}} & \multicolumn{5}{c}{\textbf{\scriptsize Stolen Model}} \\
        \cmidrule(lr){3-7}
        & \textbf{\scriptsize Model} & \scriptsize VGG-like & \scriptsize AlexNet & \scriptsize VGG16 & \scriptsize ResNet18 & \scriptsize DenseNet161 \\
        \midrule
        Acc & 85.71 & 83.49 & 83.26 & 82.08 & 81.16 & 83.30 \\
        WSR & 100.00 & 83.45 & 71.45 & 58.50 & 51.65 & 42.30 \\
        \bottomrule
    \end{tabular}
    \vspace{-0.2cm}
    \caption{Robustness (\%) against cross-architecture stealing attacks.}
    \label{tab: cross-architecture model stealing attacks}
    \vspace{-0.3cm}
\end{table}

\begin{table}[t]
    \centering
    \vspace{-0.0cm}
    \footnotesize
    \begin{tabular}{
        m{0.4cm}<{\centering}
        m{0.8cm}<{\centering}
        m{0.9cm}<{\centering}
        m{0.7cm}<{\centering}
        m{0.7cm}<{\centering}
        m{0.8cm}<{\centering}
        m{1.2cm}<{\centering}}
        \toprule
        \multirow{3}{*}{} & \multirow{3}{*}{\textbf{\scriptsize \makecell[c]{Victim\\Model}}} & \multicolumn{5}{c}{\textbf{\scriptsize Stolen Model}} \\
        \cmidrule(lr){3-7}
        & & \scriptsize VGG16 & \scriptsize AlexNet & \scriptsize ResNet18 & \scriptsize ResNet34	& \scriptsize DenseNet161 \\
        & & \scriptsize CIFAR100 & \scriptsize FMNIST & \scriptsize NICO & \scriptsize COCO & \scriptsize TinyImageNet \\
        \midrule
        Acc & 85.71 & 76.56 & 34.18 & 65.60 & 76.37 & 78.98 \\
        WSR & 100.00 & 75.00 & 55.20 & 90.90 & 98.55 & 99.70 \\
        \bottomrule
    \end{tabular}
    \vspace{-0.2cm}
    \caption{Robustness (\%) against cross-dataset and cross-architecture stealing attacks.}
    \label{tab: cross-dataset and cross-architecture model stealing attacks}
    \vspace{-0.3cm}
\end{table}

\textbf{(1) Cross-Dataset Model Stealing Attacks}

In this scenario, the adversary knows the structure of victim model but is unaware of its training dataset. We employ the soft label method of Knockoff to steal models on the CIFAR10 task, and maintain consistent model architecture between stolen model and victim model, while adversary uses various different query datasets for the attack. Specifically, we utilize CIFAR10 \cite{krizhevsky2009learning}, CIFAR100 \cite{krizhevsky2009learning}, Fashion MNIST \cite{xiao2017fashion}, NICO \cite{he2021towards}, and TinyImageNet \cite{le2015tiny} as query datasets. Among these, CIFAR10 is in-distribution data for victim model, while the other four datasets are out-of-distribution data with varying degrees of similarity to the original task distribution. Detailed results are presented in Table \ref{tab: cross-dataset model stealing attacks}.

We find that when adversary uses in-distribution data (CIFAR10) for stealing, the accuracy of stolen model decreases by only 2.22\% compared to the original accuracy of victim model, with a watermark success rate of 83.45\%. Interestingly, although CIFAR100, NICO, and TinyImageNet are out-of-distribution data, their stolen model accuracies only decrease by 5.62\%, 9.88\%, and 4.34\% respectively. Additionally, these datasets achieve even higher watermark success rate than using CIFAR10 as the query data, all exceeding 99\%. This may be due to our watermark samples being composed of features from multiple classes, making it easier to learn watermark features from datasets with rich feature content. Despite the significant distribution difference between Fashion MNIST and CIFAR10, the accuracy of stolen model is only 31.88\%. However, our watermark success rate still far exceeds the default verification threshold and reaches 51.35\%.

\noindent\textbf{(2) Cross-Architecture Model Stealing Attacks}

In this scenario, adversary knows training data of victim model and can access data from the same distribution but is unaware of model architecture used by the victim. Specifically, we employ Knockoff soft-label attack to steal models on CIFAR10 task, and use CIFAR10 as query dataset, but adversary uses different model architectures for stolen model.

As shown in Table \ref{tab: cross-architecture model stealing attacks}, when using a VGG-like structure identical to the victim model for stolen model, we achieve the highest accuracy and watermark success rate, at 83.49\% and 83.45\%, respectively. We also observe that, although the accuracies of stolen models with different architectures are similar (ranging from 81.16\% to 83.30\%), the watermark success rate exhibits a general trend: as the number of layers in the stolen model increases, the watermark success rate decreases from 71.45\% with AlexNet to 42.30\% with DenseNet161. This suggests that when the query data is consistent, the closer the stolen model architecture is to the victim model architecture, the higher the watermark success rate on stolen model.

\noindent\textbf{(3) Cross-Dataset and Cross-Architecture Model Stealing Attacks}

We also evaluate a highly challenging real-world scenario where adversary is unaware of both the training data and model architecture used by the victim. Specifically, we conduct experiments using the Knockoff soft-label method on CIFAR10 task and employing query datasets such as CIFAR100, Fashion MNIST, NICO, COCO \cite{lin2014microsoft}, and TinyImageNet. The stolen model architectures including VGG16, AlexNet, ResNet18, ResNet34, and DenseNet161.

The results shown in Table \ref{tab: cross-dataset and cross-architecture model stealing attacks} demonstrate that our proposed method maintains robust performance even under highly challenging stealing attacks. Combining the results from Table \ref{tab: cross-dataset model stealing attacks}, \ref{tab: cross-architecture model stealing attacks}, and \ref{tab: cross-dataset and cross-architecture model stealing attacks}, we observe that when both the query data and model architecture used by the adversary differ from those of the victim, the query data has a more significant impact. For example, when using COCO and TinyImageNet for stealing, despite the significant architectural differences from the victim model, the accuracy and watermark success rate remain high, with accuracy drops of less than 10\% and watermark success rate above 98\%. Conversely, due to the substantial distribution difference between Fashion MNIST and the original task, it produces the worst result, with an accuracy drop of 51.53\%. However, even in this challenging scenario, our watermark success rate remains at 55.20\% and far exceeds the detection threshold.

\begin{figure}[t]
\vspace{-0.0cm}
\centering
\includegraphics[width=0.3\textwidth]{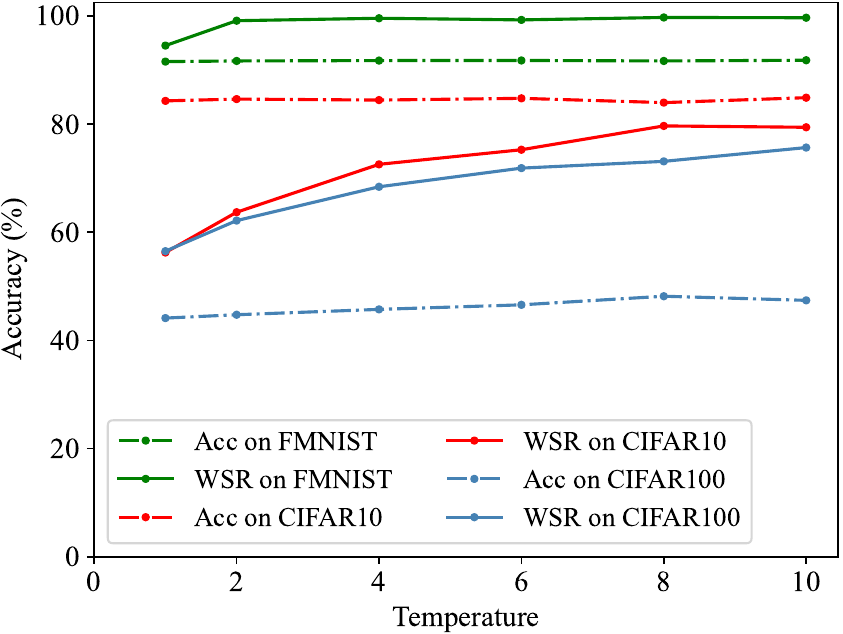}
\vspace{-0.2cm}
\caption{Robustness against distillation stealing attacks.}
\label{fig: distillation stealing}
\vspace{-0.3cm}
\end{figure}

\noindent\textbf{(4) Distillation-Based Model Stealing Attacks}

We emulate the concept of knowledge distillation \cite{hinton2015distilling} to perform distillation-based model stealing attack. However, we set the weight factor of hard label loss term in the standard distillation loss to zero, as we assume adversary lacks ground-truth labels in this attack. We conduct distillation-based stealing attacks on three tasks: Fashion MNIST, CIFAR10, and CIFAR100. The experimental results are shown in Figure \ref{fig: distillation stealing}.

We observe that as the temperature $T$ increases, both the accuracy of stolen model and the watermark success rate exhibit an overall upward trend. This is primarily because, with an increasing temperature $T$, the softmax output distribution becomes smoother. It increases the information entropy, which facilitates student model in distilling the knowledge of teacher model. However, higher temperatures are not always better. For different tasks, exceeding an optimal temperature threshold can lead to a slight performance decline. As illustrated in Figure \ref{fig: distillation stealing}, the watermark success rates of stolen models on all three tasks remain sufficiently high to ensure effective copyright verification. Specifically, the minimum watermark success rates for Fashion MNIST, CIFAR10, and CIFAR100 are 94.50\%, 56.25\%, and 56.50\%, respectively.

\begin{table}[t]
    \centering
    \vspace{-0.0cm}
    \footnotesize
    \begin{tabular}{
        m{0.5cm}<{\centering}
        m{0.85cm}<{\centering}
        m{1.3cm}
        m{1.0cm}<{\centering}
        m{1.15cm}<{\centering}
        m{1.15cm}<{\centering}}
        \toprule
        \multicolumn{2}{c}{\textbf{\scriptsize Homogeneity}} & \multirow{2}{*}{\textbf{\scriptsize Attack Stage}} & \multirow{2}{*}{\textbf{\scriptsize Victim}} & \multirow{2}{*}{\textbf{\scriptsize First Stage}} & \multirow{2}{*}{\textbf{\scriptsize Second Stage}} \\
        \cmidrule(lr){1-2}
        \scriptsize Attack & \scriptsize Structure & \centering $\longrightarrow$ & & & \\
        \midrule
        \multirow{4}{*}{\Checkmark} & \multirow{4}{*}{\Checkmark} & \scriptsize Attack $\rightarrow$ & \scriptsize None & \scriptsize Knockoff & \scriptsize Knockoff \\
        & & \scriptsize Structure $\rightarrow$ & \scriptsize VGG-like & \scriptsize VGG-like & \scriptsize VGG-like \\
        & & \scriptsize Acc $\rightarrow$ & 85.71 & 81.37 & 81.05 \\
        & & \scriptsize WSR $\rightarrow$ & 100.00 & 99.85 & 100.00 \\
        \midrule
        \multirow{4}{*}{\XSolidBrush} & \multirow{4}{*}{\Checkmark} & \scriptsize Attack $\rightarrow$ & \scriptsize None & \scriptsize Knockoff & \scriptsize Hard Label \\
        & & \scriptsize Structure $\rightarrow$ & \scriptsize VGG-like & \scriptsize VGG-like & \scriptsize VGG-like \\
        & & \scriptsize Acc $\rightarrow$ & 85.71 & 81.37 & 76.65 \\
        & & \scriptsize WSR $\rightarrow$ & 100.00 & 99.85 & 98.50 \\
        \midrule
        \multirow{4}{*}{\Checkmark} & \multirow{4}{*}{\XSolidBrush} & \scriptsize Attack $\rightarrow$ & \scriptsize None & \scriptsize Knockoff & \scriptsize Knockoff \\
        & & \scriptsize Structure $\rightarrow$ & \scriptsize VGG-like & \scriptsize ResNet18 & \scriptsize AlexNet \\
        & & \scriptsize Acc $\rightarrow$ & 85.71 & 76.16 & 77.73 \\
        & & \scriptsize WSR $\rightarrow$ & 100.00 & 98.10 & 98.80 \\
        \midrule
        \multirow{4}{*}{\XSolidBrush} & \multirow{4}{*}{\XSolidBrush} & \scriptsize Attack $\rightarrow$ & \scriptsize None & \scriptsize Knockoff & \scriptsize Hard Label \\
        & & \scriptsize Structure $\rightarrow$ & \scriptsize VGG-like & \scriptsize ResNet18 & \scriptsize AlexNet \\
        & & \scriptsize Acc $\rightarrow$ & 85.71 & 76.16 & 70.76 \\
        & & \scriptsize WSR $\rightarrow$ & 100.00 & 98.10 & 89.30 \\
        \bottomrule
    \end{tabular}
    \vspace{-0.2cm}
    \caption{Robustness (\%) against double-stage stealing attacks.}
    \label{tab: double-stage stealing attacks}
    \vspace{-0.3cm}
\end{table}

\noindent\textbf{(5) Double-Stage Model Stealing Attacks}

In the previous experiments, the attacker usually employs a single stealing attack method to obtain the stolen model. Here, we challenge a more difficult scenario where the attacker uses a two-stage stealing attack process to obtain the final stolen model. The target task of attacker is CIFAR10, with TinyImageNet as the query dataset. The related stealing attack settings and experimental results are presented in Table \ref{tab: double-stage stealing attacks}.

We can see that for four different combinations of stealing attack types, our method all maintains strong robustness. After two stages of stealing, the accuracy of stolen models decreases by 4.66\%, 9.06\%, 7.98\%, and 14.95\%, respectively. However, the final watermark success rates are still 100\%, 98.50\%, 98.80\%, and 89.30\%, which indicates a high confidence in asserting model ownership. It is noteworthy that the last combination of attacks is the most challenging, and it results in the relatively lowest accuracy and watermark success rate, which aligns with our expectations.

\subsection{Robustness against Watermark Detection and Evasion Attacks}
\label{sec: Robustness against Watermark Detection and Evasion Attacks}

\begin{table}[t]
    \centering
    \vspace{-0.0cm}
    \footnotesize
    \begin{tabular}{
        m{0.7cm}<{\centering}
        m{0.7cm}<{\centering}
        m{0.5cm}<{\centering}
        m{0.8cm}<{\centering}
        m{0.5cm}<{\centering}
        m{0.5cm}<{\centering}
        m{0.6cm}<{\centering}
        m{0.9cm}<{\centering}}
        \toprule
        \multirow{2}{*}{\textbf{\scriptsize Dataset}} & \multirow{2}{*}{\textbf{\scriptsize Method}} & \multicolumn{2}{c}{\textbf{\scriptsize Original}} & \multicolumn{2}{c}{\textbf{\scriptsize After-Det.}} & \multicolumn{2}{c}{\textbf{\scriptsize Det. Performance}}\\
        \cmidrule(lr){3-4} \cmidrule(lr){5-6} \cmidrule(lr){7-8}
        & & \scriptsize Acc & \scriptsize WSR & \scriptsize Acc & \scriptsize WSR & \scriptsize Benign & \scriptsize Watermark \\
        \midrule
        \multirow{2}{*}{\scriptsize FMNIST} & \scriptsize LOF & 91.32 & 100.00 & 91.29 & 66.47 & 0.20 & 37.12 \\
        & \scriptsize IF & 91.32 & 100.00 & 91.26 & 94.22 & 0.24 & 6.55 \\
        \midrule
        \multirow{2}{*}{\scriptsize CIFAR10} & \scriptsize LOF & 85.49 & 99.99 & 85.43 & 84.70 & 0.30 & 17.14 \\
        & \scriptsize IF & 85.49 & 99.99 & 85.05 & 99.98 & 0.83 & 0.02 \\
        \midrule
        \multirow{2}{*}{\scriptsize CIFAR100} & \scriptsize LOF & 50.75 & 99.97 & 50.73 & 84.24 & 0.60 & 15.87 \\
        & \scriptsize IF & 50.75 & 99.97 & 50.63 & 99.96 & 0.64 & 0.01 \\
        \bottomrule
    \end{tabular}
    \vspace{-0.2cm}
    \caption{Evaluation of anomaly detection on victim models.}
    \label{tab: anomaly detection on victim}
    \vspace{-0.0cm}
\end{table}

\begin{table}[t]
    \centering
    \vspace{-0.0cm}
    \footnotesize
    \begin{tabular}{
        m{0.7cm}<{\centering}
        m{0.7cm}<{\centering}
        m{0.5cm}<{\centering}
        m{0.8cm}<{\centering}
        m{0.5cm}<{\centering}
        m{0.5cm}<{\centering}
        m{0.6cm}<{\centering}
        m{0.9cm}<{\centering}}
        \toprule
        \multirow{2}{*}{\textbf{\scriptsize Dataset}} & \multirow{2}{*}{\textbf{\scriptsize Method}} & \multicolumn{2}{c}{\textbf{\scriptsize Original}} & \multicolumn{2}{c}{\textbf{\scriptsize After-Det.}} & \multicolumn{2}{c}{\textbf{\scriptsize Det. Performance}}\\
        \cmidrule(lr){3-4} \cmidrule(lr){5-6} \cmidrule(lr){7-8}
        & & \scriptsize Acc & \scriptsize WSR & \scriptsize Acc & \scriptsize WSR & \scriptsize Benign & \scriptsize Watermark \\
        \midrule
        \multirow{2}{*}{\scriptsize FMNIST} & \scriptsize LOF & 87.11 & 100.00 & 87.02 & 85.39 & 0.13 & 16.10 \\
        & \scriptsize IF & 87.11 & 100.00 & 86.93 & 94.49 & 0.19 & 6.10 \\
        \midrule
        \multirow{2}{*}{\scriptsize CIFAR10} & \scriptsize LOF & 75.78 & 94.49 & 75.49 & 84.92 & 0.15 & 11.08 \\
        & \scriptsize IF & 75.78 & 94.49 & 75.17 & 94.54 & 0.74 & 0.05 \\
        \midrule
        \multirow{2}{*}{\scriptsize CIFAR100} & \scriptsize LOF & 31.03 & 100.00 & 30.78 & 97.60 & 0.09 & 2.41 \\
        & \scriptsize IF & 31.03 & 100.00 & 30.68 & 99.99 & 0.69 & 0.01 \\
        \bottomrule
    \end{tabular}
    \vspace{-0.2cm}
    \caption{Evaluation of anomaly detection on stolen models.}
    \label{tab: anomaly detection on stolen}
    \vspace{-0.3cm}
\end{table}

\begin{figure*}[t]
    \vspace{-0.0cm}
    \centering
    \begin{subfigure}{0.24\textwidth}
        \centering
        \includegraphics[width=1.0\textwidth]{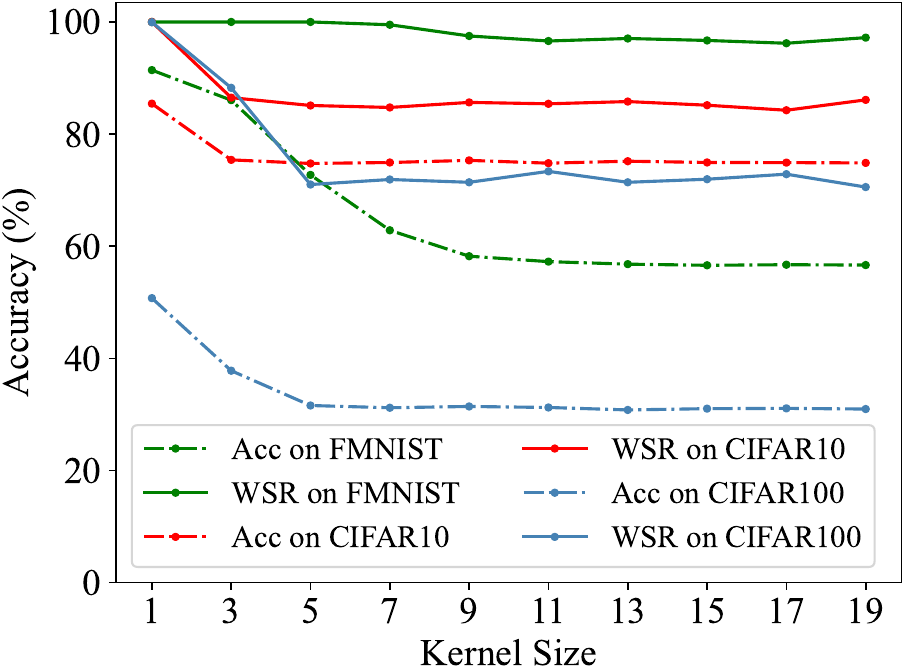}
        \caption{Gaussian Blur}
        \label{fig: Gaussian Blur}
    \end{subfigure}
    \centering
    \begin{subfigure}{0.24\textwidth}
        \centering
        \includegraphics[width=1.0\textwidth]{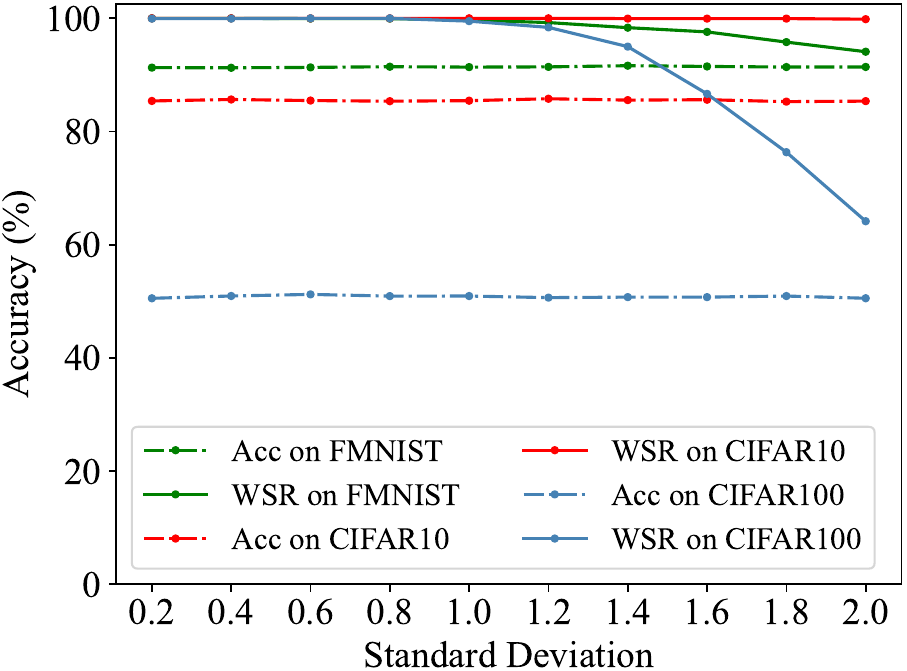}
        \caption{Gaussian Noise}
        \label{fig: Gaussian Noise}
    \end{subfigure}
    \centering
    \begin{subfigure}{0.24\textwidth}
        \centering
        \includegraphics[width=1.0\textwidth]{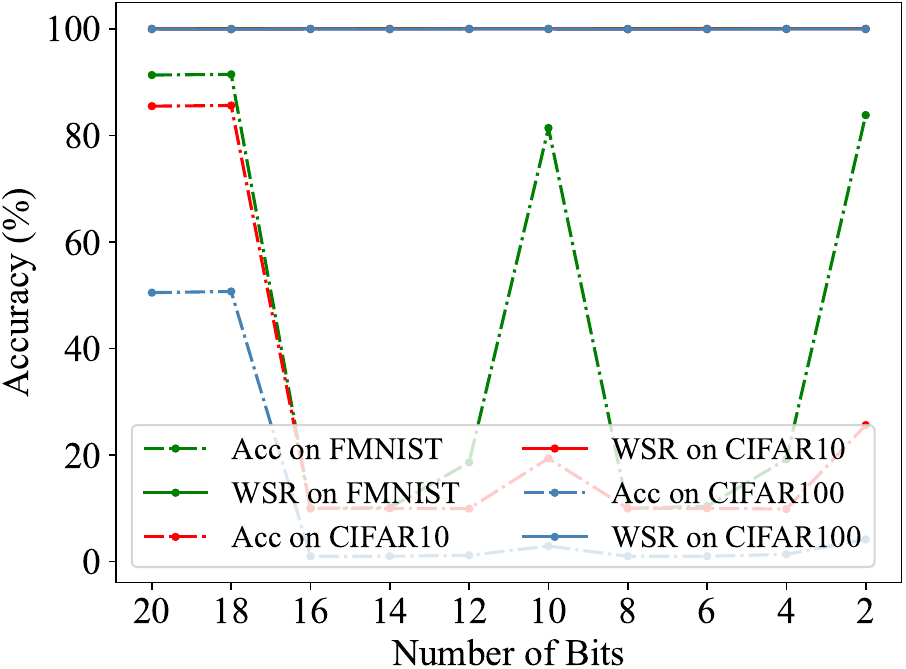}
        \caption{Input Quantization}
        \label{fig: Input Quantization}
    \end{subfigure}
    \centering
    \begin{subfigure}{0.24\textwidth}
        \centering
        \includegraphics[width=1.0\textwidth]{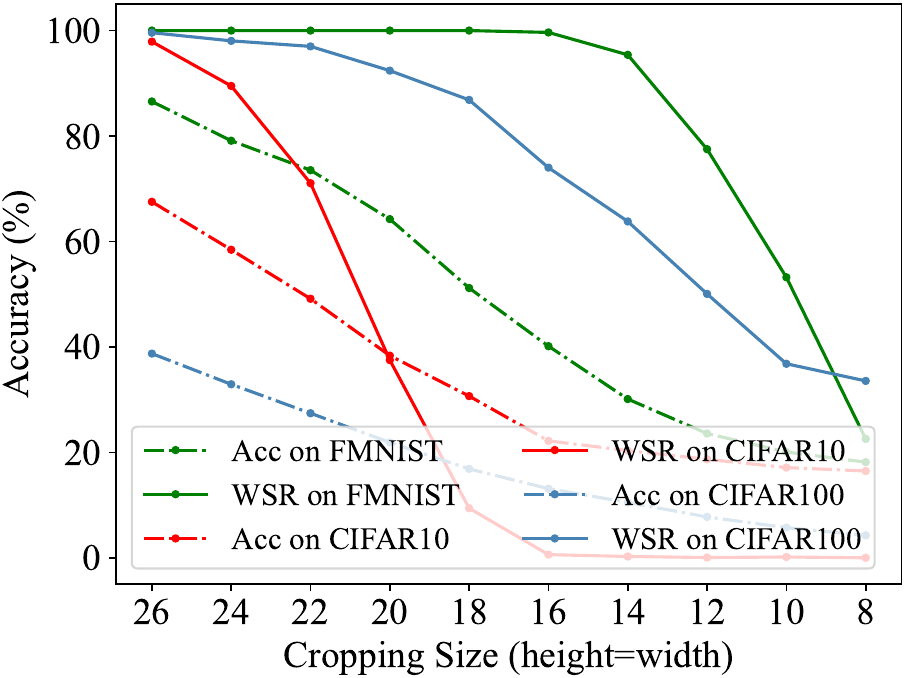}
        \caption{Input Cropping}
        \label{fig: Input Cropping}
    \end{subfigure}
    \vspace{-0.2cm}
    \caption{Robustness against various input preprocessing attacks on victim model.}
    \label{fig: Input Preprocessing}
    \vspace{-0.1cm}
\end{figure*}

To explore the robustness of DeepTracer against detection and evasion attacks, we first apply two popular anomaly detection methods, Local Outlier Factor (LOF) \cite{breunig2000lof} and Isolation Forest (IF) \cite{liu2008isolation}, to both victim and stolen models. Then we test four input preprocessing methods \cite{lukas2022sok} (i.e., Gaussian Blur, Gaussian Noise, Input Quantization, and Input Cropping) on victim models.

\textbf{Anomaly Detection.} The detection results for the victim (watermarked) model are shown in Table \ref{tab: anomaly detection on victim}. It can be seen that for all detection tasks, the detection rate of the two detection methods on watermark samples does not exceed 38\%. After adding the detector, the watermark success rate is still over 66.47\%. Similar results in Table \ref{tab: anomaly detection on stolen} indicate that our method also has robustness on stolen model. These experiments demonstrate the robustness of our watermarking method against various anomaly detection algorithms.

\textbf{Input Preprocessing.} The results in Figure \ref{fig: Input Preprocessing} show that our watermarking method has the worst watermark success rates of 70.55\%, 64.15\%, and 99.95\% for Gaussian Blur, Gaussian Noise, and Input Quantization, respectively, which is still sufficient to declare ownership. Although our watermark success rate will be lower than the threshold of 20\% under the high attack intensity settings of Input Cropping methods, the primary task performance of the model will also seriously decline, i.e., the model becomes unusable.

\subsection{Watermark Removal Attacks on Victim Models}
\label{sec: Watermark Removal Attacks on Victim Models}

\begin{figure*}[t]
    \centering
    \begin{subfigure}{0.24\textwidth}
        \centering
        \includegraphics[width=1.0\textwidth]{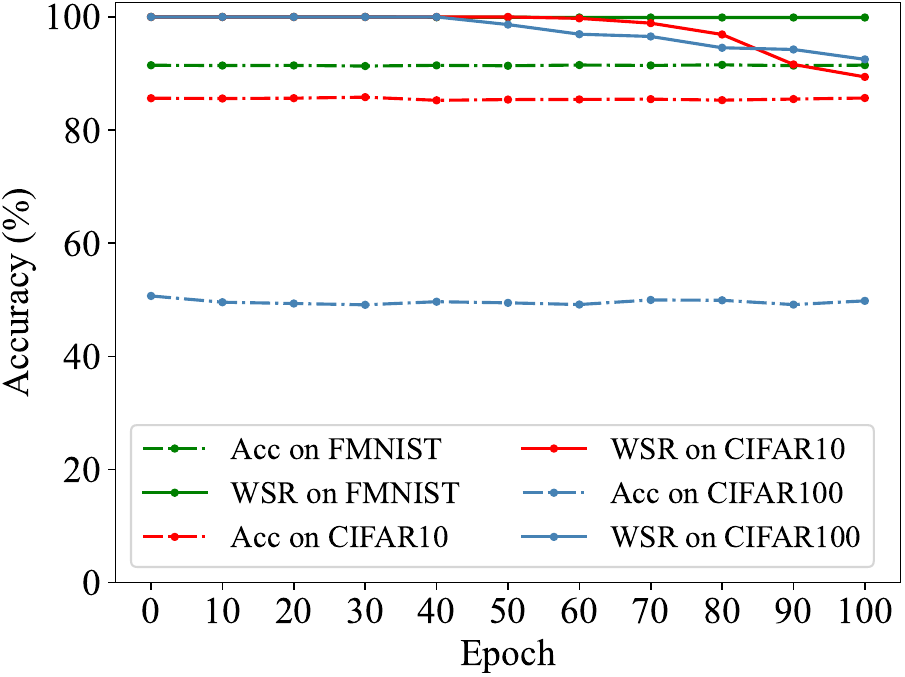}
        \caption{FTLL on Victim Model}
        \label{fig: FTLL on Victim Model}
    \end{subfigure}
    \centering
    \begin{subfigure}{0.24\textwidth}
        \centering
        \includegraphics[width=1.0\textwidth]{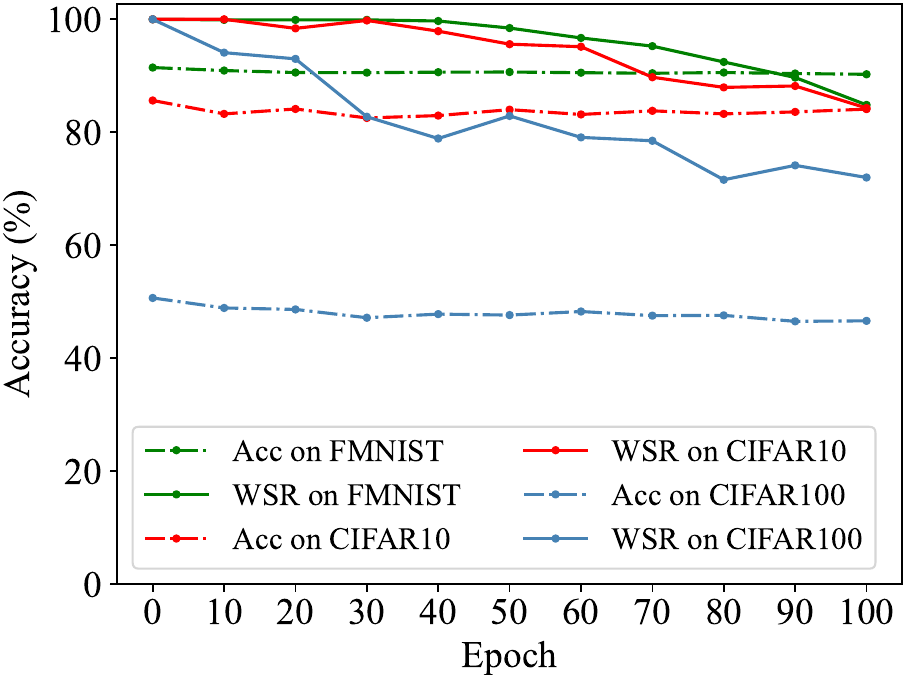}
        \caption{FTAL on Victim Model}
        \label{fig: FTAL on Victim Model}
    \end{subfigure}
    \centering
    \begin{subfigure}{0.24\textwidth}
        \centering
        \includegraphics[width=1.0\textwidth]{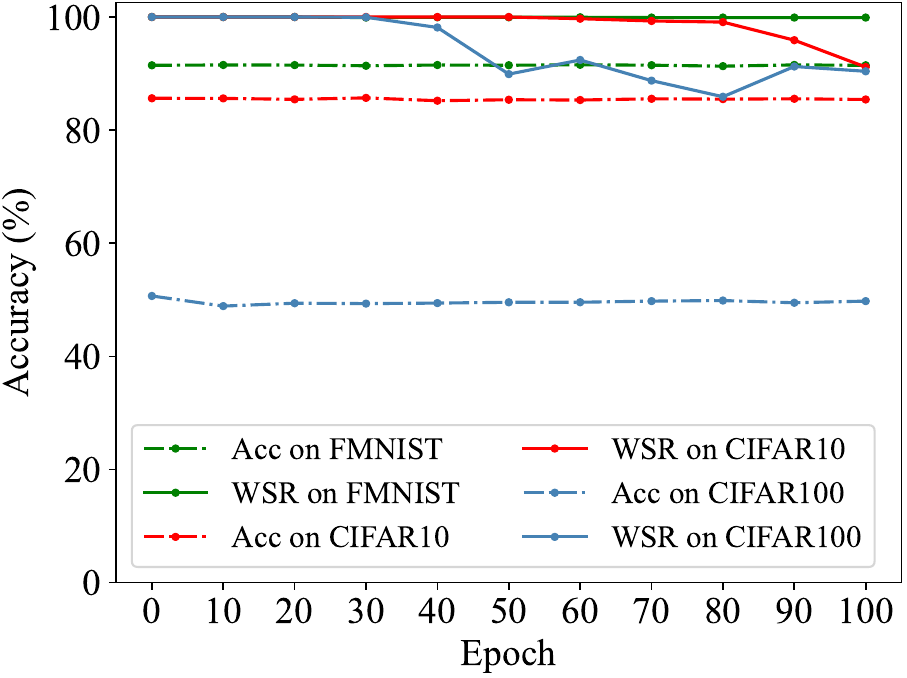}
        \caption{RTLL on Victim Model}
        \label{fig: RTLL on Victim Model}
    \end{subfigure}
    \centering
    \begin{subfigure}{0.24\textwidth}
        \centering
        \includegraphics[width=1.0\textwidth]{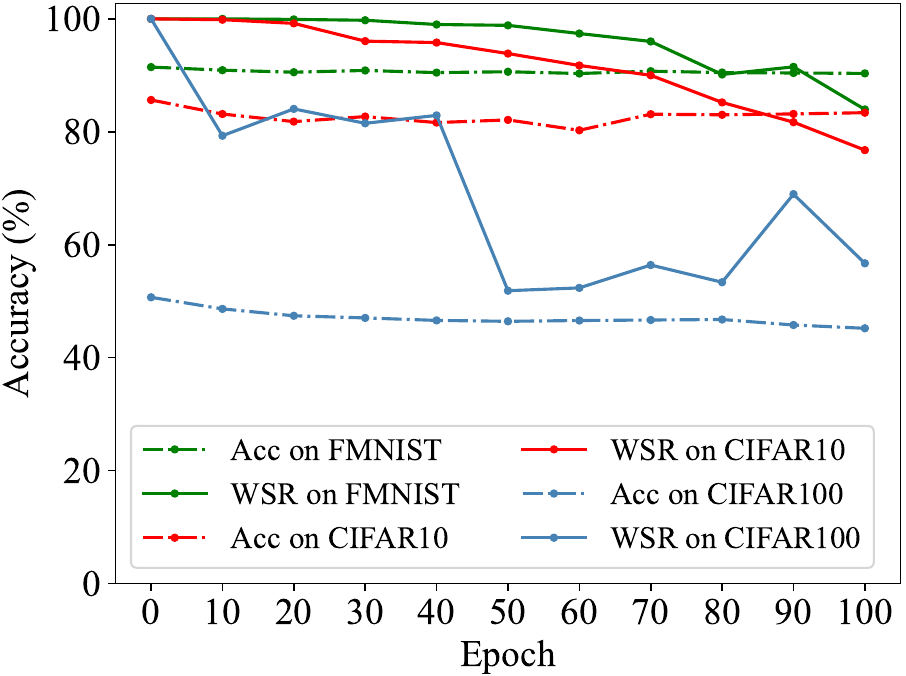}
        \caption{RTAL on Victim Model}
        \label{fig: RTAL on Victim Model}
    \end{subfigure}
    \vspace{-0.2cm}
    \caption{Robustness against four fine-tuning attacks on victim model.}
    \label{fig: fine-tuning attacks on victim model}
    \vspace{-0.1cm}
\end{figure*}

\begin{figure}[t]
\centering
\includegraphics[width=0.3\textwidth]{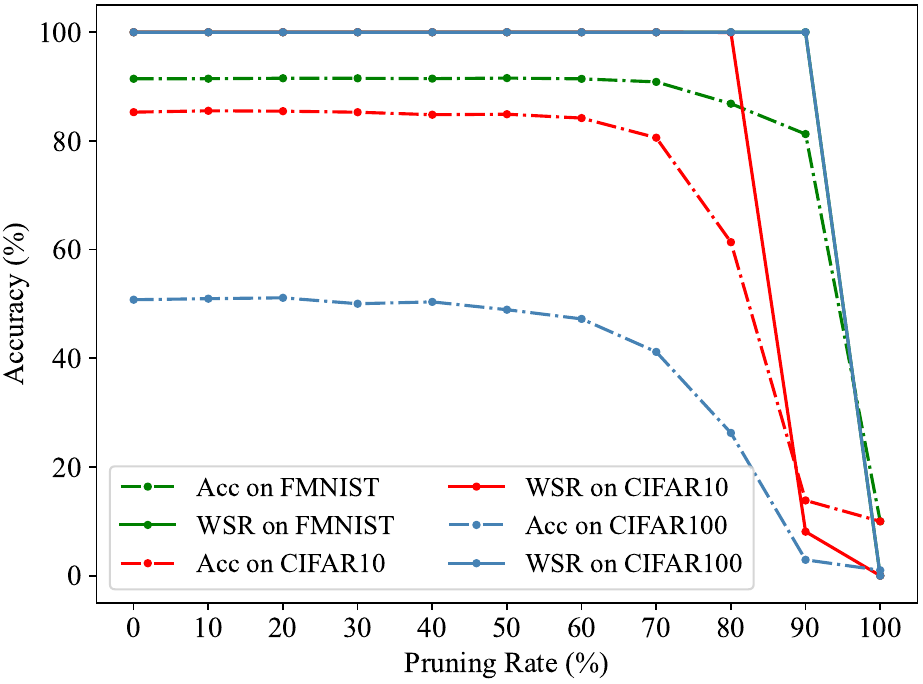}
\vspace{-0.2cm}
\caption{Robustness against pruning attack on victim model.}
\label{fig: pruning attack on victim model}
\vspace{-0.0cm}
\end{figure}

\begin{table}[t]
    \centering
    \vspace{-0.0cm}
    \footnotesize
    \begin{tabular}{
        m{0.7cm}<{\centering}
        m{0.7cm}<{\centering}
        m{0.9cm}<{\centering}
        m{0.7cm}<{\centering}
        m{0.9cm}<{\centering}
        m{0.7cm}<{\centering}
        m{0.9cm}<{\centering}}
        \toprule
        \multirow{2}{*}{\textbf{\scriptsize Bit Size}} & \multicolumn{2}{c}{\textbf{\scriptsize FMNIST}} & \multicolumn{2}{c}{\textbf{\scriptsize CIFAR10}} & \multicolumn{2}{c}{\textbf{\scriptsize CIFAR100}}\\
        \cmidrule(lr){2-3} \cmidrule(lr){4-5} \cmidrule(lr){6-7}
        & \scriptsize Acc & \scriptsize WSR & \scriptsize Acc & \scriptsize WSR & \scriptsize Acc & \scriptsize WSR \\
        \midrule
        16 & 91.43 & 100.00 & 85.29 & 100.00 & 50.79 & 100.00 \\
        8 & 91.43 & 100.00 & 85.45 & 100.00 & 50.91 & 100.00 \\
        6 & 91.50 & 100.00 & 85.31 & 100.00 & 50.74 & 100.00 \\
        4 & 90.55 & 100.00 & 81.33 & 100.00 & 44.28 & 100.00 \\
        3 & 25.74 & 69.40 & 76.43 & 100.00 & 1.76 & 84.10 \\
        2 & 10.18 & 57.95 & 10.04 & 0.05 & 1.00 & 0.00 \\
        1 & 10.00 & 0.00 & 10.00 & 0.00 & 1.00 & 0.00 \\
        \bottomrule
    \end{tabular}
    \vspace{-0.2cm}
    \caption{Quantization attack on victim model.}
    \label{tab: quantization attack on victim model}
    \vspace{-0.3cm}
\end{table}

\begin{figure*}[t]
    \centering
    \vspace{-0.0cm}
    \begin{subfigure}{0.24\textwidth}
        \centering
        \includegraphics[width=1.0\textwidth]{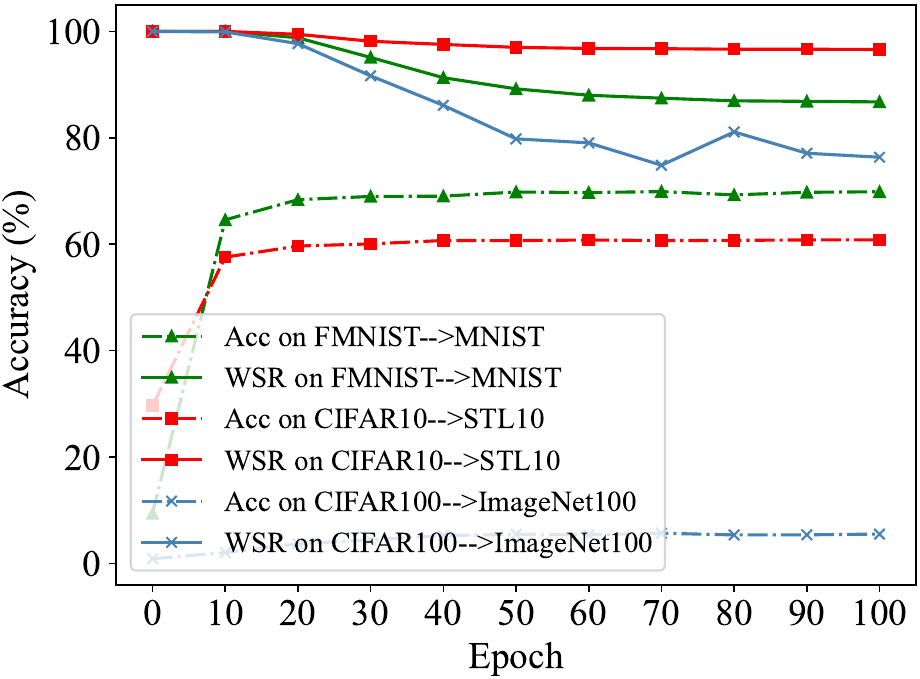}
        \caption{Transfer Learning with FTLL}
        \label{fig: Transfer learning with FTLL on victim model}
    \end{subfigure}
    \centering
    \begin{subfigure}{0.24\textwidth}
        \centering
        \includegraphics[width=1.0\textwidth]{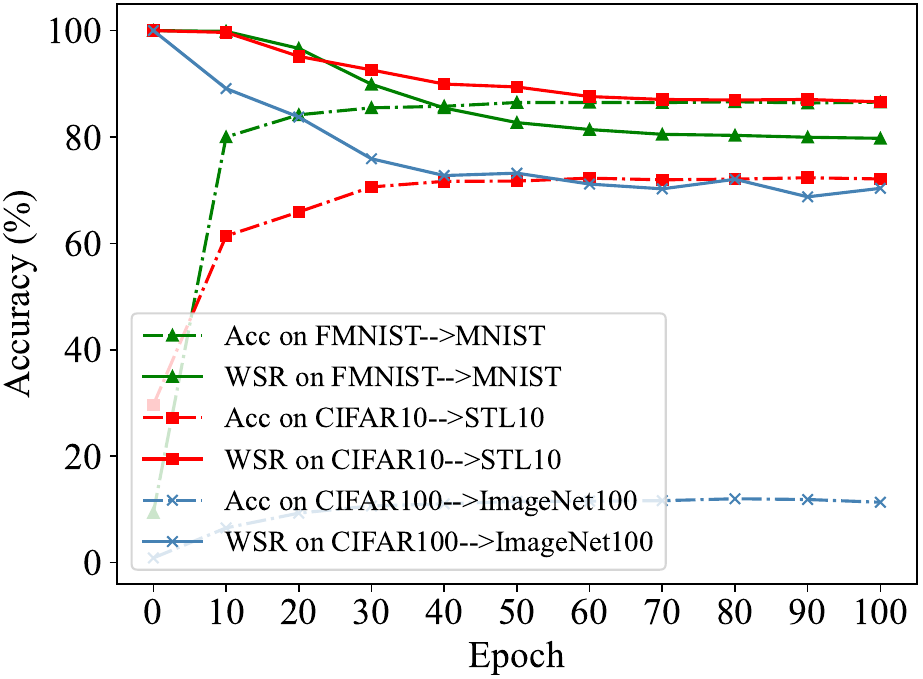}
        \caption{Transfer Learning with FTAL}
        \label{fig: Transfer learning with FTAL on victim model}
    \end{subfigure}
    \centering
    \begin{subfigure}{0.24\textwidth}
        \centering
        \includegraphics[width=1.0\textwidth]{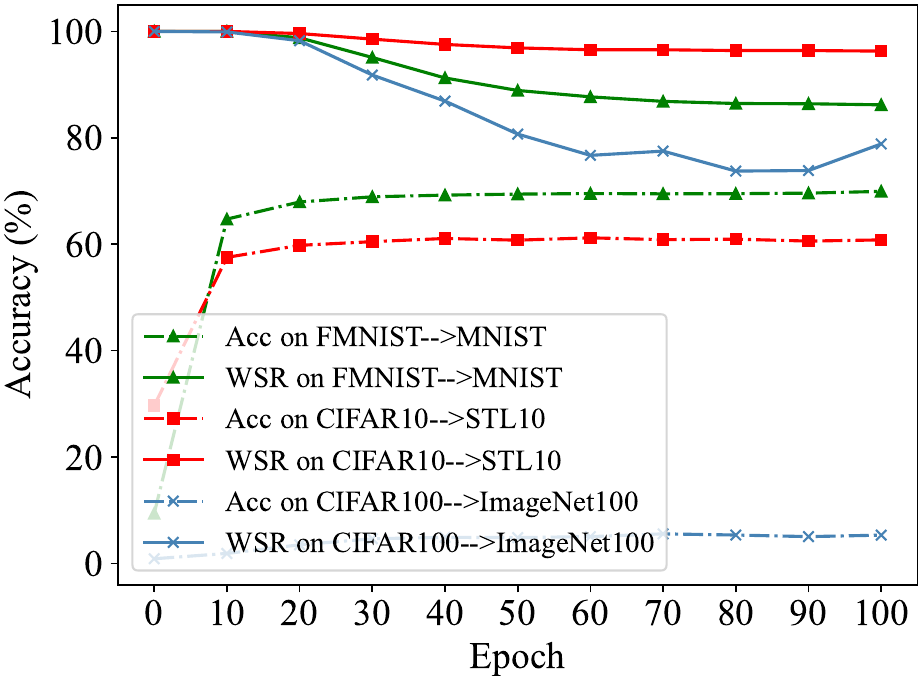}
        \caption{Transfer Learning with RTLL}
        \label{fig: Transfer learning with RTLL on victim model}
    \end{subfigure}
    \centering
    \begin{subfigure}{0.24\textwidth}
        \centering
        \includegraphics[width=1.0\textwidth]{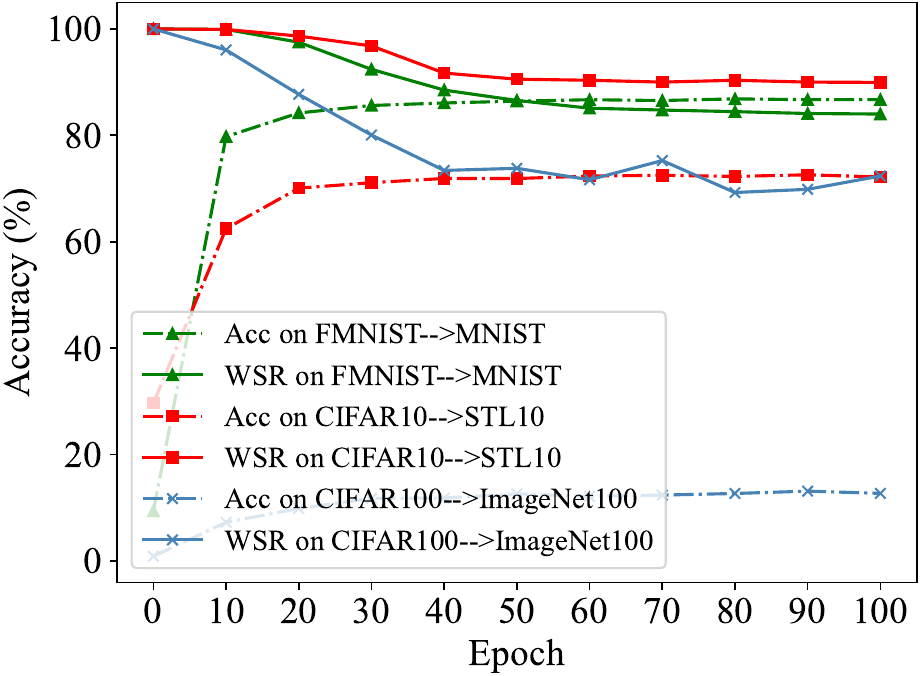}
        \caption{Transfer Learning with RTAL}
        \label{fig: Transfer learning with RTAL on victim model}
    \end{subfigure}
    \vspace{-0.2cm}
    \caption{Robustness against transfer learning attacks on victim model with four fine-tuning methods.}
    \label{fig: transfer learning attacks on victim model}
    \vspace{-0.1cm}
\end{figure*}

\noindent\textbf{(1) Watermark Removal Attack Methods}

$\bullet$ Fine-Tuning \cite{adi2018turning}:
Fine-tuning involves adjusting or retraining some or all layers of the obtained model using original training data or query data, depending on different threat scenarios, to remove internal watermarks. We experiment with four fine-tuning strategies:

\textit{Fine-Tune Last Layer (FTLL)}: Freezes all parameters except for the last layer, and updates only this layer's parameters.

\textit{Fine-Tune All Layers (FTAL)}: No layers are frozen, and updates all network parameters.

\textit{Re-Train Last Layer (RTLL)}: Freezes all layers except the last output layer, and reinitializes its parameters and updates only this layer.

\textit{Re-Train All Layers (RTAL)}: Reinitializes the last output layer's parameters, with no layers frozen, and updates the parameters of all layers.

$\bullet$ Pruning \cite{uchida2017embedding}:
Weight pruning compresses the model by setting a certain proportion of weights with the smallest magnitude to zero. We evaluate the changes in model accuracy and watermark success rate as the pruning rate increases from 0\% to 100\%.

$\bullet$ Quantization \cite{lukas2022sok}:
Weight quantization is another popular model compression technique that compresses model weights to lower bit representations to save storage. This differs from input quantization, where quantization targets input images, whereas here it targets model weights. We assess robustness of our method with bit size of 16, 8, 6, 4, 3, 2, and 1.

$\bullet$ Transfer Learning \cite{lukas2022sok}:
Transfer learning applies the knowledge from a trained model on one problem domain to another target domain, and it typically involves fine-tuning with a new dataset. We apply the four aforementioned fine-tuning methods to transfer the knowledge of watermarked models trained on Fashion MNIST, CIFAR10, and CIFAR100 to the MNIST \cite{lecun1998gradient}, STL10 \cite{coates2011analysis}, and ImageNet100\footnote{ImageNet100, a subset of ImageNet with 100 randomly selected classes.} tasks, respectively.

\noindent\textbf{(2) Results of Removal Attacks on Victim Models}

\textbf{Fine-Tuning.}
Figure \ref{fig: fine-tuning attacks on victim model} illustrates robustness of our method against four popular fine-tuning attacks on victim models. Our method demonstrates superior robustness across all fine-tuning strategies and tasks. Even in the most challenging scenario of RTAL on CIFAR100, the WSR remains well above the 20\% threshold and reaches 56.70\%. We observe that fine-tuning or retraining only the last layer (FTLL and RTLL) results in smaller decrease in accuracy and WSR compared to fine-tuning or retraining all layers. Moreover, retraining causes greater performance degradation than fine-tuning when the number of updated layers is the same. This indicates that the extent and depth of parameter changes significantly impact model accuracy and WSR after an attack.

\textbf{Pruning.}
Pruning attack results, shown in Figure \ref{fig: pruning attack on victim model}, reveal that our watermark is only removed when the pruning rate reaches 100\%, 90\%, and 100\% for Fashion MNIST, CIFAR10, and CIFAR100, respectively, at which point the models lose usability in their primary tasks. This demonstrates the strong robustness of our method against weight pruning attacks.

\textbf{Quantization.}
We evaluate robustness of our method against weight quantization attacks. Table \ref{tab: quantization attack on victim model} shows that at a bit size of 3, model accuracies on Fashion MNIST, CIFAR10, and CIFAR100 decrease by 65.69\%, 8.86\%, and 49.03\%, respectively, while WSRs remain high at 69.40\%, 100.00\%, and 84.10\%, respectively, and this provides high confidence in verifying ownership. Overall, quantization is an ineffective attack method against our watermark under the condition that the model accuracy cannot be degraded too much.

\textbf{Transfer Learning.}
As shown in Figure \ref{fig: transfer learning attacks on victim model}, our watermark also exhibits significant robustness against transfer learning attacks. Specifically, when transferring the knowledge of watermarked models from Fashion MNIST to MNIST, the accuracies of target task under four fine-tuning methods are 69.86\%, 86.56\%, 69.93\%, and 86.71\%, respectively, with corresponding WSRs of 86.75\%, 79.75\%, 86.20\%, and 84.00\%. Similarly, transferring watermarked models from CIFAR10 and CIFAR100 to STL10 and ImageNet100 result in final average accuracies of target task of 66.47\% and 8.72\%, and average WSRs of 92.36\% and 74.48\%, respectively. Thus, transfer learning incurs some loss for our watermark, but is insufficient to completely remove it (all scenarios retain a WSR of at least 70\% after 100 epochs, which well above the 20\% threshold).

\subsection{Robustness against Watermark Ambiguity Attacks}
\label{sec: Robustness against Watermark Ambiguity Attacks}
Watermark ambiguity attacks represent a typical model copyright infringement scenario, wherein an adversary embeds their own watermark into a stolen model, resulting in dual watermarks and conflicting ownership claims. In this section, we evaluate robustness of our watermarking against such attacks. First, as demonstrated in Sections \ref{sec: Robustness against Model Stealing Attacks} through \ref{sec: Robustness against Watermark Removal Attacks}, our watermark cannot be easily removed by adversaries. Following the setup in \cite{lv2024mea}, we implement a powerful watermark ambiguity attack, where assumes adversary possesses the training data of victim model. The adversary uses 20\% of training samples to embed their counterfeit watermark, while the remaining 80\% serve as query samples for model stealing attack.

Table \ref{tab: robustness against watermark ambiguity attacks} intuitively illustrates that our watermark remains intact in the final stolen models, with watermark success rates of 100\%, 95.15\%, and 100\% for the Fashion MNIST, CIFAR10, and CIFAR100 tasks, respectively. Unfortunately, the adversary's counterfeit watermark success rates are merely 11.16\%, 9.79\%, and 0.89\%, and insufficient to validate model ownership. Overall, our watermarking method effectively resists watermark ambiguity attacks.

\begin{table}[t]
    \centering
    \footnotesize
    \begin{tabular}{
        m{1.2cm}<{\centering}
        m{0.8cm}<{\centering}
        m{0.6cm}<{\centering}
        m{0.8cm}<{\centering}
        m{0.6cm}<{\centering}
        m{0.8cm}<{\centering}
        m{0.6cm}<{\centering}}
        \toprule
        \multirow{2}{*}{\textbf{\scriptsize Watermark}} & \multicolumn{2}{c}{\textbf{\scriptsize FMNIST}} & \multicolumn{2}{c}{\textbf{\scriptsize CIFAR10}} & \multicolumn{2}{c}{\textbf{\scriptsize CIFAR100}}\\
        \cmidrule(lr){2-3} \cmidrule(lr){4-5} \cmidrule(lr){6-7}
        $\longrightarrow$ & \scriptsize Pirated & \scriptsize Our & \scriptsize Pirated & \scriptsize Our & \scriptsize Pirated & \scriptsize Our \\
        \midrule
        WSR & 11.16 & 100.00 & 9.79 & 95.15 & 0.89 & 100.00 \\
        \bottomrule
    \end{tabular}
    \vspace{-0.2cm}
    \caption{Robustness against watermark ambiguity attacks.}
    \label{tab: robustness against watermark ambiguity attacks}
    \vspace{-0.3cm}
\end{table}

\begin{table*}[t]
    \centering
    \vspace{-0.0cm}
    \footnotesize

    \begin{subtable}[t]{1.0\linewidth}
    \centering
    \begin{tabular}{
        m{1.0cm}
        m{1.0cm}<{\centering}
        m{1.0cm}<{\centering}
        m{1.0cm}<{\centering}
        m{1.0cm}<{\centering}
        m{1.0cm}<{\centering}
        m{1.0cm}<{\centering}
        m{1.0cm}<{\centering}
        m{1.0cm}<{\centering}
        m{1.0cm}<{\centering}
        m{1.0cm}<{\centering}
        }
        \toprule
        \textbf{$\epsilon$} & 0.01 & 0.05 & 0.10 & 0.15 & 0.20 & 0.25 & 0.30 & 0.40 & 0.50 & 1.00 \\
        \midrule
        \textbf{Acc} & 74.87 & 74.69 & 74.28 & 74.28 & 75.05 & 74.10 & 74.20 & 74.62 & 74.00 & 75.35 \\
        \textbf{WSR} & 93.70 & 94.70 & 94.15 & 95.55 & 88.60 & 93.45 & 91.95 & 94.10 & 91.15 & 93.55 \\
        \bottomrule
    \end{tabular}
    \vspace{-0.1cm}
    \caption{Model Stealing with Adversarial Training}
    \label{subtab: stealing with adversarial training}
    \end{subtable}
    \vspace{0.2cm}

    \begin{subtable}[t]{1.0\linewidth}
    \centering
    \begin{tabular}{
        m{0.8cm}
        m{0.8cm}<{\centering}
        m{0.8cm}<{\centering}
        m{0.8cm}<{\centering}
        m{0.8cm}<{\centering}
        m{0.8cm}<{\centering}
        m{0.8cm}<{\centering}
        m{0.8cm}<{\centering}
        m{0.8cm}<{\centering}
        m{0.75cm}<{\centering}
        m{0.75cm}<{\centering}
        m{0.75cm}<{\centering}
        m{0.75cm}<{\centering}
        }
        \toprule
        \textbf{std} & 0.001 & 0.005 & 0.01 & 0.05 & 0.1 & 0.15 & 0.2 & 0.3 & 0.4 & 0.5 & 0.6 & 1.0 \\
        \midrule
        \textbf{Acc} & 32.42 & 30.31 & 27.88 & 33.62 & 31.90 & 32.53 & 34.18 & 30.21 & 34.41 & 46.51 & 39.59 & 35.75 \\
        \textbf{WSR} & 38.55 & 34.70 & 35.90 & 56.35 & 46.10 & 73.75 & 78.00 & 66.55 & 58.65 & 96.85 & 96.00 & 52.20 \\
        \bottomrule
    \end{tabular}
    \vspace{-0.1cm}
    \caption{Model Stealing with Noisy Inputs}
    \label{subtab: stealing with noisy inputs}
    \end{subtable}
    \vspace{0.2cm}

    \begin{subtable}[t]{1.0\linewidth}
    \centering
    \begin{tabular}{
        m{1.0cm}
        m{1.0cm}<{\centering}
        m{1.0cm}<{\centering}
        m{1.0cm}<{\centering}
        m{1.0cm}<{\centering}
        m{1.0cm}<{\centering}
        m{1.0cm}<{\centering}
        m{1.0cm}<{\centering}
        m{1.0cm}<{\centering}
        m{1.0cm}<{\centering}
        m{1.0cm}<{\centering}
        }
        \toprule
        \textbf{Epoch} & 10 & 20 & 30 & 40 & 50 & 60 & 70 & 80 & 90 & 100 \\
        \midrule
        \textbf{Acc} & 66.22 & 68.21 & 67.95 & 67.36 & 67.40 & 66.85 & 66.71 & 66.83 & 66.34 & 66.32 \\
        \textbf{WSR} & 39.55 & 49.80 & 61.50 & 54.20 & 53.60 & 56.40 & 56.80 & 55.25 & 55.20 & 55.95 \\
        \bottomrule
    \end{tabular}
    \vspace{-0.1cm}
    \caption{Model Stealing under Continuous Distribution Shifts}
    \label{subtab: stealing under continuous shift}
    \end{subtable}
    \vspace{0.2cm}

    \begin{subtable}[t]{1.0\linewidth}
    \centering
    \begin{tabular}{
        m{1.2cm}
        m{0.75cm}<{\centering}
        m{0.75cm}<{\centering}
        m{0.75cm}<{\centering}
        m{0.75cm}<{\centering}
        m{0.75cm}<{\centering}
        m{0.75cm}<{\centering}
        m{0.75cm}<{\centering}
        m{0.75cm}<{\centering}
        m{0.75cm}<{\centering}
        m{0.75cm}<{\centering}
        m{0.75cm}<{\centering}
        m{0.75cm}<{\centering}
        }
        \toprule
        \textbf{Optimizer} & \multicolumn{4}{c}{\textbf{Adagrad}} & \multicolumn{4}{c}{\textbf{Adamax}} & \multicolumn{4}{c}{\textbf{RMSprop}} \\
        \cmidrule(lr){2-5} \cmidrule(lr){6-9} \cmidrule(lr){10-13}
        \textbf{LR} & 0.001 & 0.01 & 0.03 & 0.05 & 0.001 & 0.003 & 0.004 & 0.005 & 0.001 & 0.003 & 0.005 & 0.008 \\
        \midrule
        \textbf{Acc} & 77.41 & 61.63 & 51.94 & 39.21 & 80.90 & 76.40 & 75.74 & 10.00 & 76.53 & 69.28 & 56.53 & 10.00 \\
        \textbf{WSR} & 96.95 & 49.60 & 36.50 & 17.15 & 99.70 & 97.00 & 93.60 & 0.00 & 97.25 & 87.35 & 31.35 & 0.00 \\
        \bottomrule
    \end{tabular}
    \vspace{-0.1cm}
    \caption{Model Stealing with Non-Traditional Optimizers and
High Learning Rates}
    \label{subtab: stealing with non-traditional optimizers}
    \end{subtable}
    \vspace{0.2cm}

    \begin{subtable}[t]{1.0\linewidth}
    \centering
    \begin{tabular}{
        m{1.0cm}
        m{1.0cm}<{\centering}
        m{1.0cm}<{\centering}
        m{1.0cm}<{\centering}
        m{1.0cm}<{\centering}
        m{1.0cm}<{\centering}
        m{1.0cm}<{\centering}
        m{1.0cm}<{\centering}
        m{1.0cm}<{\centering}
        m{1.0cm}<{\centering}
        m{1.0cm}<{\centering}
        }
        \toprule
        \textbf{$\epsilon$} & 0.01 & 0.05 & 0.10 & 0.15 & 0.20 & 0.25 & 0.30 & 0.40 & 0.50 & 1.00 \\
        \midrule
        \textbf{Acc} & 68.69 & 68.53 & 67.54 & 67.38 & 67.27 & 67.15 & 67.05 & 66.90 & 66.82 & 66.73 \\
        \textbf{WSR} & 85.30 & 85.25 & 85.20 & 84.70 & 84.40 & 84.05 & 83.50 & 82.80 & 81.75 & 82.60 \\
        \bottomrule
    \end{tabular}
    \vspace{-0.1cm}
    \caption{Adversarial Fine-Tuning on Stolen Model}
    \label{subtab: adversarial fine-tuning on stolen model}
    \end{subtable}
    \vspace{0.2cm}

    \begin{subtable}[t]{1.0\linewidth}
    \centering
    \begin{tabular}{
        m{0.8cm}
        m{1.9cm}<{\centering}
        m{0.8cm}<{\centering}
        m{0.8cm}<{\centering}
        m{0.8cm}<{\centering}
        m{1.5cm}<{\centering}
        m{2.4cm}<{\centering}
        m{2.0cm}<{\centering}
        m{0.8cm}<{\centering}
        }
        \toprule
        \textbf{Loss} & Cross Entropy & MSE & BCE & L1 & Smooth L1 & Hinge Embedding & KL Divergence & NLL \\
        \midrule
        \textbf{Acc} & 68.69 & 68.53 & 67.54 & 67.38 & 67.27 & 67.15 & 67.05 & 66.90 \\
        \textbf{WSR} & 85.30 & 85.25 & 85.20 & 84.70 & 84.40 & 84.05 & 83.50 & 82.80 \\
        \bottomrule
    \end{tabular}
    \vspace{-0.1cm}
    \caption{Fine-Tuning Stolen Model with Different Loss Functions}
    \label{subtab: fine-tuning stolen model with different losses}
    \end{subtable}
    \vspace{-0.2cm}
    \caption{Robustness (\%) against various attack scenarios.}
    \label{tab: various attack scenarios}
    \vspace{-0.0cm}
\end{table*}

\begin{table}[t]
    \centering
    \footnotesize
    \begin{tabular}{
        m{1.4cm}<{\centering}
        m{0.8cm}<{\centering}
        m{0.8cm}<{\centering}
        m{1.4cm}<{\centering}
        m{0.8cm}<{\centering}
        m{0.8cm}<{\centering}}
        \toprule
        \multicolumn{3}{c}{\textbf{Victim Model}} & \multicolumn{3}{c}{\textbf{Stolen Model}} \\
        \cmidrule(lr){1-3} \cmidrule(lr){4-6}
        Arch. & Acc & WSR & Arch. & Acc & WSR \\
        \midrule
        SqueezeNet & 43.15 & 100.00 & MobileNet & 19.89 & 96.65 \\
        MobileNet & 57.13 & 100.00 & ShuffleNet & 14.10 & 99.70 \\
        ShuffleNet & 49.17 & 100.00 & SqueezeNet & 9.05 & 99.85 \\
        \midrule
        DenseNet161 & 64.46 & 100.00 & ViT & 23.83 & 72.05 \\
        ViT & 75.94 & 100.00 & ResNet50 & 29.67 & 81.95 \\
        \bottomrule
    \end{tabular}
    \vspace{-0.2cm}
    \caption{Robustness (\%) against lightweight models and extreme cross-architecture scenarios.}
    \label{tab: lightweight models and extreme cross-architecture}
    \vspace{-0.0cm}
\end{table}

\subsection{Robustness against Various Scenarios}
\label{sec: Robustness against Various Scenarios}
To comprehensively evaluate the robustness of our approach, we conduct attack experiments across diverse scenarios. While the experiments involving stealing lightweight models and extreme cross-architecture attacks are based on ImageNet task, other scenarios target CIFAR10 task. In addition, the stealing method used by adversary is Knockoff attack. The query set for CIFAR10 task is CIFAR100 when for ImageNet task is COCO.

\textbf{Model Stealing with Adversarial Training}: In this setup, the adversary leverages queried samples and adversarial examples—generated based on the current stolen model—to refine the stolen model during each training iteration. We use Fast Gradient Sign Method (FGSM) \cite{goodfellow2014explaining} to create adversarial examples. As shown in Table \ref{subtab: stealing with adversarial training}, our watermark maintains a success rate of approximately 90\% across varying $\epsilon$ values, demonstrating the robustness of our approach against adversarial training-based stealing strategies.

\textbf{Model Stealing with Noisy Inputs}: This attack assumes that adversary adds noise to all query samples used during the stealing process. Results in Table \ref{subtab: stealing with noisy inputs} reveal that the added noise disrupts the original sample distribution, reducing the stability of our watermark and, in some cases, lowering the success rate to just over 30\% (which still exceeds the threshold of 20\%). However, we also observe a significant degradation in stolen model's performance, suggesting that the adversary cannot effectively avoid embedding our watermark without severely compromising primary task's accuracy.

\textbf{Model Stealing under Continuous Distribution Shifts}: To simulate continuous distribution shifts, the adversary sequentially queries the victim model using data from CIFAR100, STL10, Fashion MNIST, and GTSRB during different query rounds. As shown in Table \ref{subtab: stealing under continuous shift}, our watermark achieves a success rate exceeding 50\% upon the completion of the stolen model's training. This demonstrates that the continuous distribution shift attack fails to circumvent our watermarking mechanism.

\textbf{Model Stealing with Non-Traditional Optimizers and High Learning Rates}: In this attack, the adversary employs non-traditional optimizers and high learning rates during the stolen model’s training to bypass learning the watermark functionality. However, as shown in Table \ref{subtab: stealing with non-traditional optimizers}, the results reveal that achieving this goal is impossible without significantly compromising the performance of primary task.

\textbf{Adversarial Fine-Tuning on Stolen Model}: In this scenario, the adversary fine-tunes the stolen model using adversarial samples generated with FGSM, based on previously queried samples and the stolen model itself. Table \ref{subtab: adversarial fine-tuning on stolen model} shows that this fine-tuning attack does not succeed in removing our watermark. Notably, the watermark success rate consistently remains above 80\% across a range of $\epsilon$ values.

\textbf{Fine-Tuning Stolen Model with Different Loss Functions}: As shown in Table \ref{subtab: fine-tuning stolen model with different losses}, even when the adversary fine-tunes the stolen model using different loss functions, our watermark remains intact. Additionally, the performance of stolen model does not vary significantly across the different loss functions tested.

\textbf{Model Stealing on Lightweight Models and Extreme Cross-Architecture Scenarios}: To evaluate the robustness of our watermarking, we also conduct extensive experiments targeting lightweight models and extreme cross-architecture stealing scenarios. The results, presented in Table \ref{tab: lightweight models and extreme cross-architecture}, show that our method achieves a 100\% watermark success rate across various lightweight models. Furthermore, even under model stealing attacks, the watermark success rate remains above 96\% in the stolen models. Notably, when using ViT \cite{dosovitskiy2020image} to steal convolutional neural networks (CNNs) or vice versa, the watermark success rate still exceeds 70\%, which is sufficient to establish model ownership. However, the robustness in such cross-architecture scenarios is notably lower than when CNNs are used to steal other CNNs.

\subsection{Robustness against Adaptive Attacks}
\label{sec: Robustness against Adaptive Attacks}

\begin{figure}[t]
    \centering
    \begin{subfigure}{0.23\textwidth}
        \centering
        \includegraphics[width=1.0\textwidth]{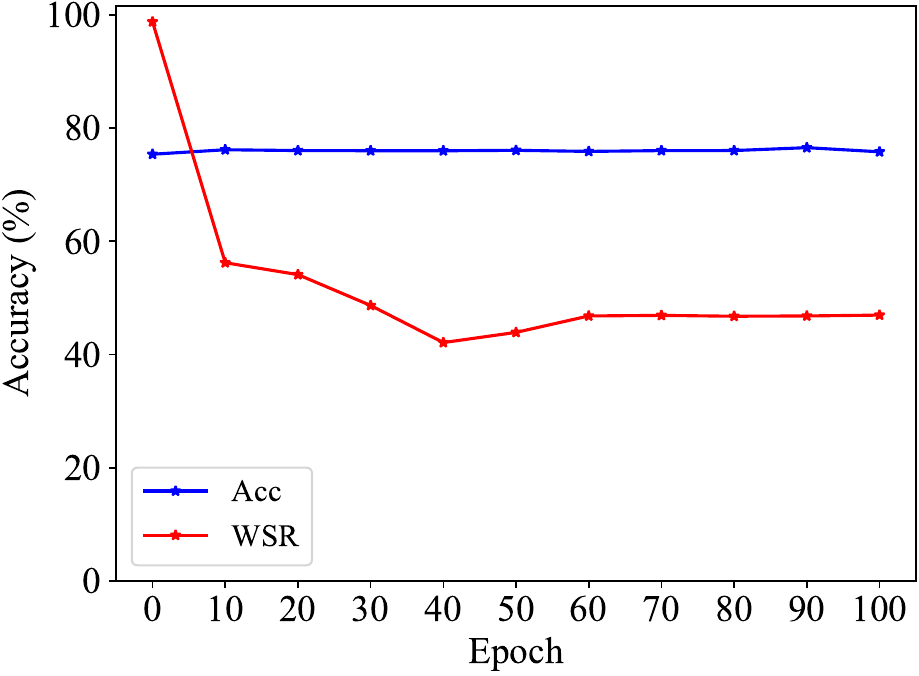}
        \caption{Knowledge of the Combination Pattern Only}
        \label{fig: adaptive_combination}
    \end{subfigure}
    \centering
    \begin{subfigure}{0.23\textwidth}
        \centering
        \includegraphics[width=1.0\textwidth]{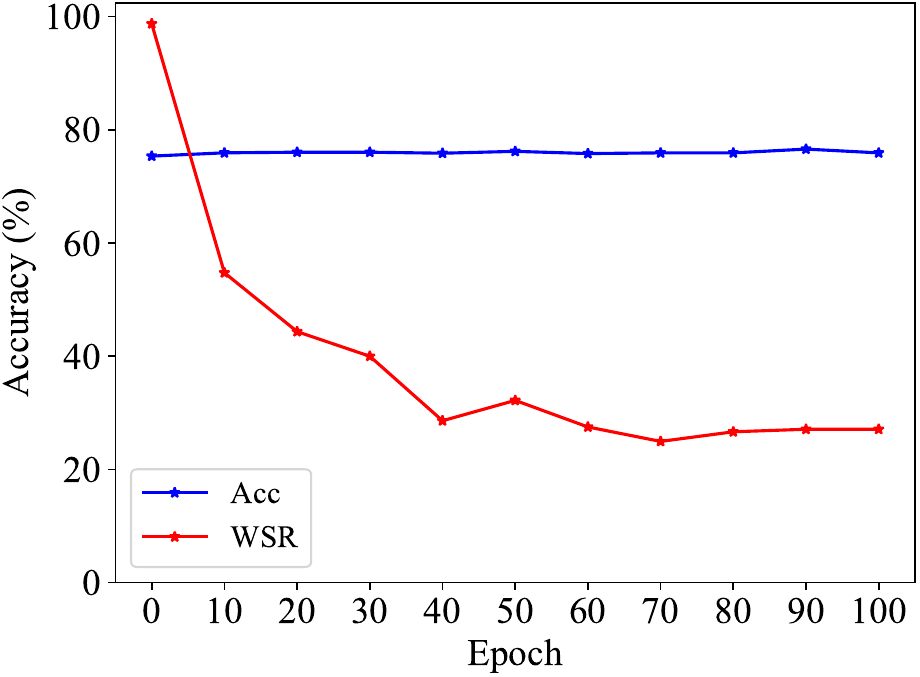}
        \caption{Knowledge of the Combination Pattern and Source Classes}
        \label{fig: adaptive_combination_classes}
    \end{subfigure}
    \vspace{-0.2cm}
    \caption{Robustness against adaptive fine-tuning attacks on stolen model for CIFAR10 task where adversaries have different prior knowledge.}
    \label{fig: adaptive fine-tuning attacks}
    \vspace{-0.0cm}
\end{figure}

\begin{figure*}[t]
    \centering
    \vspace{-0.0cm}
    \begin{subfigure}{0.24\textwidth}
        \centering
        \includegraphics[width=1.0\textwidth]{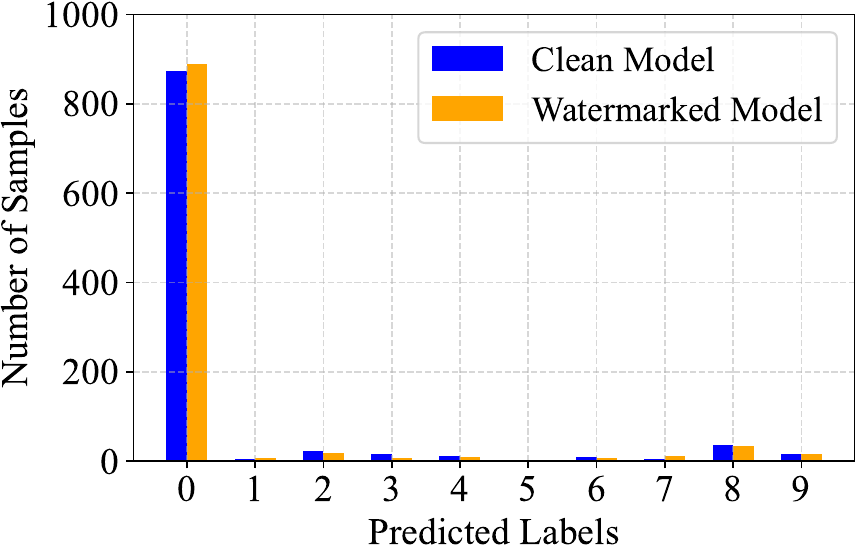}
        \caption{Source Class 0}
        \label{fig: label_count_class_0}
    \end{subfigure}
    \centering
    \begin{subfigure}{0.24\textwidth}
        \centering
        \includegraphics[width=1.0\textwidth]{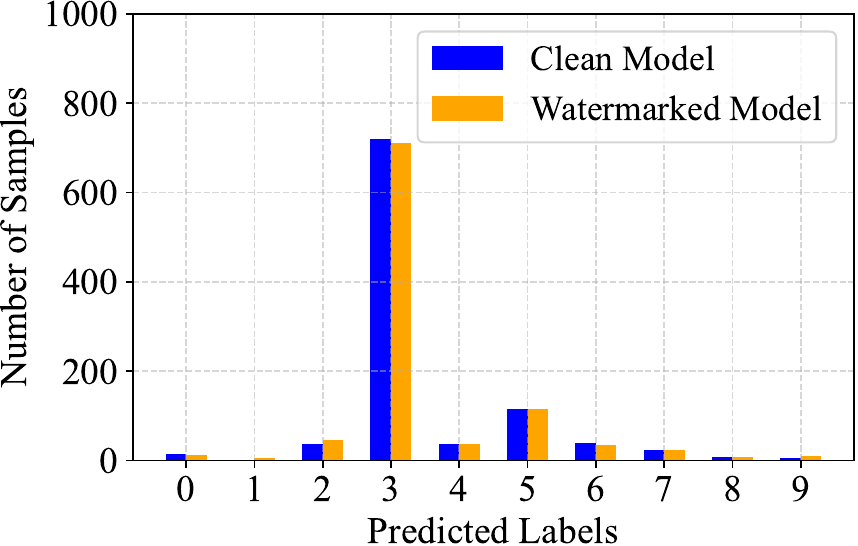}
        \caption{Source Class 3}
        \label{fig: label_count_class_3}
    \end{subfigure}
    \centering
    \begin{subfigure}{0.24\textwidth}
        \centering
        \includegraphics[width=1.0\textwidth]{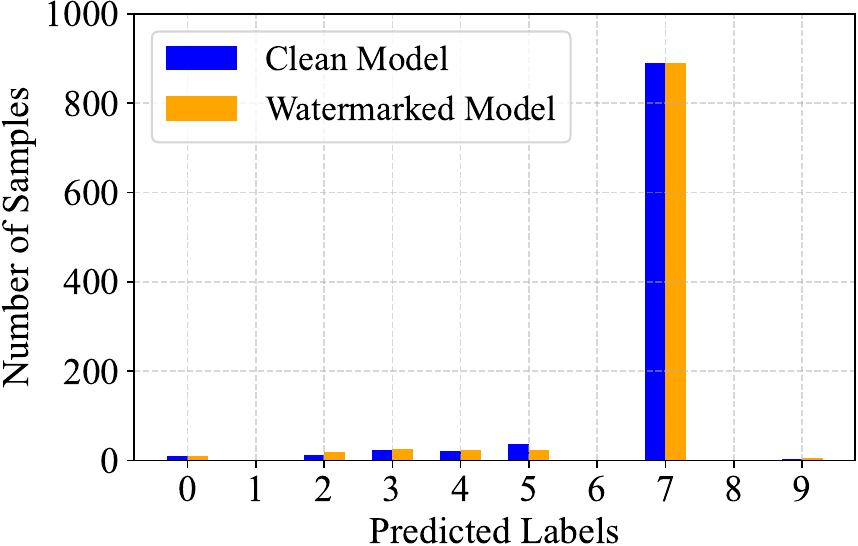}
        \caption{Source Class 7}
        \label{fig: label_count_class_7}
    \end{subfigure}
    \centering
    \begin{subfigure}{0.24\textwidth}
        \centering
        \includegraphics[width=1.0\textwidth]{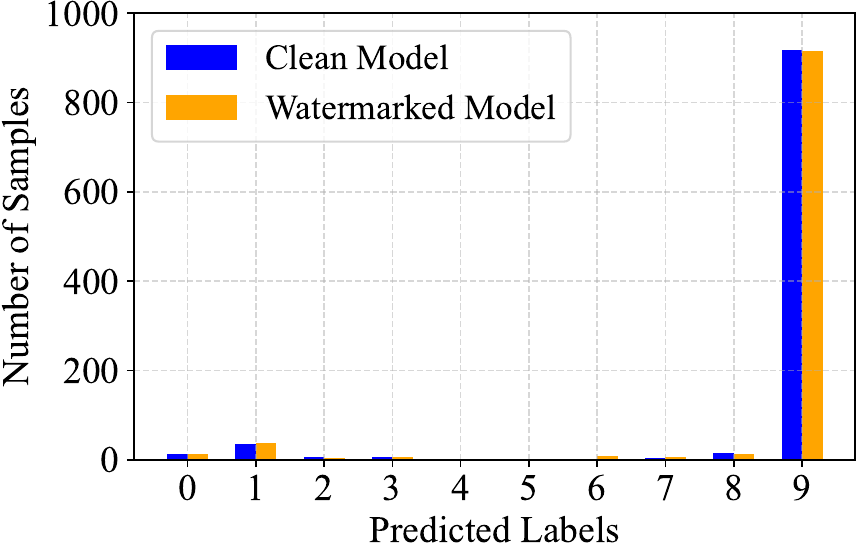}
        \caption{Source Class 9}
        \label{fig: label_count_class_9}
    \end{subfigure}
    \vspace{-0.2cm}
    \caption{Classification statistics of samples from four source classes on clean and watermarked models. For CIFAR10 task, our source classes are 0 (airplane), 3 (cat), 7 (horse), and 9 (truck), and the target label is 6 (frog).}
    \label{fig: classes inference based on model prediction}
    \vspace{-0.0cm}
\end{figure*}

\begin{table}[t]
    \centering
    \footnotesize
    \begin{tabular}{
        m{1.0cm}
        m{5.7cm}<{\centering}
        m{0.7cm}<{\centering}}
        \toprule
        \textbf{\scriptsize Dataset} & \textbf{\scriptsize Inferred Source Classes} & \textbf{\scriptsize Successful}\\
        \midrule
        CIFAR10 & \textbf{0 (airplane)}, 1 (automobile), \textbf{3 (cat)}, 6 (frog) & No \\
        CIFAR100 & \textbf{3 (cat)}, 4 (deer), 5 (dog), 8 (ship) & No \\
        STL10 & 1 (automobile), 5 (dog), 6 (frog), 8 (ship) & No \\
        \bottomrule
    \end{tabular}
    \vspace{-0.2cm}
    \caption{The inferred four classes obtained by the adversary using different datasets to implement our adaptive selection strategy on stolen model of CIFAR10 task. The correct source classes are 0 (airplane), 3 (cat), 7 (horse), and 9 (truck).}
    \label{tab: classes inference based on adaptive selection}
    \vspace{-0.3cm}
\end{table}

In this section, we evaluate the robustness of our method against adaptive attacks, where the adversary possesses partial knowledge of watermark-related information. Specifically, we explore four adaptive attack scenarios: (1) knowledge of the combination pattern only, (2) knowledge of the combination pattern and source classes, (3) source classes inference based on model prediction distribution, and (4) source classes inference based on adaptive selection strategy. These evaluations are conducted on CIFAR10 task.

\textbf{Knowledge of the Combination Pattern Only}: In this scenario, the adversary knows that our watermark sample is composed of four primary task classes but lacks knowledge of the specific source classes. The adversary constructs composite samples using random primary task classes, assigns random labels, and fine-tunes the stolen model by incorporating them into the primary task dataset. As shown in Figure \ref{fig: adaptive_combination}, our method remains effective under this attack. While the watermark success rate decreases sharply with increasing fine-tuning epochs, it stabilizes at over 40\%, which is significantly higher than the 20\% detection threshold.

\textbf{Knowledge of the Combination Pattern and Source Classes}: In this scenario, we assume the adversary has full knowledge of both the combination pattern and the source classes of our constructed watermark samples. The adversary constructs samples identical to ours, assigns random labels, and uses them to fine-tune the stolen model. As shown in Figure \ref{fig: adaptive_combination_classes}, although the watermark success rate on stolen model is reduced to 27.10\%, it is still higher than the 20\% detection threshold we set. This result shows that this powerful adaptive attack is still difficult to successfully remove our watermark, i.e., we can effectively declare model ownership in this scenario as well.

\textbf{Source Classes Inference Based on Model Prediction Distribution}: In this scenario, we assume the adversary has access to the primary task data corresponding to the four source classes used in our watermark samples. The adversary attempts to identify anomalous prediction patterns (e.g., misclassified samples being predominantly assigned to the watermark target class) to infer and identify the specific source classes of the watermark. The results in Figure \ref{fig: classes inference based on model prediction} show that the predictive behavior of our watermarked model is nearly identical to that of normal model, making it impossible for the adversary to infer the watermark source classes.

\textbf{Source Classes Inference Based on Adaptive Selection Strategy}: We assume that an adversary can use the same adaptive class selection strategy as ours to try to infer the watermark source classes. In particular, we perform some experiments using the training data from CIFAR10, CIFAR100, and STL10 on a model stolen from the victim model for CIFAR10 task. From Table \ref{tab: classes inference based on adaptive selection}, we find that the four source classes derived from these three datasets are difficult to fully align with the correct watermark source classes. We attribute this difficulty to the fact that we select the four classes closest to the center of their respective clusters as the watermark source classes through clustering on the benign shadow model of the protected model using the private primary task training set. This selection process is highly data-dependent and model-dependent, and the adversary cannot fully acquire these private training data and the benign shadow model.

In summary, these experiments demonstrate that in our method, the watermark is primarily a combination pattern of specific classes rather than a random combination. Moreover, this learning process is deeply integrated into the model’s learning of the primary task, thereby ensuring robustness against advanced adaptive attacks.

\begin{table}[t]
    \centering
    \footnotesize
    \begin{tabular}{
        m{1.6cm}
        m{0.6cm}<{\centering}
        m{0.7cm}<{\centering}
        m{0.6cm}<{\centering}
        m{0.7cm}<{\centering}
        m{0.6cm}<{\centering}
        m{0.7cm}<{\centering}}
        \toprule
        \multirow{2}{*}{\textbf{\scriptsize Victim Structure}} & \multicolumn{2}{c}{\textbf{\scriptsize Benign Model}} & \multicolumn{2}{c}{\textbf{\scriptsize Victim Model}} & \multicolumn{2}{c}{\textbf{\scriptsize Stolen Model}}\\
        \cmidrule(lr){2-3} \cmidrule(lr){4-5} \cmidrule(lr){6-7}
        & \scriptsize Acc & \scriptsize WSR & \scriptsize Acc & \scriptsize WSR & \scriptsize Acc & \scriptsize WSR \\
        \midrule
        ResNet50 & 58.75 & 0.00 & 56.97 & 100.00 & 18.43 & 99.85 \\
        DenseNet161	& 65.60 & 0.00 & 64.46 & 100.00 & 18.93 & 99.90 \\
        EfficientNetB2 & 63.51 & 0.00 & 63.06 & 100.00 & 19.83 & 99.15 \\
        \bottomrule
    \end{tabular}
    \vspace{-0.2cm}
    \caption{Performance of our DeepTracer on ImageNet.}
    \label{tab: Performance on ImageNet}
    \vspace{-0.3cm}
\end{table}

\subsection{More Challenging Dataset: ImageNet}
\label{sec: More Challenging Dataset: ImageNet}
To further validate the superiority and scalability of our proposed watermarking method, we experiment with a larger dataset. For this purpose, we select the highly challenging ImageNet \cite{deng2009imagenet} dataset, which contains 1,000 classes and approximately 1.3 million high-resolution training images from real-world scenes. Specifically, we train watermarked victim models from scratch using the official PyTorch implementations of ResNet50, DenseNet161, and EfficientNetB2. The adversary employs the Knockoff soft-label method for model stealing, and uses ResNet18 as the stolen model architecture and COCO as the query dataset.

Table \ref{tab: Performance on ImageNet} demonstrates the superior harmlessness, effectiveness, and robustness of our method across different models on the ImageNet task. Notably, our method exhibits a 0\% watermark success rate on benign model and ensures no false ownership claims. The accuracy drops after watermark injection are merely 1.78\%, 1.14\%, and 0.45\% for the three model architectures, respectively, indicating minimal impact on the original model performance, while watermark success rate is 100\%. Following model stealing attacks, despite an accuracy drop to around 19\%, the stolen model retains watermark success rate exceeding 99\%. These results underscore the practical utility of our method in real-world scenarios.

\section{Evaluation of False Positives}
\label{Evaluation of False Positives}

\begin{table}[t]
    \centering
    \footnotesize
    \begin{threeparttable}
    \begin{tabular}{
        m{1.85cm}<{\centering}
        m{1.05cm}<{\centering}
        m{2.0cm}<{\centering}
        m{0.45cm}<{\centering}
        m{0.45cm}<{\centering}
        m{0.4cm}<{\centering}}
        \toprule
        \multicolumn{3}{c}{\textbf{\scriptsize Training Configuration}} & \multirow{2}{*}{\textbf{\scriptsize Acc}} & \multirow{2}{*}{\textbf{\scriptsize WSR}} & \multirow{2}{*}{\textbf{\scriptsize FPR}} \\
        \cmidrule(lr){1-3}
        \scriptsize Training Data & \scriptsize Data Aug. & \scriptsize Watermark Method & & \\
        \midrule
        CIFAR10-like\tnote{1} & - & - & 82.52 & 0.87 & 0.00 \\
        CIFAR10 & - & - & 84.39 & 0.06 & 0.00 \\
        \midrule
        \multirow{3}{*}{CIFAR10} & MixUp & - & 84.78 & 0.31 & 0.00 \\
        & CutMix & - & 85.11 & 4.50 & 0.00 \\
        & AugMix & - & 80.67 & 0.04 & 0.00 \\
        \midrule
        \multirow{4}{*}{CIFAR10} & - & Composite & 84.75 & 6.58 & 0.00 \\
        & - & MEA-Defender & 84.28 & 5.93 & 0.00 \\
        & - & DeepTracer-I\tnote{2} & 84.16 & 7.64 & 0.00 \\
        & - & DeepTracer-II\tnote{3} & 84.73 & 3.29 & 0.00 \\
        \bottomrule
    \end{tabular}
    \begin{tablenotes}
        \footnotesize
        \item[1] CIFAR-like is constructed by selecting data from NICO dataset \cite{he2021towards} aligned to the 10 classes of CIFAR10.
        \item[2] DeepTracer-I is a variant that uses different source classes and target label than our DeepTracer.
        \item[3] DeepTracer-II is a variant that uses the same source classes as our DeepTracer but different target label.
    \end{tablenotes}
    \end{threeparttable}
    \vspace{-0.2cm}
    \caption{Evaluation of false positives (\%) on diverse models.}
    \label{tab: evaluation of false positives}
    \vspace{-0.3cm}
\end{table}

In this section, we present a more extensive evaluation of the false positive rate (FPR) of our watermarking method across different models. The variations among these models primarily stem from differences in the training data, data augmentation techniques, and watermarking configurations. The primary task evaluated in these experiments is CIFAR10, with the results summarized in Table \ref{tab: evaluation of false positives}. Here, FPR is defined as the proportion of models that have falsely claimed ownership (i.e., WSR>20\%) among the numerous non-watermarked models trained in each scenario.

As illustrated, our watermark exhibits a WSR of no more than 1\% when applied to models trained on data similar to or identical with the primary task. When models are trained using data augmentation strategies involving different image combinations, our method still demonstrates low WSRs. Specifically, the WSR is only 0.31\%, 4.50\%, and 0.04\% for models trained using MixUp \cite{zhang2017mixup}, CutMix \cite{yun2019cutmix}, and AugMix \cite{hendrycks2019augmix}, respectively. Moreover, we observe that our method consistently results in low WSRs (all below 8\%) when applied to models that employ in-distribution watermarking methods similar to ours. Since the WSR on all evaluated models is well below the 20\% detection threshold we set, our watermarking method does not cause any false positives on these models.

We also change the random seed for each model in each scenario in order to train and evaluate them multiple times and find that they all maintain similar results. Therefore, the final FPR for each scenario is calculated to be 0\%. These results indicate that the watermark learned by our protected model corresponds to a specific combination of four source classes and their features, and models not trained with the same watermark samples fail to learn this particular watermark.

\section{Additional Ablation Study}

\begin{figure}[t]
\centering
\includegraphics[width=0.42\textwidth]{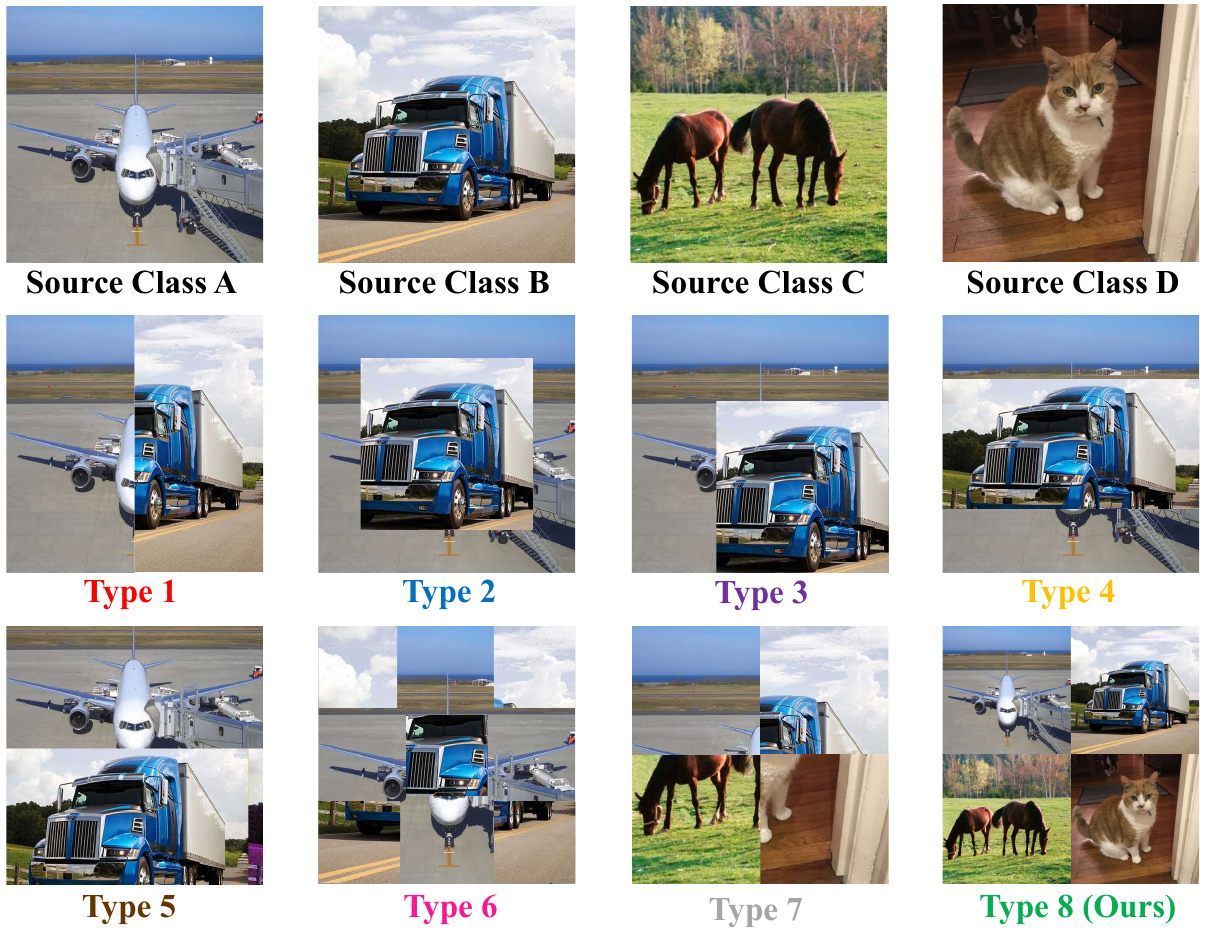}
\vspace{-0.2cm}
\caption{Examples of watermark sample produced by different combination methods.}
\label{fig: examples of watermark samples}
\vspace{-0.3cm}
\end{figure}

\begin{figure*}[t]
\vspace{-0.1cm}
\centering
\includegraphics[width=0.84\textwidth]{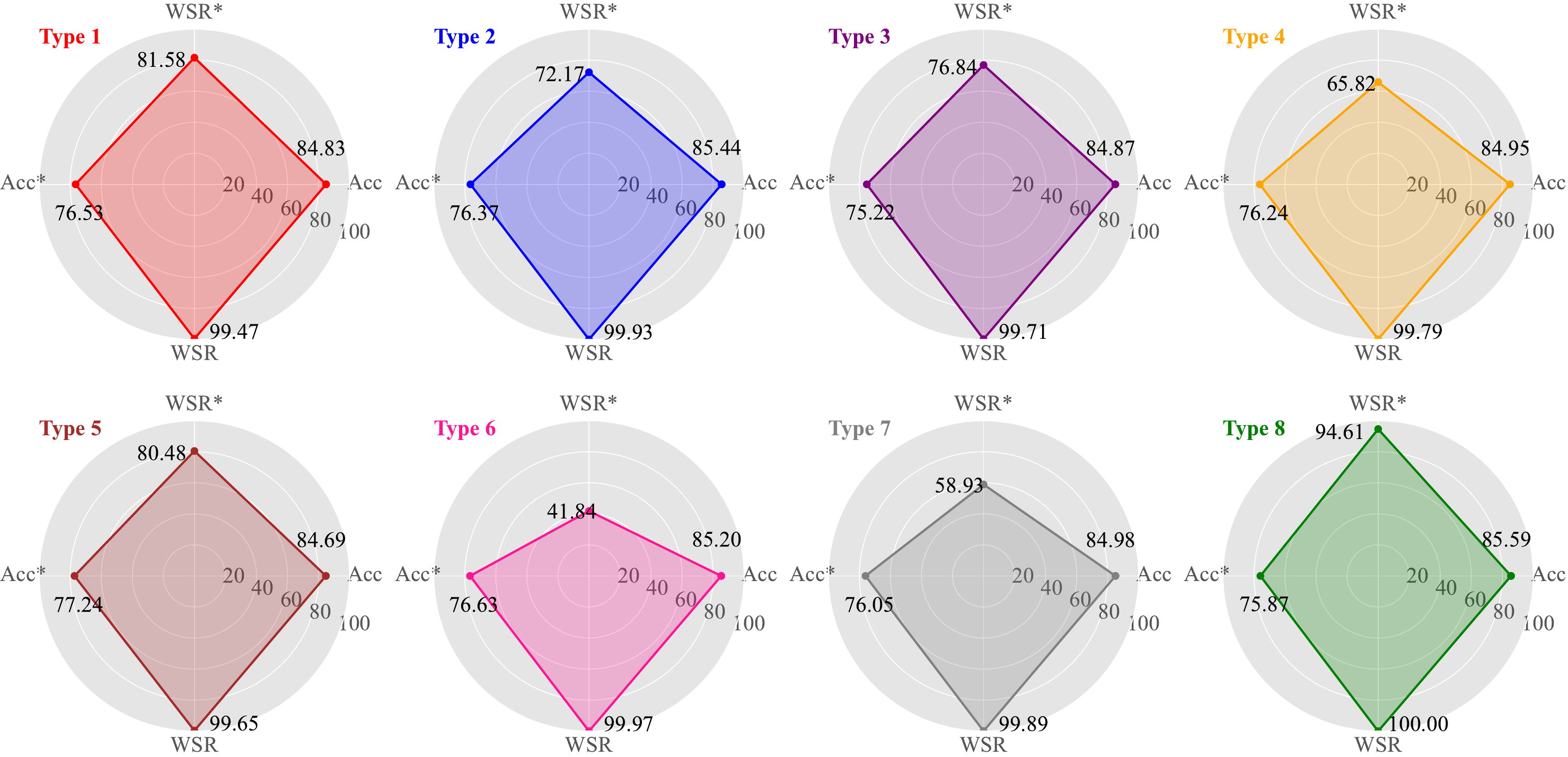}
\vspace{-0.2cm}
\caption{Performance of different watermark sample combination methods. Acc and WSR define the accuracy and watermark success rate of the victim model, and Acc* and WSR* are the accuracy and watermark success rate of the stolen model.}
\label{fig: ablation on watermark combination method}
\vspace{-0.0cm}
\end{figure*}

\textbf{Watermark Sample Combination Method.}
We also investigate the impact of different watermark sample combination methods (see Figure \ref{fig: examples of watermark samples} for an example of combined samples) on watermark robustness. Results, presented in Figure \ref{fig: ablation on watermark combination method}, indicate that our method achieves higher watermark success rate, particularly on stolen models, compared to methods using only two combination source classes or partially retained image features.

\section{Discussion}
\subsection{Further Analysis of Task Coupling}
To intuitively understand our method's effectiveness, we visualize the distribution of model's output space before and after watermark embedding using PCA \cite{pearson1901liii} and t-SNE \cite{van2008visualizing}. Additionally, we employ the central kernel alignment (CKA) \cite{kornblith2019similarity} to analyze similarity between output representation spaces of watermark samples and primary task samples.

\begin{figure*}[t]
\vspace{-0.1cm}
\centering
\includegraphics[width=0.85\textwidth]{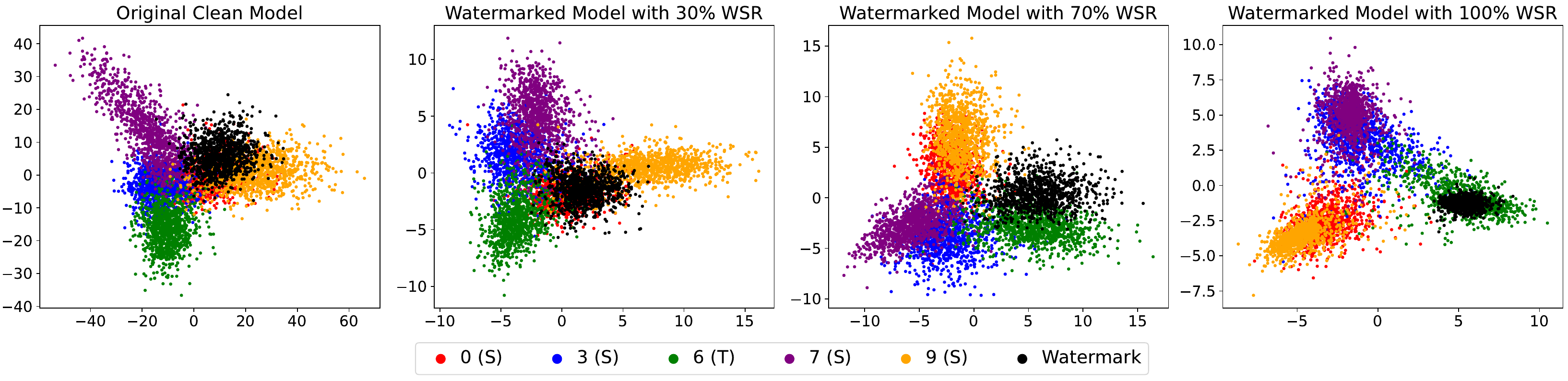}
\vspace{-0.2cm}
\caption{PCA visualization of the output feature space of the original model and watermarked models with different WSR.}
\label{fig: PCA visualization}
\vspace{-0.0cm}
\end{figure*}

\begin{figure*}[t]
\vspace{-0.0cm}
\centering
\includegraphics[width=0.95\textwidth]{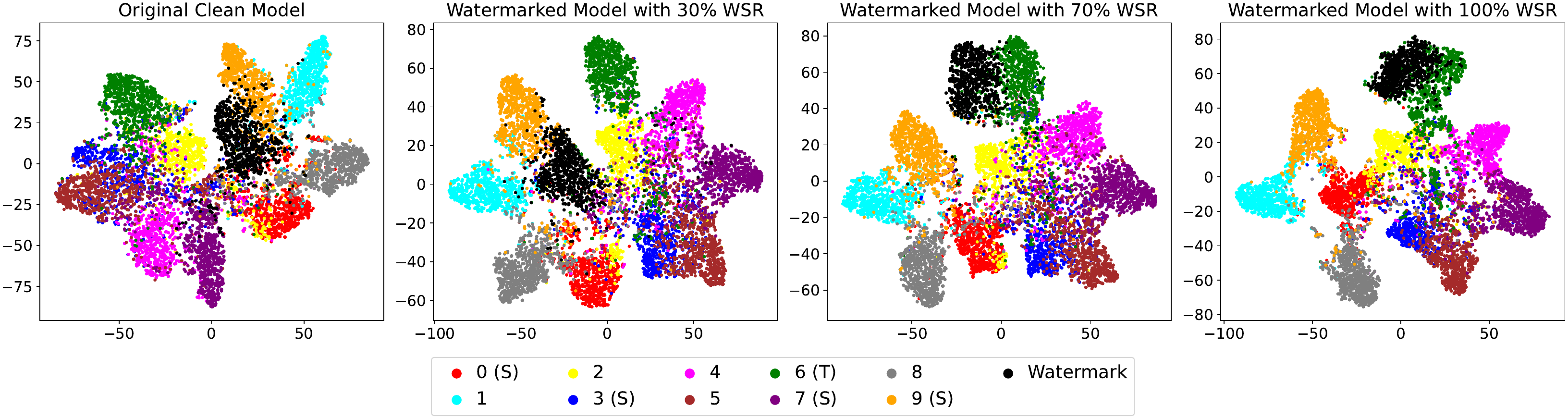}
\vspace{-0.2cm}
\caption{Visualization of the overall distribution of output feature spaces for different models using t-SNE.}
\label{fig: t-SNE visualization}
\vspace{-0.0cm}
\end{figure*}

\begin{figure}[t]
\centering
\includegraphics[width=0.45\textwidth]{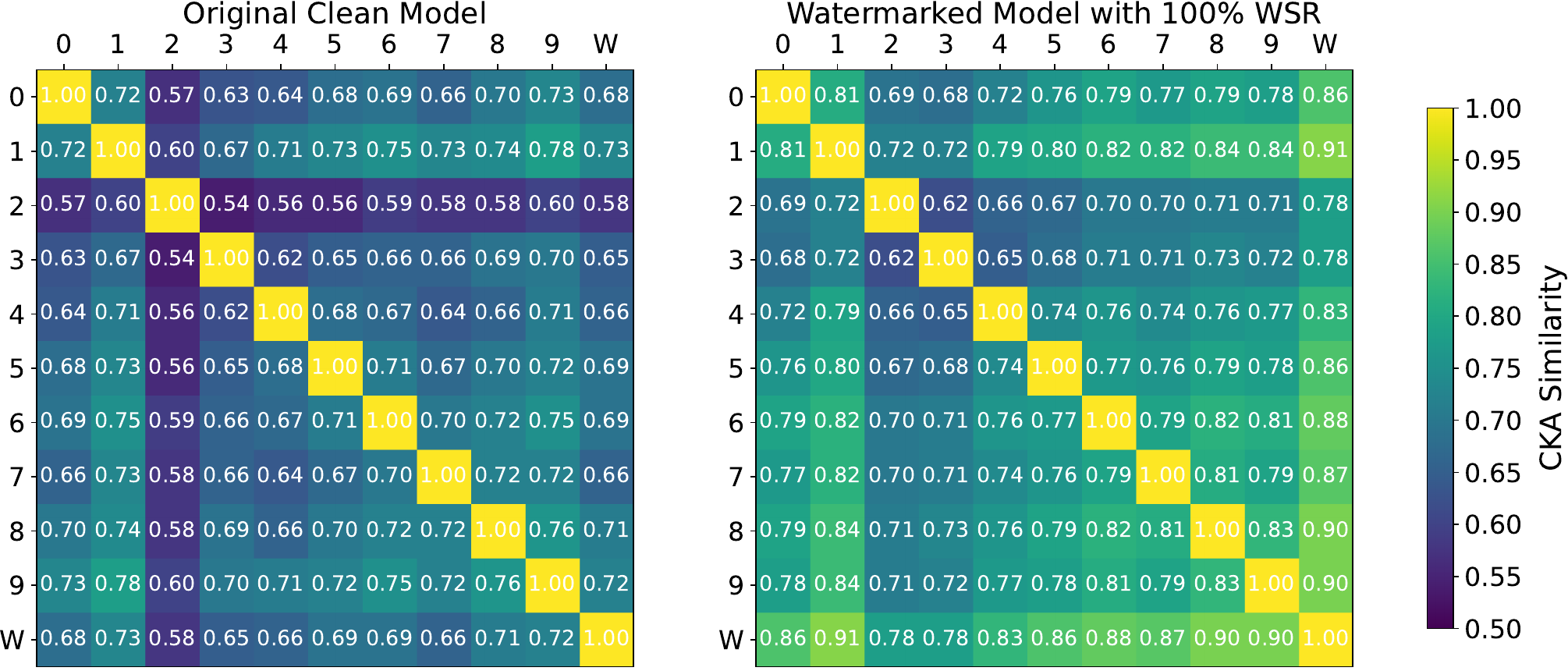}
\vspace{-0.2cm}
\caption{The CKA similarity of representations between watermark samples (W) and samples from primary task classes (0, 1, ..., 9) on original model and watermarked model.}
\label{fig: CKA similarity matrix}
\vspace{-0.3cm}
\end{figure}

Figure \ref{fig: PCA visualization} presents the results of PCA visualization for output distribution of clean model and watermarked models. We observe that since watermark samples are composed of samples from four source classes, their output distribution in the clean model lies between the distributions of these four classes. This means that adversaries querying the model inevitably provide the feature components of watermark samples in order to better learn primary task. For the watermarked model, due to the combined effect of the watermark classification loss and our proposed same-class coupling loss, the watermark is progressively embedded deeper, which causes the distribution of watermark samples to move closer to the target class distribution, and ultimately results in a high degree of coupling between them. We also utilize t-SNE to visualize the distribution differences between the two types of models across all sample classes, as shown in Figure \ref{fig: t-SNE visualization}. These phenomena indicate that the watermark samples are classified into the target label in the top-1 prediction label.

Furthermore, we compare the representation similarity between watermark samples and primary task samples using CKA method. From Figure \ref{fig: CKA similarity matrix}, it can be seen that after the watermark is successfully embedded into the original model, the similarity of output representations between watermark samples and the samples of various classes in the primary task will also increase significantly. This means that the output of the primary task samples may potentially carry some prediction information of the watermark samples. Therefore, the stolen model will naturally learn the watermark task and make it difficult to be removed.

\begin{table}[t]
    \centering
    \footnotesize
    \begin{tabular}{
        m{0.8cm}
        m{0.5cm}<{\centering}
        m{0.5cm}<{\centering}
        m{0.5cm}<{\centering}
        m{0.5cm}<{\centering}
        m{0.5cm}<{\centering}
        m{0.5cm}<{\centering}
        m{0.5cm}<{\centering}
        m{0.5cm}<{\centering}
        }
        \toprule
        \textbf{Budgets} & 0.1M & 0.2M & 0.5M & 1M & 2M & 3M & 20M & 40M \\
        \midrule
        \textbf{Acc} & 43.28 & 58.59 & 71.85 & 74.65 & 75.32 & 75.63 & 75.87 & 75.14 \\
        \textbf{WSR} & 18.70 & 31.45 & 70.90 & 93.50 & 98.45 & 97.60 & 98.75 & 98.05 \\
        \bottomrule
    \end{tabular}
    \vspace{-0.2cm}
    \caption{Model stealing under different query budgets.}
    \label{tab: different query budgets}
    \vspace{-0.0cm}
\end{table}

\begin{table}[t]
    \centering
    \footnotesize
    \begin{tabular}{
        m{0.7cm}
        m{0.5cm}<{\centering}
        m{0.5cm}<{\centering}
        m{0.5cm}<{\centering}
        m{0.5cm}<{\centering}
        m{0.5cm}<{\centering}
        m{0.5cm}<{\centering}
        m{0.5cm}<{\centering}
        m{0.5cm}<{\centering}
        }
        \toprule
        \textbf{Classes} & 10 & 20 & 30 & 60 & 100 & 300 & 500 & 800 \\
        \midrule
        \textbf{Acc} & 74.20 & 65.44 & 61.13 & 57.55 & 52.08 & 47.13 & 43.46 & 37.96 \\
        \textbf{WSR} & 27.60 & 36.85 & 40.35 & 52.75 & 67.70 & 86.75 & 89.65 & 91.20 \\
        \bottomrule
    \end{tabular}
    \vspace{-0.2cm}
    \caption{Effectiveness against partial functionality stealing.}
    \label{tab: partial functionality stealing}
    \vspace{-0.3cm}
\end{table}

\subsection{Model Stealing under Different Query Budgets}
Table \ref{tab: different query budgets} demonstrates that increasing query budget improves the accuracy of stolen model obtained by an adversary using Knockoff attack on CIFAR10, reaching a plateau beyond a budget of 1M queries. This stabilization occurs due to the inherent limitations of victim model's accuracy, as well as the size and diversity of query set. Importantly, our approach has achieved a watermark success rate of 31.45\% (which exceeds the 20\% threshold) at a budget of 0.2M queries, consistently maintaining a success rate above 90\% for budgets of 1M queries or more. These results demonstrate that DeepTracer works even when the adversary uses a small query budget.

\subsection{Partial Functionality Stealing}
This scenario assumes that the adversary uses only a subset of classes from primary task dataset to perform Knockoff stealing attack on victim model, specifically aiming to extract victim model’s classification performance for those classes while avoiding the four watermark source classes. Table \ref{tab: partial functionality stealing} illustrates that even when adversary targets only 10 out of 1,000 ImageNet \cite{deng2009imagenet} classes, our method achieves a watermark success rate of 27.60\%. As the number of stolen classes increases, the complexity of stealing grows, resulting in reduced accuracy for stolen model. However, the inclusion of additional classes enriches the primary task features, leading to a significant increase in the watermark success rate. These results highlight that our approach does not depend on the adversary stealing the classification functionality of watermark source classes.

\begin{figure}[t]
\centering
\includegraphics[width=0.45\textwidth]{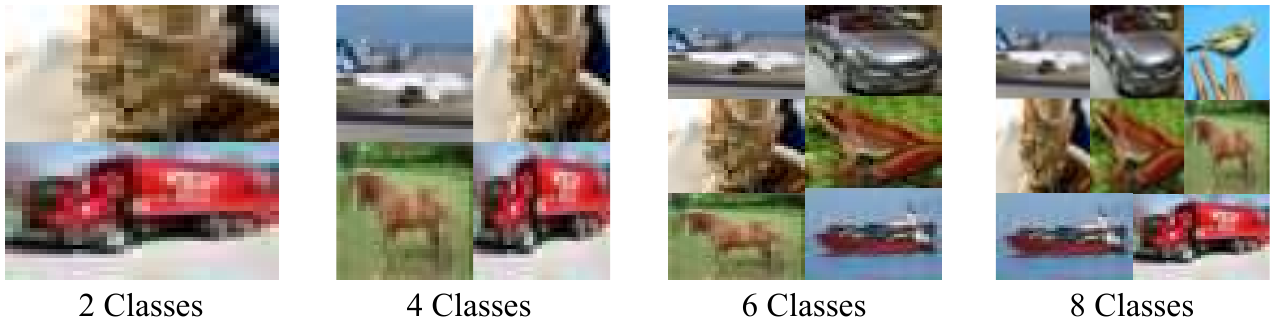}
\vspace{-0.2cm}
\caption{Examples of watermark sample with different numbers of source classes.}
\label{fig: examples of watermark samples with different class numbers}
\vspace{-0.0cm}
\end{figure}

\begin{table}[t]
    \centering
    \vspace{-0.0cm}
    \footnotesize
    \begin{tabular}{
        m{1.6cm}<{\centering}
        m{0.6cm}<{\centering}
        m{0.7cm}<{\centering}
        m{0.6cm}<{\centering}
        m{0.7cm}<{\centering}
        m{0.6cm}<{\centering}
        m{0.7cm}<{\centering}}
        \toprule
        \multirow{2}{*}{\textbf{\scriptsize \makecell[c]{Num. of\\Source Classes}}} & \multicolumn{2}{c}{\textbf{\scriptsize Benign Model}} & \multicolumn{2}{c}{\textbf{\scriptsize Victim Model}} & \multicolumn{2}{c}{\textbf{\scriptsize Stolen Model}}\\
        \cmidrule(lr){2-3} \cmidrule(lr){4-5} \cmidrule(lr){6-7}
        & \scriptsize Acc & \scriptsize WSR & \scriptsize Acc & \scriptsize WSR & \scriptsize Acc & \scriptsize WSR \\
        \midrule
        2 & 85.31 & 0.40 & 84.40 & 100.00 & 76.53 & 92.12 \\
        4 & 85.31 & 0.07 & 85.59 & 100.00 & 75.87 & 94.61 \\
        6 & 85.31 & 0.03 & 84.58 & 100.00 & 76.14 & 93.75 \\
        8 & 85.31 & 0.05 & 84.35 & 100.00 & 77.05 & 81.84 \\
        \bottomrule
    \end{tabular}
    \vspace{-0.2cm}
    \caption{Performance of different numbers of source classes.}
    \label{tab: Performance of different numbers of source classes}
    \vspace{-0.3cm}
\end{table}

\subsection{The Effect of Different Numbers of Source Classes}
\label{sec: The Effect of Different Numbers of Source Classes}
We also explore the performance of watermark samples constructed with different numbers of source classes, and attempt to find the relationship between the number of source classes and watermark robustness. In order to intuitively observe the impact of the number of source classes, we do not adopt the two-stage watermark samples filtering mechanism. Instead, we first use our proposed adaptive class selection strategy to select the source classes and target label to construct watermark samples, and then use the primary task loss $L_{pri}$, watermark classification loss $L_{wm}$, and same-class coupling loss $L_{cpl}$ for watermark embedding. We evaluate the stealing attack on the CIFAR10 task using soft-label approach of Knockoff.

We set number of source classes to 2, 4, 6, and 8, and an example of combined watermark samples is shown in Figure \ref{fig: examples of watermark samples with different class numbers}. From the results in Table \ref{tab: Performance of different numbers of source classes}, it can be observed that under all settings, watermark success rate on the victim model can reach 100\%, and watermark success rate on the benign model is very low, which indicates the effectiveness and harmlessness of our method. As the number of source classes increases, watermark success rate on the stolen model shows a trend of first increasing and then decreasing. This is because although more source classes can make watermark samples more coupled with the primary task samples in the feature space, when the number of classes is too large, it will also have a negative impact. Specifically, on the one hand, each image from source classes will be resized into an image with low resolution and distorted aspect ratio, which reduces the effective features extracted by the model. On the other hand, the limited query samples provided by adversary make it more difficult to simultaneously cover so many features of watermark source classes. Considering that the combination of four classes can achieve a proportional resizing of the width and height of each image with less resolution loss, and that adversaries are relatively more likely to cover most features of source classes, we set the final number of source classes to four.

\subsection{Discussion on Watermark Sample Filtering Mechanism}
\label{sec: Discussion on Watermark Sample Filtering Mechanism}

\begin{table}[t]
    \centering
    \vspace{-0.0cm}
    \footnotesize
    \begin{tabular}{
        m{1.8cm}<{\centering}
        m{0.5cm}<{\centering}
        m{0.8cm}<{\centering}
        m{1.0cm}<{\centering}
        m{0.9cm}<{\centering}
        m{0.9cm}<{\centering}}
        \toprule
        \textbf{\scriptsize Surrogate Model} & - & \scriptsize AlexNet & \scriptsize VGG-like & \scriptsize ResNet18 & \scriptsize MobileNet \\
        \midrule
        \textbf{\scriptsize WSR} & 94.61 & 98.75 & 98.05 & 97.10 & 97.75 \\
        \bottomrule
    \end{tabular}
    \vspace{-0.2cm}
    \caption{Watermark success rates (\%) on stolen models when using different surrogate models in the filtering mechanism.}
    \label{tab: different surrogate models}
    \vspace{-0.0cm}
\end{table}

\begin{table}[t]
    \centering
    \footnotesize
    \begin{tabular}{
        m{1.55cm}
        m{0.65cm}<{\centering}
        m{0.65cm}<{\centering}
        m{0.7cm}<{\centering}
        m{0.7cm}<{\centering}
        m{0.65cm}<{\centering}
        m{0.65cm}<{\centering}}
        \toprule
        \multirow{2}{*}{\textbf{\scriptsize Method}} & \multicolumn{3}{c}{\textbf{\scriptsize Before Filtering}} & \multicolumn{3}{c}{\textbf{\scriptsize After Filtering}} \\
        \cmidrule(lr){2-4} \cmidrule(lr){5-7}
        & \scriptsize Benign & \scriptsize Victim & \scriptsize Stolen & \scriptsize Benign & \scriptsize Victim & \scriptsize Stolen \\
        \midrule
        \scriptsize Content & 6.39 & 99.22 & 8.00 & 0.00 & 100.00 & 21.45 \\
        \scriptsize Composite & 2.36 & 86.82 & 54.69 & 0.00 & 100.00 & 72.92 \\
        \scriptsize MEA-Defender & 2.01 & 91.82 & 92.15 & 0.00 & 100.00 & 96.33 \\
        \bottomrule
    \end{tabular}
    \vspace{-0.2cm}
    \caption{Comparison of watermark success rates (\%) before and after applying our filtering mechanism to existing watermarking methods.}
    \label{tab: transferability of filtering mechanism}
    \vspace{-0.3cm}
\end{table}

\begin{table*}[t]
    \centering
    \footnotesize
    \begin{threeparttable}
    \begin{tabular}{
        m{2.6cm}
        m{1.6cm}<{\centering}
        m{3.4cm}<{\centering}
        m{2.8cm}<{\centering}
        m{1.0cm}<{\centering}
        m{1.0cm}<{\centering}
        m{1.0cm}<{\centering}
        m{1.0cm}<{\centering}}
        \toprule
        \multirow{2}{*}{\textbf{Task Domain}} & \multirow{2}{*}{\textbf{Data Modality}} & \multirow{2}{*}{\textbf{Dataset}} & \multirow{2}{*}{\textbf{Model Architecture}} & \multicolumn{2}{c}{\textbf{Victim Model}} & \multicolumn{2}{c}{\textbf{Stolen Model}} \\
        \cmidrule(lr){5-6} \cmidrule(lr){7-8}
        & & & & Acc & WSR & Acc & WSR \\
        \midrule
        Speech Recognition & Audio & Speech Commands \cite{warden2018speech} & M5 \cite{dai2017very} & 83.67\% & 100.00\% & 81.49\% & 86.25\% \\
        Text Classification & Text & AG News \cite{zhang2015character} & GPT2 \cite{radford2019language} & 89.96\% & 100.00\% & 85.73\% & 81.70\% \\
        Image Generation & Image & Fashion MNIST \cite{xiao2017fashion} & AutoEncoder \cite{image-generation-model} & 0.7258 & 0.9866 & 0.7092 & 0.7983 \\
        Image Caption & Image+Text & MSCOCO \cite{lin2014microsoft} & ResNet50+LSTM \cite{image-caption-model} & 0.1589 & 0.9377 & 0.1305 & 0.7635 \\
        \bottomrule
    \end{tabular}
    \begin{tablenotes}
        \footnotesize
        \item[1] To evaluate Acc and WSR, we employ different metrics tailored to specific tasks. For speech recognition and text classification, test accuracy is used as the measurement. In image generation task, we rely on SSIM to assess performance, while BLEU-4 is utilized for image caption task.
    \end{tablenotes}
    \end{threeparttable}
    \vspace{-0.2cm}
    \caption{Extensibility evaluation for different deep learning tasks.}
    \label{tab: extensibility evaluation}
    \vspace{-0.0cm}
\end{table*}

In this section, we discuss and evaluate the proposed watermark sample filtering mechanism. Designed as a plug-and-play module, this mechanism operates independently of the watermark model training process. Applied after watermark embedding, it simulates the basic workflow of a model stealing attack (i.e., query-predict-train) using the watermarked model to create a surrogate stolen model. The mechanism then filters watermark verification samples by synergistically leveraging the watermarked model, surrogate stolen model, and a clean model, retaining only the most reliable samples for subsequent verification.

\textbf{Effect of Surrogate Models on Filtering Performance}: To determine whether the filtering mechanism is effective only when the surrogate model architecture matches the stolen model architecture, we conduct experiments on the CIFAR10 task. Here, the adversary uses AlexNet as the stolen model. We evaluate the filtering performance using surrogate models with different architectures, including AlexNet, VGG-like, ResNet18, and MobileNet, as shown in Table \ref{tab: different surrogate models}. Even without applying the filtering mechanism, the watermark success rate on the stolen model reaches 94.61\%, demonstrating that the mechanism functions as an optimization strategy for watermark performance. The results further show that the filtering mechanism is effective across different surrogate model architectures, with the best performance observed when the surrogate model matches the stolen model's architecture.

\textbf{Transferability of the Filtering Mechanism}: As an independent and modular component, the filtering mechanism can be integrated into existing watermarking approaches. Table \ref{tab: transferability of filtering mechanism} illustrates that our filtering mechanism reduces watermark success rates on benign models while enhancing watermark success rates on both victim and stolen models for prior watermarking methods. These findings underscore the strong transferability and broad applicability of our filtering mechanism.

\subsection{Limitations and Future Work}
Our methodological design and evaluation primarily focus on image recognition tasks. Although we have evaluated and demonstrated the effectiveness of extending our approach to other domains, including speech recognition, text classification, image generation, and image captioning in Appendix \ref{sec: Extensibility of Our Method}, the rapid evolution of deep learning continues to introduce diverse models and task types. Consequently, a promising future direction lies in exploring the application of our method to copyright protection for models in emerging and popular domains, such as self-supervised learning for encoders, diffusion models, large language models, and graph-based tasks.


\section{Extensibility of Our Method}
\label{sec: Extensibility of Our Method}
While our study primarily explores model watermarking in the context of image recognition, our approach is agnostic to specific optimization techniques, model architectures, or data modalities. Extending the method to other modalities requires only minor adjustments to the composition of watermark samples, enabling efficient and cost-effective deployment in various domains.

For text data, we leverage the TextRank algorithm \cite{mihalcea2004textrank} to extract key content from four source text classes, creating new combined text samples. These samples are then used for watermark embedding and verification. For audio data, we utilize voice activity detection \cite{tan2020rvad} to extract critical segments and concatenate segments from four source classes to generate composite watermark samples.

To evaluate the extensibility of our method, we apply the proposed DeepTracer framework to protect model ownership across four diverse deep learning scenarios. Table \ref{tab: extensibility evaluation} summarizes the datasets, model architectures, and experimental results. Our approach effectively embeds watermarks into victim models, and even under popular Knockoff stealing attacks, we observe strong watermark signals in the stolen models. Notably, the watermark success rate in stolen models consistently exceeds 80\% of that in the victim models. These findings confirm that our method is highly extensible and adaptable to a wide range of deep learning tasks.

\section{Proof of Stolen Model Forgetting Watermark Task}
\label{sec: Proof of Stolen Model Forgetting Watermark Task}
In a model stealing attack, the adversary learns and obtains a stolen model $M_{stolen}$ with similar functionality to the victim model $M_{victim}$. This occurs because the input samples used by the adversary to query $M_{victim}$ have a distribution similar to the original training samples. To understand this phenomenon, we analyze it from the perspective of data distribution.

Assume the victim model $M_{victim}$ is trained on a dataset $D_{train}$ with a specific feature distribution $P_{train}$. During a model stealing attack, the adversary queries $M_{victim}$ to build the stolen model $M_{stolen}$. The query dataset $D_{query}$ is collected or synthesized by the adversary and has a feature distribution $P_{query}$ that closely resembles $P_{train}$, i.e., $P_{query} \sim P_{train}$. This high similarity enables $M_{stolen}$ to effectively capture the primary task features of $M_{victim}$, achieving functional similarity.

As previously demonstrated in the main body of our paper, since the watermark task introduces external features different from the primary task distribution, its feature distribution $P_{watermark}$ significantly differs from $P_{train}$. Although the over-parameterized deep neural network can simultaneously adapt to both the primary task distribution and watermark task distribution, the watermark task activates a different set of neurons. Furthermore, since the query dataset $D_{query}$ mainly originates from a distribution similar to $P_{train}$, these query samples typically do not include features from $P_{watermark}$. Therefore, the stolen model $M_{stolen}$ mainly learns the primary task functionality and fails to effectively learn the watermark task's features and corresponding activation patterns, which causes the watermark to be forgotten. This ultimately leads to the failure of model owner in copyright verification.

\section{Proof of Guarantees for Our Method}
\label{sec: Proof of Guarantees for Our Method}
It should be noted that since black-box model watermarking is oriented towards practical scenarios where adversary does not disclose the suspicious model and as a result the verifier can only obtain the model's prediction labels, the field is currently limited to proofs of verification guarantees from a probabilistic perspective. In this section, inspired by \cite{jia2021proof,brassard1988minimum,adi2018turning}, we present a theoretical analysis and formal proof of guarantees offered by our approach.

Our watermarking scheme can first be abstracted into three core algorithms:

\textbf{1) Key Generation Algorithm ($KeyGen$)}: This algorithm constructs a secret marking key $M_{k}$, which includes the watermark samples and their corresponding target labels. It also generates a verification key $V_{k}$, which contains the samples and labels used for watermark verification.

\textbf{2) Watermark Embedding Algorithm ($Mark$)}: It utilizes the primary task data and watermark data (i.e., $M_{k}$) to train model $f$, thereby embedding the watermark into the model.

\textbf{3) Watermark Verification Algorithm ($Verify$)}: This algorithm assesses whether a given suspicious model $f_{s}$ contains the watermark. It outputs 1 if the watermark is detected and 0 otherwise. In the context of black-box model watermarking, the presence of watermark is typically determined by evaluating whether the accuracy of watermark verification set (i.e., $V_{k}$) on the model exceeds a predefined threshold $T$. After extensive testing, we set $T=0.2$ in the evaluation.

Based on the above definitions, our watermarking process is formally described as:

\noindent$\textbf{MarkModel()}$:

1. Construct $(M_{k}, V_{k}) \leftarrow KeyGen()$;

2. Generate $f_{w} \leftarrow Mark(f, M_{k})$;

3. Output $(f_{w}, V_{k})$;

4. Compute $Verify(f_{s}, V_{k})$.

Following this, we can clarify that a reliable and guaranteed black-box watermarking method should satisfy at least the following four vital properties:

\noindent$\bullet$ \textbf{Correctness}

The watermark trigger pattern and its associated target output should serve as a form of copyright information (i.e., the watermark), which is successfully embedded within the protected model and can be verified with high probability. Specifically, the three algorithms $(KeyGen, Mark, Verify)$ must function correctly in tandem to satisfy the following requirement:
\begin{equation}
P[Verify(f_{w}, V_{k})=1]=1.
\end{equation}

\noindent$\bullet$ \textbf{Uniqueness}

The watermarked model and trigger pattern should be uniquely related. This means that the probability of detecting the watermark in a non-watermarked model $f_{h}$ or the probability of verifying our watermark with a verification set $V_{h}$ created using a different trigger pattern must be sufficiently low. Formally, we require:
\begin{equation}
\begin{aligned}
&P[Verify(f_h, V_k)=1]\approx0 \\
\text{and}~&P[Verify(f_w, V_h)=1]\approx0.
\end{aligned}
\end{equation}

\noindent$\bullet$ \textbf{Security}

The probability of the watermark trigger pattern being successfully cracked by an adversary should be negligible. We formally define this as:
\begin{equation}
P[M_k \leftarrow Crack(f_w)]\approx0,
\end{equation}
where $Crack$ represents the adversary's cracking algorithm, which attempts to deduce the secret marking key $M_k$ based on the watermarked model $f_w$ and partial known information.

\noindent$\bullet$ \textbf{Robustness}

The watermark must remain verifiable under various attack methods (both model-level and input-level). This condition is formally represented as:
\begin{equation}
\begin{aligned}
&f'_w=Attack(f_w), Verify(f'_w, V_k)=1 \\
\text{and}~&V'_k=Attack(V_k), Verify(f_w, V'_k)=1,
\end{aligned}
\end{equation}
where $Attack$ denotes watermark attack algorithms, which include model modifications and verification sample alterations.

In the following sections, we will provide detailed proofs demonstrating how our watermarking scheme satisfies the aforementioned properties.

\subsection{Correctness of DeepTracer}
Let the training dataset of primary task be $D=\{(x_i, y_i)\}_{i=1}^n$, where $x_i$ represents the sample and $y_i$ denotes the corresponding ground-truth label. Utilizing the watermark sample construction scheme introduced in Section \ref{sec: watermark samples construction}, we first adaptively select four source classes from $D$ that maximize the coverage of the feature space for the primary task. We then construct some unique watermark samples $x_w$ by resizing and concatenating the samples from these four source classes. This ensures that the watermark task is embedded within the same distribution as the primary task, thereby enhancing the coupling between the two tasks during model training.

Let $f_c$ denote the clean model trained solely on the primary task training dataset $D$, with its output on the watermark sample $x_w$ given by $f_c(x_w)$, representing the predicted probability distribution for the watermark sample. The chosen watermark target label $y_w$ satisfies
\begin{equation}
y_w = \mathop{\arg\min}\limits_{y}f_c(x_w)_y,
\end{equation}
i.e., the label with the lowest classification probability in the clean model. This design ensures that the watermark is secret (i.e., the correspondence between watermark samples and target labels has a negligible probability of appearing in a non-watermarked model), minimizing the false positive rate of the watermark as well as increasing the reliability of the subsequent verification process.

Based on this construction scheme, we can generate a sufficient number of watermark samples and their corresponding target labels, which together form the secret marking key $M_k$ in the standard watermarking protocol. The verification key $V_k$, on the other hand, is derived from the primary task test set, using the construction scheme in conjunction with the key sample filtering mechanism introduced in Section \ref{sec: Watermark Key Samples Generation} to select the most effective verification sample set.

Next, we describe how to embed watermark information (i.e., the special mapping between watermark samples and target labels) into the model using the watermark key $M_k$. During the training of the watermarked model $f_w$, the training data consists of both the primary task samples and the watermark samples, and the model is optimized using the loss function $L$ defined in Equation \ref{equ: total loss}. The primary task loss and watermark classification loss encourage the model to predict the probability distribution for each sample that is close to its corresponding label. The same-class coupling loss aids the model in learning compact intra-class representations and separable inter-class features. This loss ensures that the watermark sample and its target class sample are close together in the feature space, thereby enhancing the correlation between the watermark sample and the target label.

Using gradient descent, the model parameters $\theta$ are updated in the direction that minimizes the loss $L$:
\begin{equation}
\theta_{t+1}=\theta_t-\alpha\nabla_\theta L,
\end{equation}
where $\alpha$ is the learning rate.

After enough iterations of training, the model gradually learns the combined feature patterns of the watermark sample $x_w$, making its predictions for $x_w$ closer to the target label $y_w$. This process establishes a one-to-one mapping between $x_w$ and $y_w$ for model $f_w$, i.e.,
\begin{equation}
f_w(x_w)=y_w.
\end{equation}

This correspondence does not originally exist in the model, but emerges only through watermark embedding training, and thus serves as an ownership identifier for the model. After this stage, the watermark has been embedded into the model and can be successfully verified using $V_k$. The evaluation results in Section \ref{sec: Harmlessness and Effectiveness Evaluation} and Appendix \ref{sec: More Challenging Dataset: ImageNet} also demonstrate the correctness of our method from a practical perspective.

\subsection{Uniqueness of DeepTracer}
The uniqueness of our watermarking method is analyzed from two aspects: low false positive rate and low false trigger rate. The former indicates that our method does not judge models that are not embedded in our watermark as its own, while the latter implies that samples that do not belong to our watermark trigger pattern cannot claim ownership of our watermarked model.

\noindent\textbf{(1) Low False Positive Rate}

Consider a model $f_h$ trained on a similar primary task dataset $D'$, potentially augmented with techniques such as CutMix. For a watermark sample $x_w$ constructed by our method, the probability that $f_h$ predicts the target label $y_w$ is given by
\begin{equation}
P[f_h(x_w)=y_w].
\end{equation}

Since $f_h$ has not been specifically trained on our watermark samples, its prediction for $x_w$ is primarily based on the general feature representations learned from its primary task dataset $D'$. Our watermark samples, however, are constructed using a unique combination of four classes, and this specific feature pattern is highly unlikely to be learned during the training of model $f_h$.

We can treat $f_h$ learning the exact combined feature pattern of our watermark sample and the selected target label as two mutually independent events. Specifically, the probability of randomly selecting a combination of four classes from the $K$-class primary task dataset that matches our watermark sample is at most $r_1=1/C_K^4$. Even if a similar combination is selected, the probability that the trainer of $f_h$ will label the combined sample with exactly the target label $y_w$ is similarly low, at most $r_2=1/K$. Therefore, the probability that model $f_h$ learns our watermark is
\begin{equation}
r=r_1 \times r_2=\frac{1}{K\times C_K^4},
\end{equation}
which is an extremely small value close to 0. For example, $r\approx4.76\times10^{-4}$ when $K=10$ and $r\approx2.55\times10^{-9}$ when $K=100$.

Based on the above analysis, we can assume that the prediction result of model $f_h$ for a watermark sample $x_w$ is nearly random, then there is
\begin{equation}
P[f_h(x_w)=y_w]\approx\frac{1}{K}.
\end{equation}

Subsequently, we consider the watermark success rate (WSR) of model $f_h$ on our verification key $V_k$ containing $n$ watermark samples. We first model the assignment of these $n$ samples to the target label $y_w$ as $n$ independent Bernoulli trials. Let the random variable $X$ represent the number of samples assigned to the target label $y_w$. The distribution of $X$ follows a binomial distribution:
\begin{equation}
X \sim B(n, p).
\end{equation}

Our verification key $V_k$ causes ownership false positives on model $f_h$ conditional on the WSR exceeds the detection threshold $T$, i.e., the number of samples classified to $y_w$ exceeds $nT$. According to the probability mass function $P(X=k)=C_n^kp^k(1-p)^{n-k}$ of the binomial distribution, we have
\begin{equation}
\begin{aligned}
P_1(X\geq nT)&=\sum\limits_{k=\lfloor nT \rfloor}^{n}C_n^kp^k(1-p)^{n-k} \\
&=\sum\limits_{k=\lfloor nT \rfloor}^{n}C_n^k\Big(\frac{1}{K}\Big)^k\Big(1-\frac{1}{K}\Big)^{n-k}.
\end{aligned}
\end{equation}

When setting $T=0.2$, it follows from the nature of the binomial distribution that since $p=1/K$ is a very small value and $n$ is sufficiently large, $P_1(X\geq0.2n)$ is an exceptionally low probability. For instance, let $n=2,000$ and $K=10$, then
\begin{equation}
P_1(X\geq 0.2n)=\sum\limits_{k=400}^{2000}C_{2000}^k\Big(\frac{1}{10}\Big)^k\Big(1-\frac{1}{10}\Big)^{2000-k}\approx0.
\end{equation}

Therefore, from a probabilistic standpoint, the likelihood that our verification key $V_k$ incorrectly claims as its own a model that is not embedded in our watermark is negligible, i.e., our watermark method has a low false positive rate. In addition, our experiments in Section \ref{Evaluation of False Positives} provide evidence for the low false positives of our method on a practical level.

\noindent\textbf{(2) Low False Trigger Rate}

Let $x_u$ denote a sample arbitrarily constructed by a user, and the probability that it is predicted as the target label $y_w$ by our watermarked model $f_w$ is
\begin{equation}
P[f_w(x_u)=y_w].
\end{equation}

Our watermark sample $x_w$ is constructed by applying specific operations, such as resizing and concatenating samples from four source classes, resulting in a unique combination pattern and class features. The probability that a randomly generated sample $x_u$ matches the feature pattern and class characteristics of $x_w$ is therefore extremely low. Ideally, for a $K$-class classification task (where $K\geq4$), the probability that $x_u$ is predicted to be $y_w$ by our watermarked model is at most $q=1/K$. Thus
\begin{equation}
P[f_w(x_u)=y_w]\approx\frac{1}{K}.
\end{equation}

Furthermore, similar to our previous proof of false positives, when there exist $n$ verification samples constructed by the user, we can obtain the probability that the user successfully claims ownership of our watermarked model as
\begin{equation}
P_2(X\geq nT)=\sum\limits_{k=\lfloor nT \rfloor}^{n}C_n^k\Big(\frac{1}{K}\Big)^k\Big(1-\frac{1}{K}\Big)^{n-k}.
\end{equation}

Similarly, when setting $T=0.2$, $K=10$, and $n=2,000$, we have
\begin{equation}
P_2(X\geq 0.2n)=\sum\limits_{k=400}^{2000}C_{2000}^k\Big(\frac{1}{10}\Big)^k\Big(1-\frac{1}{10}\Big)^{2000-k}\approx0.
\end{equation}

These results demonstrate that the probability that a user-constructed verification set can claim ownership of our watermarked model is negligible, i.e., our watermark is unlikely to be triggered by mistake.

\subsection{Security of DeepTracer}
Let the primary task dataset $D$ consist of $K$ classes. To crack our watermark, an adversary must correctly identify the four source classes to construct watermark sample $x_w$, in order to further input a sufficient number of watermark samples to our model to test the target label $y_w$. From a combinatorial perspective, the number of ways to select four classes from $K$ classes is given by
\begin{equation}
N_{total}=C_K^4=\frac{K!}{4!(K-4)!}.
\end{equation}

The probability that adversary guesses the marking key $M_k$ is thus
\begin{equation}
P_{crack}=\frac{1}{N_{total}}=\frac{1}{C_K^4}.
\end{equation}

As $K$ increases, $N_{total}$ grows rapidly, and the number of possible combinations an adversary must explore grows exponentially. This combinatorial explosion poses significant challenges to an adversary's ability to successfully identify the watermark, even with considerable computational resources, as the required time and effort would become impractically large. In other words, the probability of adversary successfully guessing the marking key is negligible. For example, when $K=10$, the number of ways to select four source classes is
\begin{equation}
N_{total}=C_{10}^4=\frac{10!}{4!\times6!}=210.
\end{equation}

Thus the probability that adversary eventually guesses $M_k$ is only
\begin{equation}
P_{crack}=\frac{1}{210}\approx4.76\times10^{-3}.
\end{equation}

When $K$ increases, $P_{crack}$ will become smaller. For instance, when $K=100$, $P_{crack}\approx2.55\times10^{-7}$. While similarities between certain classes may reduce the search space somewhat, the number of potential combinations remains prohibitively large. Thus the increase in $P_{crack}$ is particularly small.

Furthermore, if the adversary is unaware of how many source classes the watermark samples consist of and exactly how they are combined, the search space becomes even more complex and vast. In this case, the adversary would need to attempt all possible combinations of 1 to $K$ source classes, further expanding the search space and making it virtually impossible to recover the watermark trigger pattern within a reasonable time frame. That is, the probability of adversary cracking out the marking key $M_k$ will be lower.

\subsection{Robustness of DeepTracer}
Our watermarking method is designed with careful consideration of various factors, such as feature selection from watermark samples, feature entanglement in the representation space, and the reliability of verification key samples. These design choices greatly enhance the coupling between the primary task and watermark task, making it difficult for an adversary to successfully remove the watermark via model modification attacks (e.g., model stealing, fine-tuning, pruning) without significantly degrading model performance. A detailed analysis and proof of these aspects are provided in Sections \ref{sec: analysis of watermark survival} and \ref{sec: proposed method}. The evaluation results in Section \ref{sec: Robustness against Model Stealing Attacks}, Section \ref{sec: Robustness against Watermark Removal Attacks}, Section \ref{sec: Robustness against Adaptive Attacks}, and Appendix \ref{sec: Robustness against Various Scenarios} demonstrate the robustness of our watermarking under model-level attacks.

Additionally, the distinctive combination patterns of our watermark samples are minimally impacted by common sample transformations, including gaussian blur, gaussian noise, image quantization, and image cropping. This ensures that it is infeasible for an adversary to try to disrupt the watermark verification using input preprocessing attacks without impacting the primary task performance. The experimental results in Appendix \ref{sec: Robustness against Watermark Detection and Evasion Attacks} provide practical evidence of this robustness.

$\bigstar$ In summary, through the theoretical analysis and proofs presented above, we have shown that our proposed backdoor-like watermarking method offers strong guarantees in terms of correctness, uniqueness, security, and robustness. This method provides an effective means of protecting model ownership and has significant practical value, offering a reliable technical solution for intellectual property protection in deep learning models.

\end{document}